\RequirePackage{fix-cm}
\documentclass[smallextended, natbib]{svjour3}       
\smartqed  
\usepackage{graphicx}
\usepackage{mathptmx}      
\usepackage{amssymb}
\usepackage{amsmath}
\usepackage{xcolor}
\usepackage{media9}
\usepackage{hyperref}
\def \beq{\begin{equation}}
\def \eeq{\end{equation}}
\def \bear{\begin{eqnarray}}
\def \ear{\end{eqnarray}}
\def \curlB {\vec{\nabla}\times \vec{B}}

\newcommand{\dd}{\mathrm{d}}

\newcommand{\gcc}{\mbox{g~cm$^{-3}$}}

\newcommand{\dive} {\vec{\nabla}\cdot}

\newcommand{\derparn}[2] {\frac{\partial #2}{\partial #1}}

\newcommand{\de} {{\rm d}}

\journalname{Living Reviews in Computational Astrophysics}

\begin{document}

\title{Magnetic, thermal and rotational evolution of isolated neutron stars}



\author{Jos\'e A. Pons \and
        Daniele Vigan\`o
}


\institute{J. A. Pons \at 
  Departament de F\'{\i}sica Aplicada \at
  Universitat d'Alacant, Spain \\
  \email{jose.pons@ua.es}
  \and
  D. Vigan\`o \at
  Departament  de  F\'{\i}sica \at Universitat  de  les  Illes Balears, \\
  Institut d'Estudis Espacials de Catalunya and Institut d'Aplicacions
  Computacionals de Codi Comunitari (IAC3), \\
  Palma de Mallorca, Baleares E-07122, Spain
}

\date{Received: date / Accepted: date}

\maketitle

\begin{abstract}
The strong magnetic field of neutron stars is intimately coupled to the observed temperature and spectral properties, as well as to the observed timing properties (distribution of spin periods and period derivatives). Thus, a proper theoretical and numerical study of the magnetic field evolution equations, supplemented with detailed calculations of microphysical properties (heat and electrical conductivity, neutrino emission rates) is crucial to understand how the strength and topology of the magnetic field vary as a function of age, which in turn is the key to decipher the physical processes behind the varied neutron star phenomenology.  In this review, we go through the basic theory describing the magneto-thermal evolution models of neutron stars, focusing on numerical techniques, and providing a battery of benchmark tests to be used as a reference for present and future code developments. We summarize well-known results from axisymmetric cases, give a new look at the latest 3D advances, and present an overview of the expectations for the field in the coming years.

\end{abstract}

\newpage
\setcounter{tocdepth}{3}
\tableofcontents

\section{Introduction}
\label{sec:1}

Neutron stars (NSs), the endpoints of the evolution of massive stars, are fascinating astrophysical sources that display  a bewildering variety of manifestations.  They are arguably the only stable environment in the present Universe where extreme physical conditions of density, temperature, gravity, and magnetic fields, are realized simultaneously. Thus, they are ideal laboratories to study the properties of matter and the surrounding plasma under such extreme limits. 
NSs were first discovered as rotation-powered radio {\it pulsars} (standing for pulsating stars, due to their periodic signal), sometimes called {\it standard pulsars}, the most numerous class with about three thousand identified members.\footnote{See the online Australia Telescope National Facility catalog, \url{http://www.atnf.csiro.au/research/pulsar/psrcat/}} The number is continuously increasing thanks to new extended surveys and the use of high-sensitivity instruments like LOFAR \citep{lofar13}, and a few more thousand sources are expected to be observed by the soon-available Square Kilometre Array.
To a lesser extent, NSs have also been observed in $X$ rays (about one hundred NSs so far), as persistent or transient sources, and/or as $\gamma$-ray pulsars (over two hundred and fifty so far). In most cases, this high-energy radiation is non-thermal, originated by particle acceleration (synchro-curvature emission, \citealt{zhang97,vigano15}) or Compton up-scattering of lower-energy photons 
by the particles composing the magnetospheric plasma \citep{lyutikov06}. 

A particularly intriguing class of  isolated NSs are the {\it magnetars} \citep{2015SSRv..191..315M,2015RPPh...78k6901T,2017ARA&A..55..261K}, relatively slow rotators with typical spin periods of several seconds and ultra-strong magnetic fields ($10^{13}$--$10^{15}$ G). In most cases, they show a relatively high persistent (i.e., constant over many years) X-ray luminosity ($L_x \approx 10^{33}$--$10^{35} $ erg/s), well exceeding their rotational energy losses, in contrast with radio (standard) and $\gamma$-ray pulsars. This leads to the conclusion that the main source of energy is provided by the strong magnetic field, instead of rotational energy, in agreement with the high values of the surface dipolar magnetic field  inferred from the timing properties.
Magnetars are also identified for their complex transient phenomenology in high energy X-rays and $\gamma$-rays, including short (tenths of a second) bursts, occasional energetic outbursts with months-long afterglows \citep{2011ASSP...21..247R,coti18} and, much more rarely (only three observed so far), giant flares \citep{hurley99,palmer05}. During giant flares, the energy release is as large as $10^{46}$ erg in less than a second. The source of energy of such transient, violent behavior is also generally agreed to be of magnetic origin, as proposed in \citealt{1995MNRAS.275..255T,1996ApJ...473..322T}. Alternative or complementary power sources, such as accretion, nuclear reactions, or residual cooling from the interior, are less effective to account for the transient activity.

Although isolated NSs have been historically differentiated in sub-classes, mostly based on observational grounds (detectability in X and/or radio,  
transient vs. persistent properties, and presence/absence of pulsations), there is no sharp boundary 
between classes, and the distributions of their physical properties, such as the inferred magnetic field, partially overlap.
Indeed, the evidence accumulated in the last decade has shown that the presence of a strong dipolar field is not a sufficient condition to trigger observable magnetar-like 
events and, conversely, there has been an increasing number of {\it low-magnetic-field magnetars} discovered in the recent past \citep{2010Sci...330..944R,2012ApJ...754...27R,2014ApJ...781L..17R,2013ApJ...770...65R}. They are NSs with relatively low values of the inferred surface dipolar magnetic fields, showing nevertheless magnetar-like activity. Similar activity has been displayed by a couple of {\it high-magnetic-field radio pulsars} with inferred $B\sim 10^{13}$ G \citep{gavriil08,gogus16}, and by a puzzling young, extremely slowly spinning NS \citep{rea16}, belonging to the so-called sub-class of  {\it central compact objects}, a handful of young NSs surrounded by a supernova remnant, detectable due to a persistent, mostly non-pulsating X-ray emission \citep{deluca17}. 

It seems now clear that the non-linear, dynamical interplay between the internal and external magnetic field evolution plays a key role to understand the observed
phenomenology, and their study requires numerical  simulations. Particularly important issues are the transfer of energy between toroidal and poloidal 
components and between different scales, the location and distribution of long-lived electrical currents within the star,  how magnetic helicity can be generated and transferred to the exterior to sustain magnetospheric currents (i.e., how to twist the magnetic field lines), and how instabilities leading to outbursts and flares are triggered.
In order to answer all these questions, 2D and 3D numerical simulations are required. The problem is similar to other scenarios in plasma physics or solar physics, but with extreme conditions and additional ingredients (strong gravity and possibly superconductivity).
The goal of this paper is to provide an overview of the subject of modeling NS evolution accessible not only to specialists on the subject, 
but to a wider community including astrophysicists in general, and particularly students. For this purpose, we will review the basic equations and the numerical techniques
applied to each part of the problem, with a special focus on the distinctive features of NSs, compared to other stellar sources.

This work is organized as follows.
In  Sect.~\ref{sec:cooling}  the theory of the cooling of NSs is reviewed;  the magnetic field evolution is described in detail in Sect.~\ref{sec:magnetic_evolution}, where we discuss the physical processes in different parts of the star. In Sect.~\ref{sec:magnetic_methods} we review the specific numerical methods and techniques used to model the magnetic evolution. They can be implemented and tested with the benchmark cases presented in Sect.~\ref{sec:tests}. In Sect.~\ref{sec:magnetosphere} we discuss the challenging coupling between the slowly evolving interior and the force-free magnetosphere, and how it determines the evolution of the spin period. Some examples of realistic evolution models from the recent literature are presented in Sect.~\ref{sec:examples}. Finally, in Sect.~\ref{sec:conclusions} we comment on future developments and open issues.

\section{Neutron star cooling}
\label{sec:cooling}

For a few tens of isolated NSs, the detected X-ray spectra show a clear thermal contribution directly originated from a relatively large fraction of the star surface. For the cases in which an independent estimate of the star age is also available, one can study how temperatures correlate with age, which turns out to be an indirect method to test the physics of the NS interior.
The evolution of the temperature in a NS was theoretically explored even before the first detections, in the 1960s \citep{tsuruta64}. Today, \emph{NS cooling} is the most widely accepted terminology for the research area studying how temperature evolves as NSs age and their observable effects. We refer the interested reader to the introduction in a
recent review  \citep{potekhin_rev15a} for a thorough historical overview of the foundations of the NS cooling theory. 

According to the standard theory, a proto-NS is born as extremely hot and liquid, with $T\gtrsim 10^{10}$ K, and a relatively large radius, $\sim 100$ km. Within a minute, it becomes transparent to neutrinos and shrinks to its final size, $R\sim 12$ km \citep{burrows86,keil1995,pons99}. Neutrino transparency marks the  starting point of the \emph{long-term cooling}. At the initially high temperatures, there is a copious production of thermal neutrinos that abandon the NS core draining energy from the interior. In a few minutes, the temperature drops by another order of magnitude to $T \sim 10^{9}$ K, below the melting point of a layer where matter begins to crystallize,  forming the crust. Since the melting temperature depends on the local value of density, the gradual growth of the crust takes place from hours to months after birth. The outermost layer (the envelope, sometimes called the ocean) with a typical thickness ${\cal O}(10^2~ {\rm m})$,  remains liquid and possibly wrapped by a very thin ${\cal O({\rm cm})}$ gaseous atmosphere. In the inner core, a mix of neutrons, electrons, protons and plausibly more exotic particles (muons, hyperons, or even deconfined quark matter), the thermal conductivity is so large that the dense core quickly becomes  isothermal. 
 
The central idea of NS cooling studies is to produce
realistic evolution models that, when confronted with  observations of the thermal emission of NSs with different ages
\citep{2004ApJS..155..623P,2004ARA&A..42..169Y,2008AIPC..983..379Y,2009ASSL..357..247P,2009ASSL..357..289T,potekhin_rev15b}, 
provide useful information about the chemical composition, the magnetic field strength and topology of the regions where this radiation is produced,  or even the properties of matter at higher densities deeper inside the star.
Two interesting examples are the low temperature (and thermal luminosity) shown by the Vela pulsar, arguably a piece of evidence for fast neutrino emission associated to higher central densities or exotic matter, or the controversial observational evidence for fast cooling of the supernova remnant in 
Cassiopeia A \citep{2010ApJ...719L.167H,posselt18},  proposed to be a signature of the core undergoing a superfluid transition \citep{2011PhRvL.106h1101P,2011MNRAS.412L.108S,2015PhRvC..91a5806H,wijngaarden19}. 
We now review the theory of NS cooling, beginning with a brief revision of the stellar structure equations and by introducing notation.

\subsection{Neutron star structure}

The first NS cooling studies (and most of the recent works too) considered a spherically symmetric 1D background star, in part for simplicity, and in part motivated by the small deviations expected. The matter distribution can be assumed to be spherically symmetric to a very good approximation, except for the extreme (unobserved) cases of structural deformations due to spin values close to the breakup values ($P \lesssim 1$ ms) or ultra-strong magnetic fields ($B \gtrsim 10^{18}$~G, unlikely to be realized in nature). Therefore, using spherical coordinates $(r,\theta,\varphi)$, the space-time structure is accurately described by the Schwarzschild metric
\begin{equation}
\dd s^2 = - \mathrm{e}^{2\nu(r)} c^2 \dd t^2 + \mathrm{e}^{2\lambda(r)} \dd r^2 
+ r^2 (\dd \theta^2 + \sin^2\theta \dd \varphi^2),
\label{Schw}
\end{equation}
where  $ \lambda(r) = - \frac{1}{2} \ln \left[ 1-  \frac{2 G}{c^2} \frac{m(r)}{r^2}\right]$ accounts for the space-time curvature, $$m(r)=4\pi\int_0^r\rho(\tilde{r})\tilde{r}^2\dd \tilde{r}$$ 
is the gravitational mass inside a sphere of radius $r$, $\rho$ is the mass-energy density, $G$ is the gravitational constant, and $c$ is the speed
of light. The \emph{lapse} function $e^{2\nu(r)}$ is determined by the equation
\begin{equation}
   \frac{\dd \nu(r)}{dr} =  \frac{G}{c^2} \frac{m(r)}{r^2}\,
      \left( 1 + \frac{4\pi r^3 P}{c^2 m(r)} \right)\,
      \left( 1 - \frac{2 G}{c^2} \frac{m(r)}{r} \right)^{-1}, 
\end{equation}
with the boundary condition $\mathrm{e}^{2\nu(R)}=1-2GM/c^2R$
at the stellar radius $r=R$.  Here, $M\equiv m(R)$ is the total gravitational mass of the star.
The pressure profile, $P(r)$, is determined by  the Tolman-Oppenheimer-Volkoff equation
\begin{equation}
\frac{\dd P(r)}{dr} =   - \left(\rho +\frac{P}{c^2}\right)   \frac{\dd \nu(r)}{dr}. 
\label{Phi}
\end{equation}
Throughout the text, we will keep track of the metric factors for consistency, unless indicated. The Newtonian limit can easily be recovered by setting $\mathrm{e}^{\nu}=\mathrm{e}^{\lambda}=1$ in all equations.

To close the system of equations, one must provide the equation of state (EoS), i.e., the dependence of the pressure on the other variables $P=P(\rho,T, Y_i)$ ($Y_i$ indicating the particle fraction of each species).
Since the Fermi energy of all particles is much higher than the thermal energy (except in the outermost layers) the dominant contribution is given by degeneracy pressure. The thermal and magnetic contributions to the pressure, for typical conditions, are negligible in most of the star volume. Besides, the assumptions of charge neutrality and beta-equilibrium uniquely determine the composition at a given density. Thus, one can assume an effective barotropic EoS, $P=P(\rho)$, to calculate the background mechanical structure. Therefore, the radial profiles describing the energy-mass density and chemical composition can be calculated once and kept fixed as a background star model for the thermal evolution simulations. 

\begin{figure}
	\centering
	\includegraphics[width=0.7\textwidth]{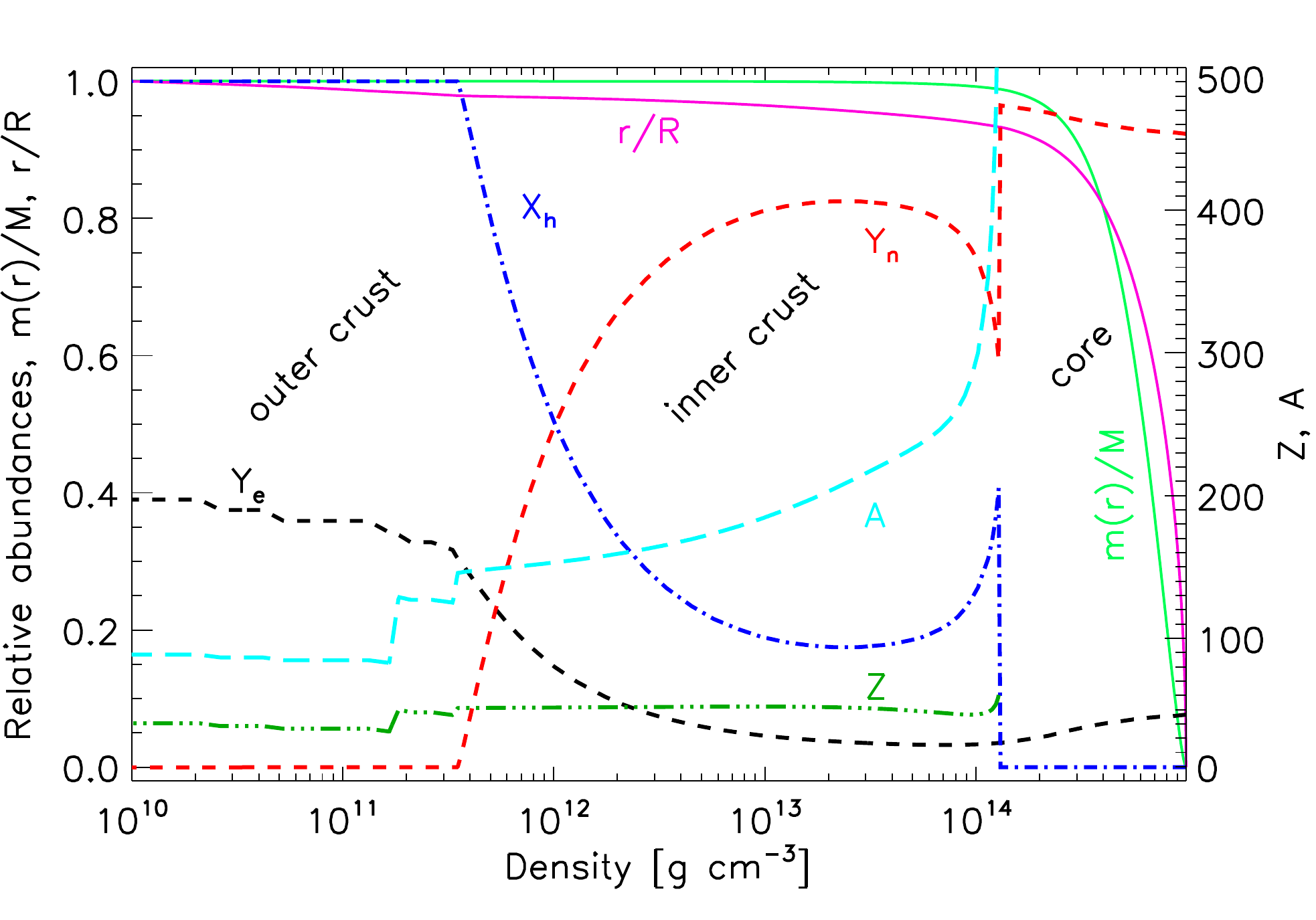}
	\caption{Structure and composition of a $1.4\,M_\odot$ NS, with SLy EoS. The plot shows, as a function of density from the outer crust to the core, the following quantities: mass fraction in the form of nuclei $X_h$ (blue dot-dashed line), the fraction of electrons per baryon $Y_e$ (black dashes), the fraction of free neutrons per baryon $Y_n$ (red dashes), the atomic number $Z$ (dark green triple dot-dashed), the mass number $A$ (cyan long dashes), radius normalized to $R$  (pink solid), and the corresponding enclosed mass normalized to the star mass (green solid).
}
	\label{fig:ns_profile}
\end{figure} 
 In Fig.~\ref{fig:ns_profile} we show a typical profile of a NS, obtained with the EoS SLy4 \citep{douchin01}, which is among the realistic EoS supporting a maximum mass compatible with the observations, $M_{\max}\sim 2.0$--$2.2\,M_\odot$ \citep{demorest10,antoniadis13,margalit17,ruiz18,radice18,cromartie19}. We show the enclosed radius and mass, and the fractions of the different components, as a function of density, from the outer crust to the core. For densities $\rho\gtrsim 4\times 10^{11}~\gcc$, neutrons drip out the nuclei and, for low enough temperatures, they would become superfluid. Note that the core contains about 99\% of the mass and comprises 70--90\% of the star volume (depending on the total mass and EoS). Envelope and atmosphere are not represented here. For a more detailed discussion we refer to, e.g., \cite{2007ASSL..326.....H,potekhin_rev15a}.

\subsection{Heat transfer equation}

Spherical symmetry was also assumed  in most NS cooling studies during the 1980s and 1990s.
However, in the 21st century, the unprecedented amount of data collected by soft X-ray observatories such as Chandra and XMM-Newton, provided evidence that  most nearby NSs whose thermal emission is visible in the X-ray band of the electromagnetic spectrum show some anisotropic temperature distribution \citep{haberl,2007Ap&SS.308..171P,2011ApJ...736..117K} . 
This observational evidence made clear the need to build multi-dimensional models
 and gave a new impulse to the development of the cooling theory including 2D effects \citep{geppert04,geppert06,2007Ap&SS.308..403P,aguilera08a,aguilera08b,vigano13}. The cooling theory builds upon the heat transfer equation, which includes both flux transport and source/sink terms.

\begin{figure}[t]
	\centering
	\includegraphics[width=0.45\textwidth]{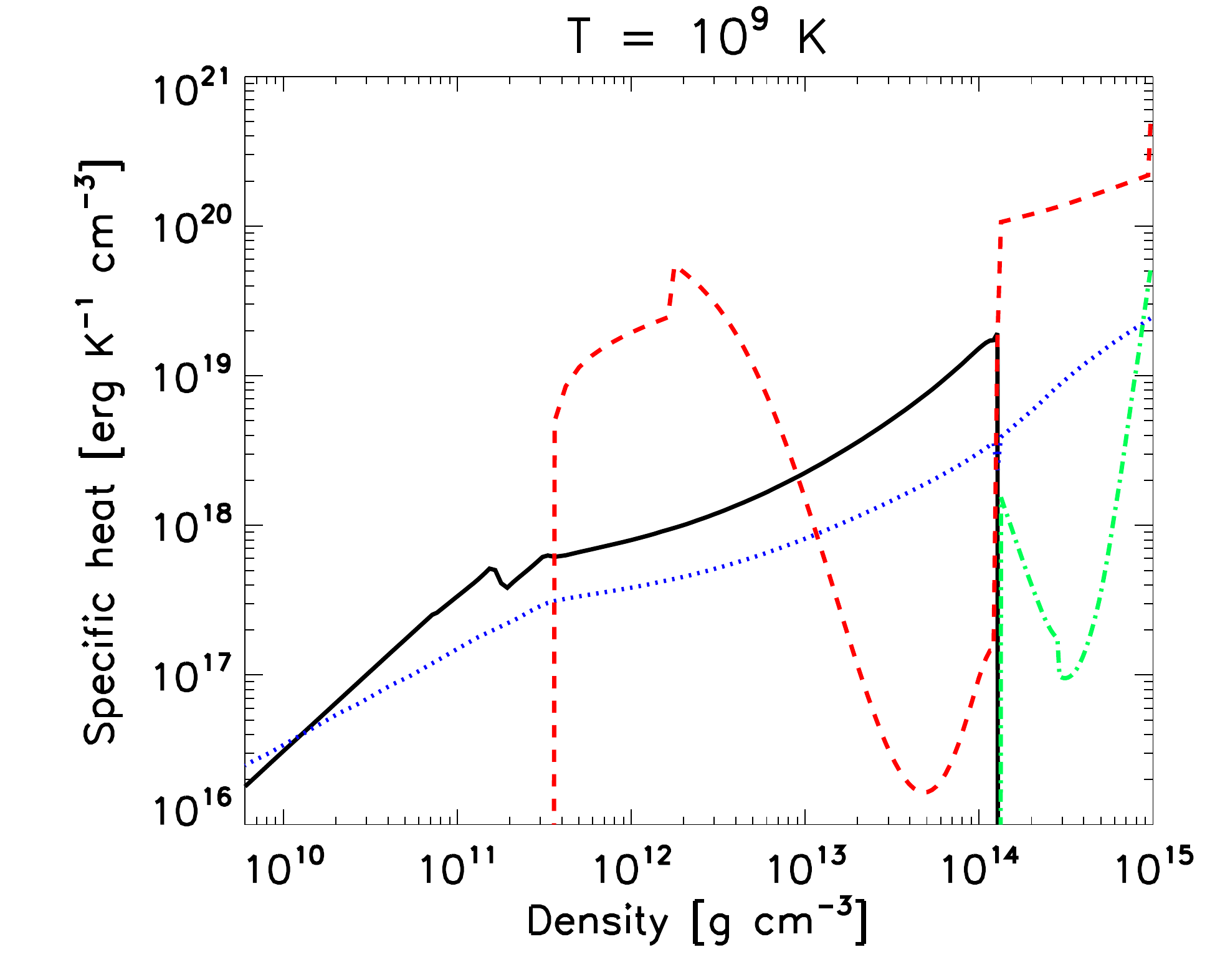}
	\includegraphics[width=0.45\textwidth]{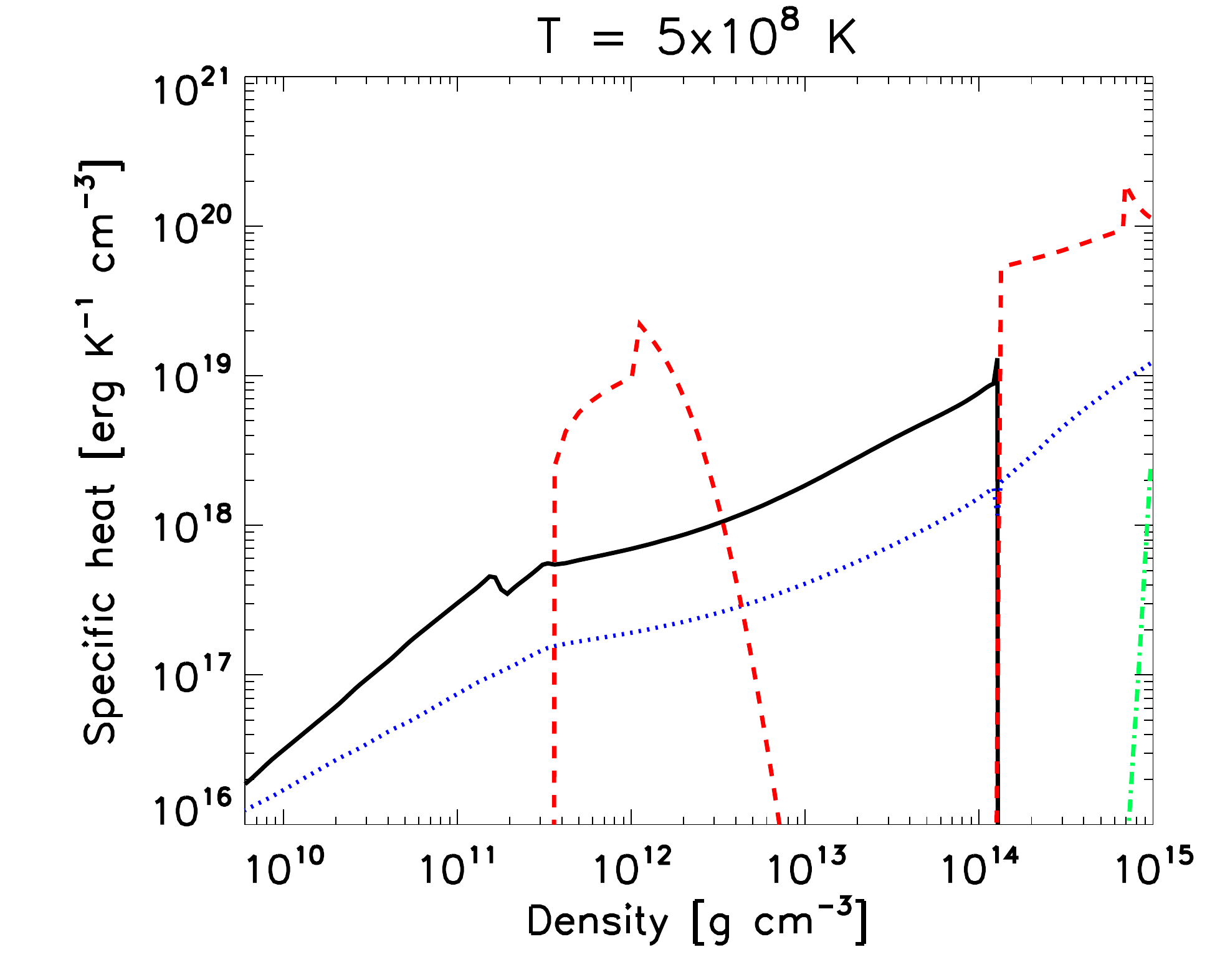}\\
	\includegraphics[width=0.45\textwidth]{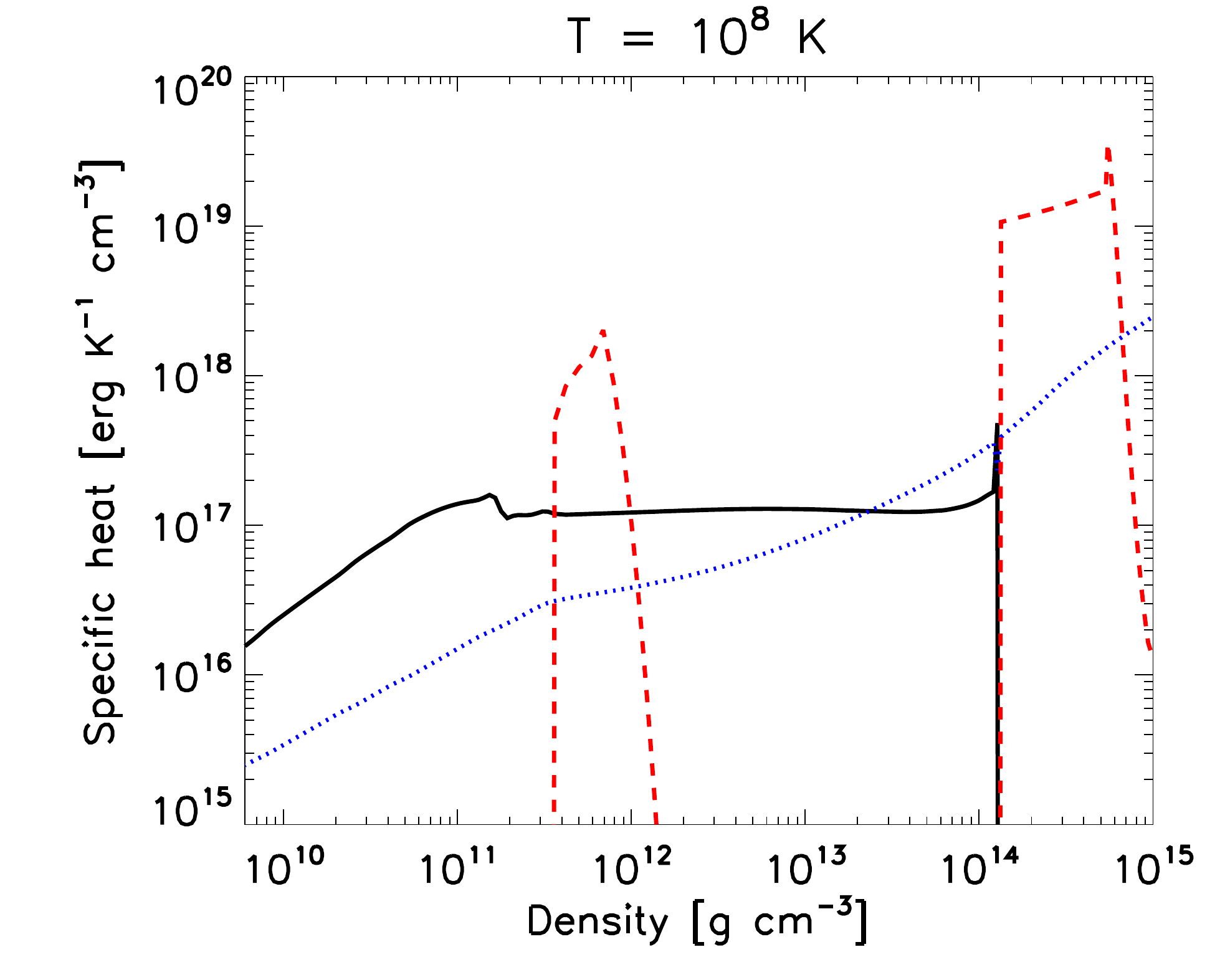}
	\includegraphics[width=0.45\textwidth]{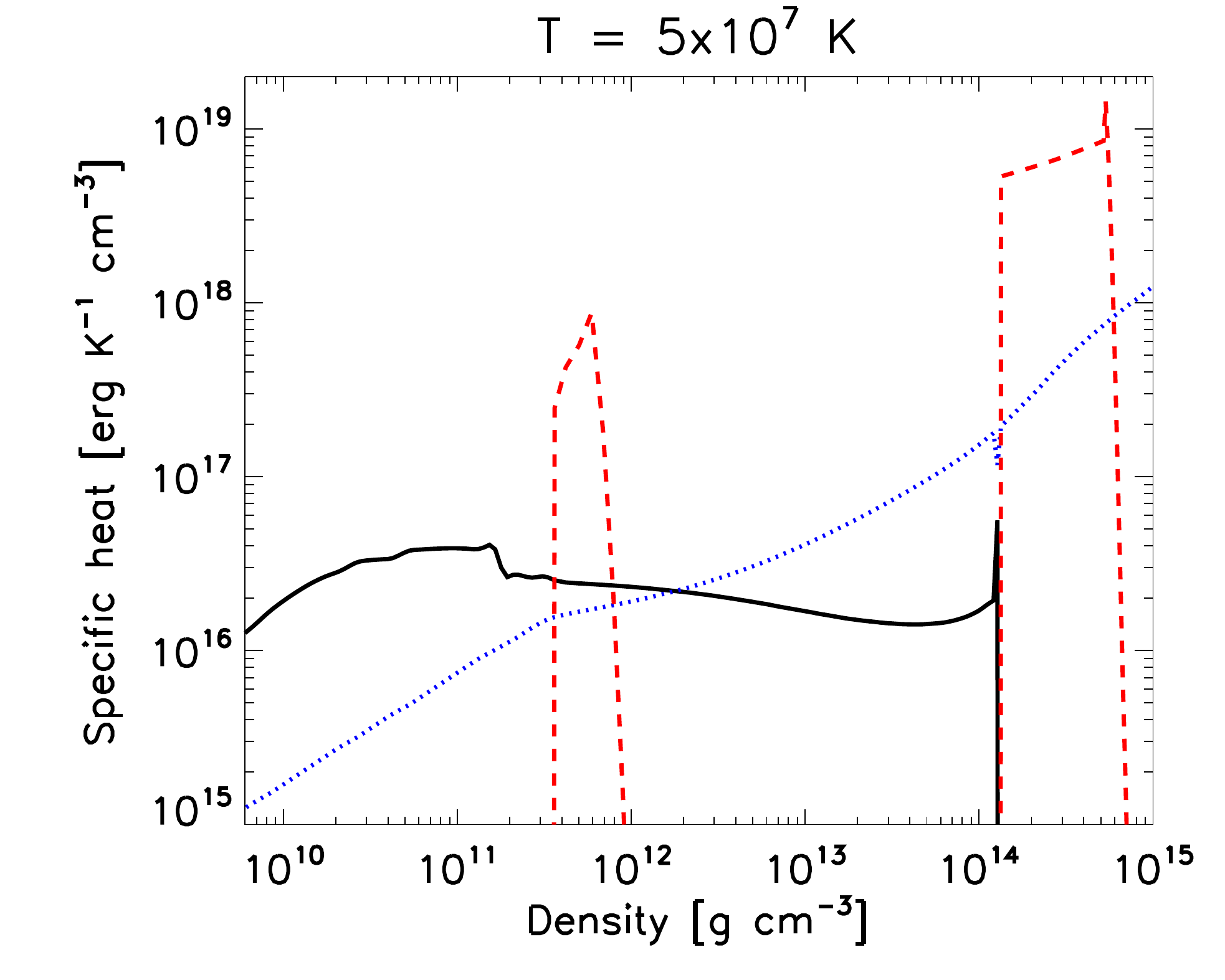}
	\caption{Contributions to the specific heat from neutrons (red dashes), protons (green dot-dashed), electrons (blue dots), and ions (black solid line) as a function of density, from the outer crust to the core, and for different temperatures in each panel (as indicated). The superfluid gaps employed are the same as in \citep{Ho2012}. }
	\label{fig:cv}
\end{figure}

The equation governing the temperature evolution at each point of the star's interior reads:
\begin{equation}
  c_\mathrm{v}\,\frac{\partial (T \mathrm{e}^\nu)}{\partial t}
      + \nabla\cdot(\mathrm{e}^{2\nu} \vec{F}) = 
           \mathrm{e}^{2\nu} (H - Q)~,
\label{Tbalance}
\end{equation}
where $c_\mathrm{v}$ is specific heat, and the source term is given by the neutrino emissivity $Q$ (accounting for energy losses by neutrino emission), and the heating
power per unit volume $H$, both functions of temperature, in general. The latter can include contributions from accretion and, more relevant for this paper, Joule heating by magnetic field dissipation. All these quantities (including the temperature) vary in space and are measured in the local frame, with the metric (redshift) corrections accounting for the change to the observer's frame at infinity.\footnote{Throughout the text, we will use the $\nabla$ operator for conciseness, but we note that it must include the metric factors, e.g., using the metric (\ref{Schw}), the gradient
would be $\nabla\equiv \left( \mathrm{e}^{-\lambda} \frac{\partial}{\partial r},\, \frac{1}{r} \frac{\partial}{\partial  \theta},\,  \frac{1}{r \sin\theta} \frac{\partial}{\partial \varphi} \right).$}

The heat flux density $\vec{F}$ is given by 
\begin{equation}
 \vec{F} = - \mathrm{e}^{-\nu} \hat{\kappa}\cdot\nabla(\mathrm{e}^\nu T)~,
\end{equation}
with $\hat{\kappa}$ being the thermal conductivity tensor.  In Fig.~\ref{fig:cv} we show the different contributions to the specific heat by ions, electrons, protons and neutrons, for $T=\{10,5,1,0.5\} \times 10^8$ K, respectively, computed again with SLy EoS. For the superfluid/superconducting gaps we use the phenomenological formula for 
the momentum dependence of the energy gap at zero temperature employed in \cite{Ho2012}, in particular their \emph{deep neutron triplet} model.

The bulk of the total heat capacity of a NS is given by the core, where most of the mass is contained. The regions with superfluid nucleons are visible as deep drops of the specific heat. The proton contribution is always negligible. Neutrons in the outer core are not superfluid, thus their contribution is dominant.
The crustal specific heat is given by the dripped neutrons, the degenerate electron gas and the nuclear lattice \citep{vanriper91}. The specific heat of the lattice is generally the main contribution, except in parts of the inner crust where neutrons are not superfluid, or for temperatures $\lesssim 10^8$ K, when the electron contribution becomes dominant. In any case, the small volume of the crust implies that its heat capacity is small in comparison to the core contribution. For a detailed computation of the specific heat and other transport properties, we recommend the codes publicly available at \url{http://www.ioffe.ru/astro/EIP/}, describing the 
EoS for a strongly magnetized, fully ionized electron-ion plasma \citep{potekhin10}.

The second ingredient needed to solve the heat transfer equation is the thermal conductivity (dominated by electrons, due to their larger mobility). For weak magnetic fields, the conductivity is isotropic: the tensor becomes a scalar quantity times the identity matrix. Since the background is spherically symmetric, at first approximation, the temperature gradients are essentially radial throughout most of the star. In this limit, 1D models are accurately representing reality, at least in the core and inner crust.
However, for strong magnetic fields (needed to model magnetars), the electron thermal conductivity tensor becomes anisotropic also in the crust: in the direction perpendicular to the magnetic field the conductivity is strongly suppressed, which reduces the heat flow orthogonal to the magnetic field lines. 

In the relaxation time approximation,
the ratio of conductivities parallel ($\kappa^\parallel$) and orthogonal ($\kappa^\perp$)  to the magnetic field is
\begin{equation}\label{eq:omegatau}
\frac{\kappa^\parallel}{\kappa^\perp}  \approx 1 + (\omega_B \tau_e)^2~.
\end{equation}
Here we have introduced the so-called \emph{magnetization parameter} \citep{1980SvA....24..425U}, $\omega_B \tau_e$, where $\tau_e$ is the electron relaxation time and $\omega_B = eB/m^*_ec$ is the gyro-frequency of electrons with charge $-e$ and  effective mass $m^*_e$ moving in a magnetic field with intensity $B$.
Equation~(\ref{eq:omegatau}) is only strictly valid in the classical approximation (see \citealt{potekhin2018} for a recent
discussion of quantizing effects), but this dimensionless quantity is always a good indicator of the suppression of the thermal conductivity
in the transverse direction. We will see later that this is also the relevant parameter to discriminate between different regimes for the magnetic field evolution.

\begin{figure}[ht]
	\centering
	\includegraphics[width=0.49\textwidth]{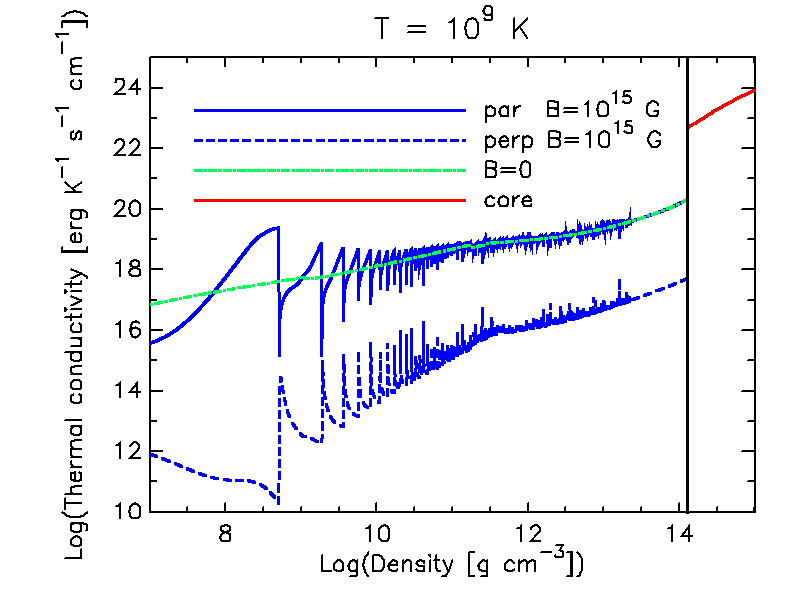}
	\includegraphics[width=0.49\textwidth]{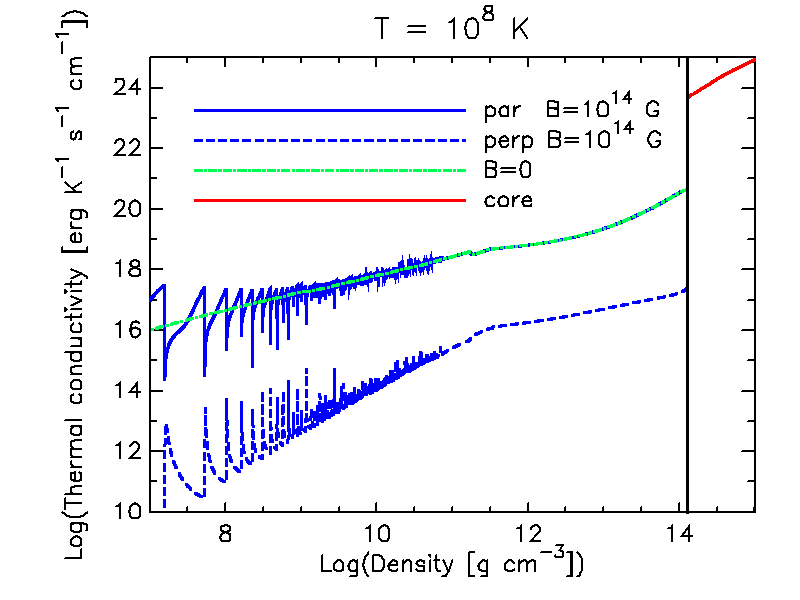}
	\caption{Thermal conductivity in the directions parallel (solid lines) and perpendicular (dashes) to the magnetic field, including quantizing effects. We show the cases $T=10^9$ K, $B=10^{15}$ G (left panel) and $T=10^8$ K, $B=10^{14}$ G (right panel). For comparison, the $B=0$ values are shown with green lines in both figures.}
	\label{fig:cond_class_quant}
\end{figure}

Figure~\ref{fig:cond_class_quant} shows the thermal conductivity including the contributions of all relevant carriers, for two different combinations of temperatures and magnetic field, roughly corresponding to a recently born magnetar ($T=10^9$ K, $B=10^{15}$ G), or after $\sim 10^4$ yr ($T=10^8$ K, $B=10^{14}$ G). Note that the thermal conductivity of the core is several orders of magnitude higher than in the crust, which results in a nearly isothermal core.  
Thus, the precise value of the core thermal conductivity becomes unimportant, and thermal gradients can only be developed and maintained in the crust and the envelope.
In the crust, the dissipative processes responsible for the finite thermal conductivity include all the mutual interactions between electrons, lattice phonons (collective motion of ions in the solid phase), impurities (defects in the lattice), superfluid phonons (collective motion of superfluid neutrons) or normal neutrons. The mean free path of free neutrons, which is limited by the interactions with the lattice, is expected to be much shorter than for the electrons, but a fully consistent calculation is yet to be done \citep{chamel08}.
Quantizing effects due to the presence of a strong magnetic field become important only in the envelope, or in the outer crust for very large magnetic fields ($B \gtrsim 10^{15}$ G). 
For comparison, we also plot the $B=0$ values. The quantizing effects are visible as oscillations around the classical (non-magnetic) values, corresponding to the gradual filling of Landau levels.
More details about the calculation of the microphysics input ($\hat{\kappa}, c_v, Q$) can be found in Sect.~2 of \cite{potekhin_rev15a}. 

We can understand how and where anisotropy becomes relevant by considering electron conductivity in the presence of a strong magnetic field (and for now, ignoring quantizing effects).
The heat flux is then reduced to the compact form \citep{perez06}: 
\begin{equation}
\vec{F} = -  e^{-\nu} \kappa^\perp \left[ \vec{\nabla} (e^\nu T) + (\omega_B \tau_e)^2 (\vec{b}\cdot  \vec{\nabla} (e^\nu T)) \vec{b} + \omega_B \tau_e (\vec{b} \times  \vec{\nabla} (e^\nu T)) \right]~,
\label{hflux}
\end{equation}
where $\vec{b} \equiv \vec{B}/B$ is the unit vector in the local direction of the magnetic field. The heat flux is thus explicitly decomposed in three parts: heat flowing in the direction of the redshifted temperature gradient, $\vec{\nabla} (e^\nu T)$, heat flowing along magnetic field lines (direction of $\vec{b}$), and heat flowing in the direction perpendicular to both.

In the low-density region (envelope and atmosphere), radiative equilibrium will be established much faster than the interior evolves. 
The difference by many orders of magnitude of the thermal relaxation timescales 
between the envelope and the interior (crust and core)  makes computationally unpractical to perform cooling simulations in a numerical grid including all layers up to the star surface.
Therefore, the outer layer is effectively treated as a boundary condition. It relies on a separate calculation of stationary envelope models to obtain a functional fit giving a relation between the surface temperature $T_s$, which determines the radiation flux, and the temperature $T_b$ at the crust/envelope boundary. This $T_s - T_b$ relation provides the outer boundary condition to the heat transfer equation. The radiation from the surface is usually assumed to be blackbody radiation, although the alternative possibility of more elaborated atmosphere models, or anisotropic radiation from a condensed surface, have also been
studied \citep{turolla04,vanadelsberg05,perez05,potekhin12}.
A historical review and modern examples of such envelope models are discussed in Sect.~5 of \cite{potekhin_rev15a}. Models include different values for the curst/envelope boundary density, 
magnetic field intensity and geometry,  and chemical composition (which is uncertain).

The first 2D models of the stationary thermal structure in a realistic context (including the comparison to 
observational data) were obtained by \cite{geppert04,geppert06} and \cite{perez06}, paving the road for subsequent 2D simulations of the time evolution of
temperature in strongly magnetized NS \citep{aguilera08a,aguilera08b,2014MNRAS.442.3484K}.
In all these works, the magnetic field was held fixed, as a background, exploring different
possibilities, including superstrong ($B\sim10^{15}$\,--\,$10^{16}$~G) toroidal magnetic fields in the crust  to explain the strongly non-uniform distribution
of the surface temperature. Only recently \citep{vigano13}, the fully coupled evolution of temperature
and magnetic field has been studied with detailed numerical simulations. In the remaining of this section, we focus on the main aspects of the
numerical methods employed to solve Eq.~(\ref{Tbalance}) alone, and we will return to the specific problems originated by the coupling with the
magnetic evolution in the following sections.

\subsection{Numerical methods for 2D cooling}
\label{sect:cooling 2D}

There are two general strategies to solve the heat equation: spectral methods and finite-difference schemes. Spectral methods are well known to
be elegant, accurate and efficient for solving partial differential equations with parabolic and elliptic terms, where Laplacian (or similar) operators are present. However, they are much more
tedious to implement and to be modified, and usually require some strong previous mathematical understanding. On the contrary, finite-difference schemes are very easy to implement and do not
require any complex theoretical background before they can be applied. On the negative side, finite-difference schemes are less efficient and accurate, when compared to spectral methods 
using the same amount of computational resources. The choice of one over the other is mostly a matter of taste. However, in realistic problems with ``dirty'' microphysics (irregular or discontinuous
coefficients, stiff source-terms, quantities varying many orders of magnitude, etc), simpler finite-difference schemes are usually more robust and more flexible  than
the heavy mathematical machinery normally carried along with spectral methods, which are often derived for constant microphysical parameters. 
For this last reason, here we will discuss the use of finite-difference methods to solve our particular problem.

Let us consider the energy balance equation (\ref{Tbalance}), with the flux given by Eq.~(\ref{hflux}). 
We first note that, in axial symmetry, the $\varphi-$component of the flux is generally non-zero but need not to be evaluated since it is independent of $\varphi$, 
so that its contribution to the flux divergence vanishes. 
For example, in the case of a purely
poloidal field (only $r,\theta$ components), we can ignore the last term in Eq.~(\ref{hflux}) because it does not result in the time variation of the temperature. 
However, in the presence of a significant toroidal component $B_\varphi$, the last term gives
a non-negligible contribution to the heat flux in the direction perpendicular to $\vec{\nabla} (e^\nu T)$ (it acts as a Hall-like term).

\begin{figure}[h]
 \centering
\includegraphics[width=.5\textwidth]{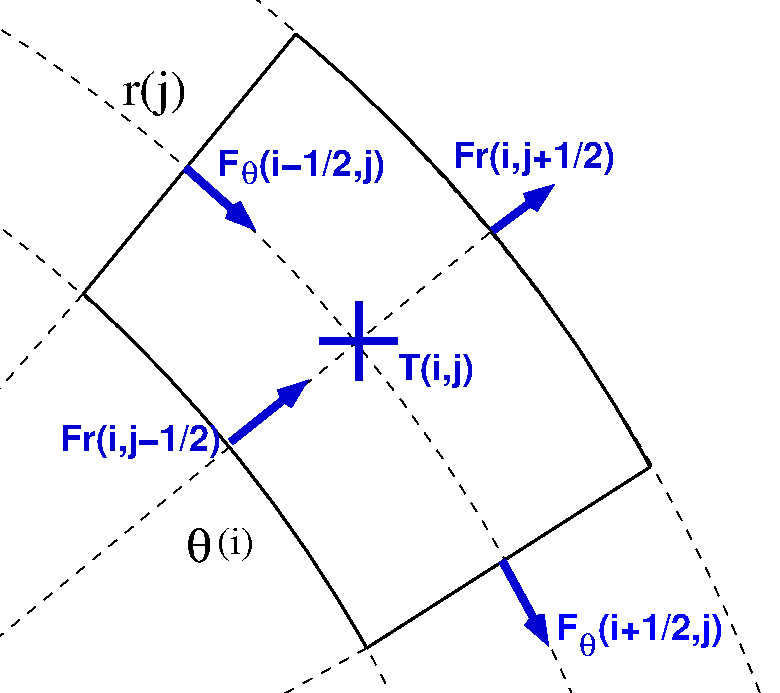}
\caption{Schematic illustration of the allocation of temperatures (cell centers) and fluxes (cell interfaces) in a typical grid in polar coordinates.}
\label{fig_staggered_T}
\end{figure}

In \cite{aguilera08a,aguilera08b,vigano13} and related works, they assume axial symmetry and adopt a finite-differences numerical scheme.  Values of temperature are defined at the center of each cell, where also the heating rate and the neutrino losses are evaluated, while fluxes are calculated at each cell-edge, as illustrated in Fig.~\ref{fig_staggered_T}. 
The boundary conditions at the center ($r=0$) are simply $\vec{F}=0$, while on the axis the non-radial components of the flux must vanish. As an outer boundary, 
they consider the crust/envelope interface, $r=R_b$, where the outgoing radial flux, $F_{\rm out}$, is given by a formula depending on the values of $T_b$ and $\vec{B}$ in the last numerical cell. For example, assuming blackbody emission from the surface, for each outermost numerical cell, characterized by an outer surface $\Sigma_r$ and a given value of $T_b$ and $\vec{B}$, one has $F_{\rm out}=\sigma_B \Sigma_r T_s^4$ where $\sigma_B$ is the Stefan-Boltzmann constant, and $T_s$ is given by the $T_s - T_b$ relation (dependent on $\vec{B}$), as discussed in the previous subsection.

To overcome the strong limitation on the time step in the heat equation, $\Delta t \propto (\Delta x)^2$, the diffusion equation can be discretized in time in a semi-implicit or fully implicit way, which results in a linear system of equations described by a block tridiagonal matrix \citep{bookRM67}.
The ``unknowns'' vector, formed by the temperatures in each cell, is advanced by inverting the matrix with standard numerical techniques for linear algebra problems, like
the lower-upper (LU) decomposition, a common Gauss elimination based method for general matrices, available in open source packages like \textsc{LAPACK}.  However,  this  is not the most efficient method for large matrices.
A particular adaptation of the Gauss elimination to the block-tridiagonal systems, known as Thomas algorithm \cite{Thomas1949} or matrix-sweeping algorithm, is much more efficient,  but its parallelization is limited to the operations within each of the block matrices. A new idea that has been proposed to overcome parallelization restrictions is to combine the Thomas method with a different decomposition of the block tridiagonal matrix \citep{Belov2017}.

A word of caution is in order regarding the treatment of the source term. The thermal evolution during the first Myr is strongly dominated by neutrino emission processes, which enter the evolution equation through a very stiff source term, typically a power-law of the temperature with a high index ($T^8$ for modified URCA processes, $T^6$ for direct URCA processes). These source terms cannot be handled explicitly without reducing the time step to unacceptable small values but, since they are local rates, linearization followed by a fully implicit discretization is straightforward and  results in the redefinition of the source vector and the diagonal terms of the matrix. 
A very basic description to deal with stiff source terms can be found in Sect.~17.5 of \cite{NumRec}.
This procedure is stable, at the cost of losing some precision, but it can be improved by using more elaborated implicit-explicit Runge--Kutta algorithms \citep{IMEX}.

\subsection{Temperature anisotropy in a magnetized neutron star}\label{sec:T_anis}

\begin{figure}[h]
 \centering
\includegraphics[width=.9\textwidth]{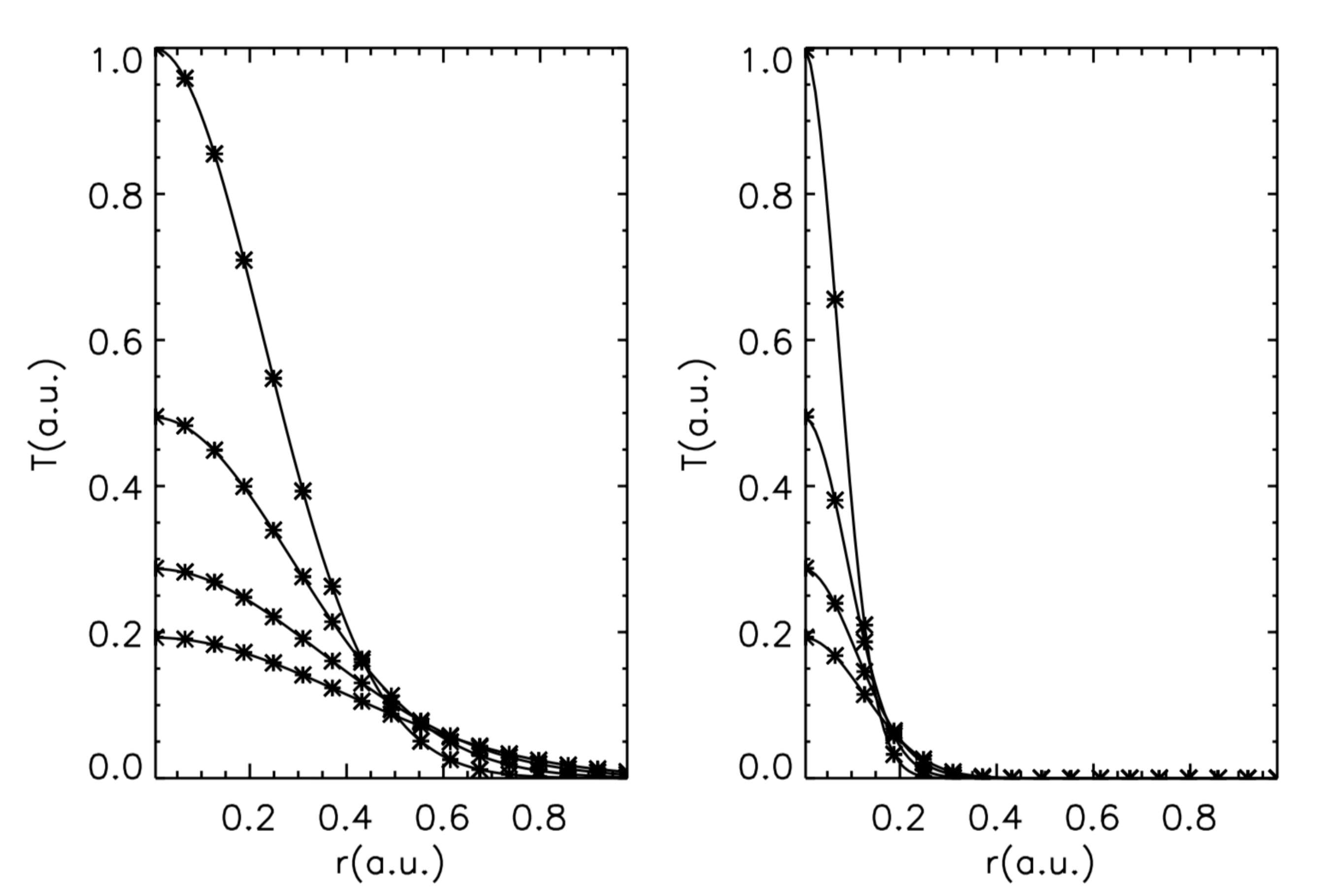}
\caption{Temperature profiles at different times comparing the analytic solution (solid) and the numerical evolution (stars) of a thermal pulse
in a medium embedded in a homogeneous magnetic field. The left (right) panel shows four different times during the evolution of polar (equatorial) profiles in arbitrary units. The simulation has been done with a fully implicit scheme and the linear system is solved with the Thomas algorithm.
Figure courtesy of \cite{perez06}.} 
 \label{fig_test_cool}
\end{figure}

\begin{figure}[h]
 \centering
\includegraphics[width=.9\textwidth]{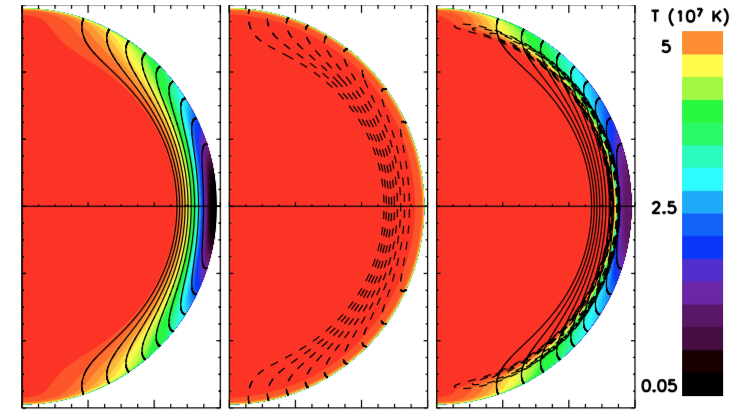}
\caption{Temperature anisotropy induced in the NS crust by the presence of a strong magnetic field confined into the crust. The projections of the poloidal field lines are shown with solid lines in the left and right panels, and dashed lines in the central panel. The left panel corresponds to a model without toroidal field, the central panel to a force-free configuration (toroidal magnetic flux contours and poloidal magnetic field lines are aligned), and the right panel shows a model with a toroidal component confined to a narrow region of the crust represented by dashed lines. Figure courtesy of \cite{perez06}.}
\label{tanis}
\end{figure}

An analytical solution that can be used to test numerical codes in multi-dimensions is the  evolution of a  thermal pulse  in an infinite  medium, embedded in a homogeneous 
magnetic field oriented along the $z$-axis, which causes the anisotropic diffusion of heat.
Assuming constant conductivities, and neglecting relativistic effects, the following analytical  solution  for  the  temperature  profile 
can be obtained for $t>t_0$:
\begin{equation}
T(t, r, \theta) = T_0 \left(\frac{t_0}{t}\right)^{3/2} \exp\left[ - \frac{r^2}{4  t {\kappa^\perp} } \left( {\sin^2\theta}+ 
\frac{\cos^2\theta}{1 + (\omega_B \tau_e)^2} \right) \right]~,
\end{equation}
where $T_0$ is the  central temperature at the initial time $t_0$.
In Fig. \ref{fig_test_cool} we show the comparison between
the  analytical  (solid)  and  numerical  (stars)  solution  for  a
model with $t_0=10^{-4}$, $T_0=1$, $\kappa^\perp = 10^2$ and $\omega_B \tau_e = 3$.  The boundary conditions employed are $F=0$ at the center and  the  temperature  corresponding to the 
analytical solution at the surface ($r=1$). \cite{perez06} found deviations from the analytical solution to be less than 0.1\% in any particular cell within the entire domain, even with a relatively low grid resolution of 100 radial  zones and 40 angular zones.

To conclude this section, the induced anisotropy in a realistic NS  reported by \cite{perez06} is shown in Fig.~\ref{tanis}.
The figure shows  equilibrium thermal solutions, in the absence of heat sources and sinks.  The core temperature is kept at $5\times 10^7$ K, and the surface boundary condition is given by the $T_s-T_b$ relation, assuming blackbody radiation. The poloidal component is the same in all models ($B_p = 10^{13}$ G).
The effect of the magnetic field on the temperature distribution can be easily understood by examining the expression of the heat flux (\ref{hflux}). 
When $\omega_B \tau_e \gg 1$, the dominant contribution 
to the flux is parallel to the magnetic field and proportional to  $\vec{b}\cdot \nabla (e^\nu T)$. Thus, in the stationary regime (i.e., $\nabla\cdot(e^{2\nu}\vec{F})=0$ if no sources are present), the temperature distribution must be such that 
$\vec{b} \perp \nabla (e^\nu T)$: magnetic field lines are tangent to surfaces of constant temperature. This is explicitly visible in the left panel, 
which corresponds to the stationary solution for a purely poloidal configuration with a core temperature of $5 \times 10^7$ K. 
Only near the surface, the large temperature gradient can result in a significant heat flux across the magnetic field lines.
When we add a strong toroidal component, the Hall term (proportional to $\omega_B \tau_e$) 
in Eq.~(\ref{hflux}), activates meridional heat fluxes which lead to a nearly isothermal crust. 
The central panel shows the temperature distribution for a force-free magnetic field with a global toroidal component, present in both the crust and the envelope. The right panel shows a third model with a strong toroidal component confined to a thin crustal region (dashed lines). It acts as an insulator maintaining a temperature gradient between both sides of the toroidal field.

\section{Magnetic field evolution in the interior of neutron stars: theory review}
\label{sec:magnetic_evolution}

The interior of a NS is a complex multifluid system, where different species coexist and may have different average hydrodynamical velocities. 
In most of the crust, for instance, nuclei have very restricted mobility and form a solid lattice. Only  the ``electron fluid'' can flow, providing the currents that sustain the magnetic field.
In the inner crust superfluid neutrons are partially decoupled from the heavy nuclei, providing a third neutral component.
In the core, the coexistence of superfluid neutrons and superconducting protons makes the situation even less clear. Since a full multifluid, reactive MHD-like description of the system is far from being affordable, one must rely on different levels of approximation that gradually incorporate the relevant physics. In this section we give an overview of the theory, trying to capture the most relevant processes governing the magnetic field evolution in a relatively simple mathematical form. For consistency with the previous section, we assume the same spherically symmetric background metric and we keep track of the most important relativistic corrections.

The evolution of the magnetic field is given by Faraday's induction law:
\begin{equation}
\frac{\partial \vec{B}}{\partial t} = - c \vec{\nabla}\times ( \mathrm{e}^{\nu} \vec{E} )~,
\label{induction0}
\end{equation}
which needs to be closed by the prescription of the electric field $\vec{E}$ in terms of the other variables (constituent component velocities and the magnetic field itself), either using simplifying assumptions (e.g., Ohm's law) or solving additional equations. Very often, this prescription involves the electrical current density, which in many MHD variations can be obtained from Amp\'ere's law, neglecting the displacement currents 
\begin{equation}\label{eq:current_mhd}
\vec{j} = \mathrm{e}^{-\nu} \frac{c}{4 \pi} \vec{\nabla}\times \left(\mathrm{e}^{\nu} \vec{B}  \right)~. 
\end{equation}
In a complete multi-fluid description of plasmas, the set of hydrodynamic equations complements Faraday's law.  From the multi-fluid hydrodynamics equations, a generalized Ohm's law -- in which the electrical conductivity is a tensor -- can be derived  \citep{1990SvAL...16...86Y,1995MNRAS.273..643S} $$ \vec{j} = \hat{\sigma}  \vec{E} . $$
Expressing the tensor components in a basis referred to the magnetic field orientation, one can identify longitudinal, perpendicular and Hall components, that give rise to a complex structure when the equation is inverted to express $\vec{E}$ as a function of $\vec{j}$, and $\vec{B}$. However, in some regimes,  one can make simplifications to make the problem affordable \citep{1980SvA....24..425U,jones1988,goldreich92}. 
The three main processes are Ohmic dissipation, Hall drift (only relevant in the crust) and ambipolar diffusion (only relevant in the core) \citep{goldreich92,1995MNRAS.273..643S,cumming04},
although additional terms could in principle be also included in the induction equation. 
For instance, there are theoretical arguments proposing additional slow-motion dynamical terms, such as  plastic flow \citep{beloborodov14,lander16,lander19}, magnetically induced superfluid flows \citep{ofengeim18} or vortex buoyancy \citep{muslimov1985,konenkov2000,Elfritz2016,dommes}. Typically, all these effects are introduced as advective terms, of the type $\vec{E} = -\vec{v}\times \vec{B}$, with $\vec{v}$ being some effective velocity. Thermoelectric effects have also been proposed to become significant in regions with large temperature gradients \citep{geppert91,wiebicke91,wiebicke92,wiebicke95,geppert95,wiebicke96}; 
These additional terms are not included in most of the existing literature, and no detailed numerical simulations are known so far. However, some of them may play a more important role than expected and should be carefully revisited. Here, we review the principal characteristics of the most standard and better understood physical processes.

\subsection{Ohmic dissipation}

In the simplest case, the electric field in the reference frame comoving with matter is simply related to the 
electrical current density, $\vec{j}$, by:
\begin{equation}\label{eq:ohm}
\vec{E} = \frac{\vec{j}}{\sigma} ~,
\end{equation}
where the conductivity $\sigma$, dominated by electrons, must take into account all the (usually temperature-dependent) collision processes of the charge carriers. 
Here, $\sigma$ actually represents the longitudinal (to the magnetic field) component of the general conductivity tensor $\hat{\sigma}$. In the weak field limit, the tensor becomes a scalar ($\sigma \equiv \sigma_\parallel $) times 
the identity, and possible anisotropic effects are absent.

\begin{figure}[t]
	\centering
	\includegraphics[width=0.49\textwidth]{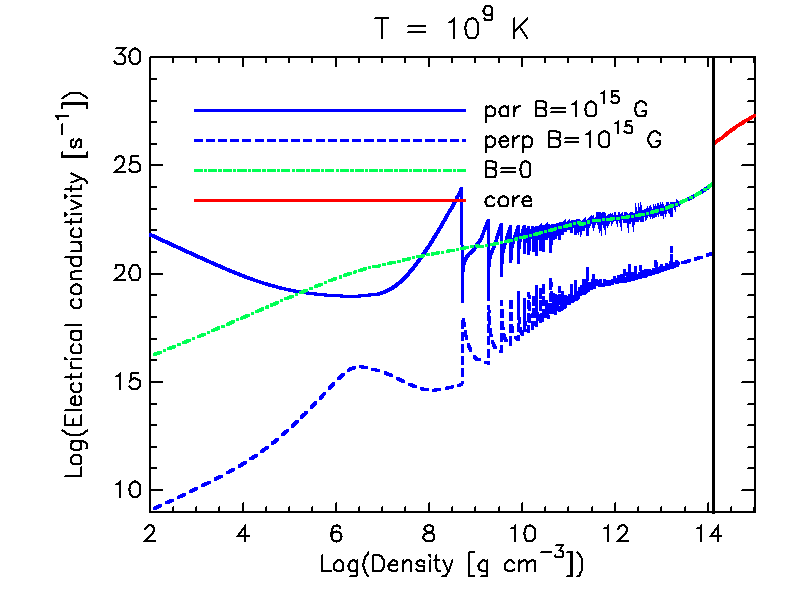}
	\includegraphics[width=0.49\textwidth]{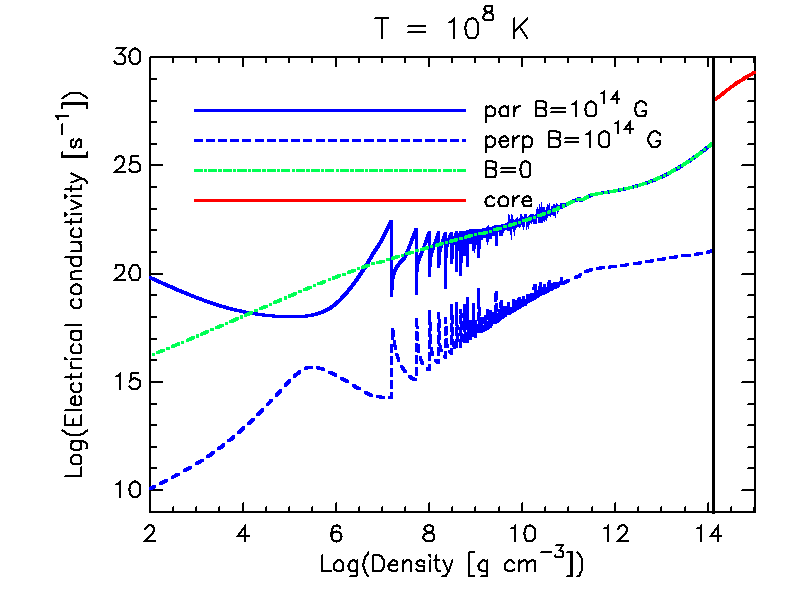}
	\caption{Electrical conductivity in the directions parallel (solid lines) and perpendicular (dashes) to the magnetic field, including quantizing effects. We show the cases $T=10^9$ K, $B=10^{15}$ G (left panel) and $T=10^8$ K, $B=10^{14}$ G (right panel). For comparison, the $B=0$ values are shown with green lines in both figures.}
	\label{fig:cond_el}
\end{figure}

The induction equation, when we have only Ohmic dissipation, conforms a {\it vector diffusion equation}:
\begin{equation}
	\frac{\partial \vec{B}}{\partial t} +  \vec{\nabla}\times \left( \eta \vec{\nabla}\times  (\mathrm{e}^{\nu}\vec{B} ) \right) = 0~,
	\label{ohm0}
\end{equation}
where we have defined the magnetic diffusivity $\eta \equiv \frac{c^2}{4 \pi \sigma}$. In the relaxation time approximation, the electrical conductivity parallel to the magnetic field, 
$\sigma=e^2 n_e \tau_e/m^*_e$, with $n_e$ being the electron number density.
Typical values of the electrical conductivity in the crust are $\sigma \sim 10^{22}$--$10^{25}$ s$^{-1}$, several orders of magnitude larger than in the most conductive terrestrial metals described 
by the band theory in solid state physics. In the core, the even larger electrical conductivity ($\sigma \sim 10^{26}$--$10^{29}$ s$^{-1}$) results in much longer Ohmic timescales, thus potentially affecting the magnetic field evolution only at a very late stage ($t \gtrsim 10^8$ yr), when isolated NSs are too cold to be observed. In Fig.~\ref{fig:cond_el} we show typical profiles of the electrical conductivity, for the same combinations of $T$ and $B$ shown for the thermal conductivity in Fig.~\ref{fig:cond_class_quant}. Since, neglecting inelastic scattering, both thermal and electrical conductivities are proportional to the collision time $\tau_e$, they share some trends: the suppression of the conduction in the direction orthogonal to a strong magnetic field, and the quantizing effects visible as oscillations around the classical value \citep{potekhin_rev15a,potekhin2018}. We note that, if inelastic scattering contributes significantly, $\tau_e$ can be different for thermal and electrical conductivities. 

\subsection{The Hall drift}

At the next level of approximation, one must consider not only Ohmic dissipation but also advection of the magnetic field lines by the charged component of the fluid, say the electrons, with velocity $\vec{v_e}$. 
The electric field has the following form 
\begin{equation}\label{eq:induction}
\vec{E} =   \frac{\vec{j}}{\sigma} - \frac{\vec{v_e}}{c} \times \vec{B}   ~.
\end{equation}
In the crust, the electron velocity is simply proportional to the electric current
\begin{equation}
\vec{v}_e = - \frac{ \vec{j} }{e n_e} ~,
\end{equation}
and the Hall--MHD (or electron--MHD) induction equation reads
\begin{equation}\label{eq:hall_induction}
\frac{\partial \vec{B}}{\partial t} =
- \vec{\nabla}\times \left\{ \eta \vec{\nabla}\times (\mathrm{e}^{\nu}\vec{B} )
+ \frac{c }{4\pi e n_e} \left[ \vec{\nabla}\times (\mathrm{e}^{\nu}\vec{B}) \right] 
\times \vec{B}\right\}~.
\end{equation}
Here, the first term on the right-hand side is the same as in Eq.~(\ref{ohm0}) and accounts for Ohmic dissipation, while the second term is the nonlinear Hall term. Note that the latter does not depend on the temperature, but it varies by orders of magnitude in the crust due to the inverse dependence with density.
We can factor out the magnetic diffusivity and express the Hall induction equation in the  form
\begin{equation} 
\frac{\partial\vec B}{\partial t}= - {\nabla \times}\left(\eta 
\left\{\nabla \times (\mathrm{e}^{\nu}\vec{B} ) + \omega_B\tau_e [(\nabla \times (\mathrm{e}^{\nu}\vec{B} )) \times \vec b ]\right\} \right).
\label{Hallind} 
\end{equation} 
This form of the induction equation makes explicit that the magnetization parameter $\omega_B \tau_e$, which also determined the degree of anisotropy 
in the heat transfer, Eq.~(\ref{eq:omegatau}), plays the role of the magnetic Reynolds number: it gives the relative weight of the 
Hall and Ohmic dissipation terms. Generally speaking, as we approach the surface from the interior, $\omega_B \tau_e$ increases.
We note that, given these considerations, one has to be careful interpreting analytical estimates of the Ohmic or Hall timescales, since both vary by many orders of magnitude depending on the local conditions.

The vast majority of the existing studies of magnetic field evolution in NS crust
\citep{HR2002,HR2004,PonsGeppert2007,Reise2007,Pons2009,Kondi2011,vigano12a,vigano13,gourgouliatos13,marchant14,gourgouliatos14a,gourgouliatos15a,gourgouliatos15b,wood15}
are restricted to 2D simulations, but the few recent 3D simulations suggest that the main aspects of 2D results partially hold:  although the Hall term itself conserves energy, 
the creation of small-scale structures results in an enhanced Ohmic dissipation. Some distinctive 3D features are the Hall-induced, small scale, azimuthal magnetic
structures that seem to persist on long timescales (see Sect.~\ref{sec:examples}).

\subsection{Plasticity and crustal failures}

The main idea for the Hall--MHD description of the crust is that ions are locked in the crustal lattice and only electrons are mobile. However, molecular dynamics simulations \citep{horowitz09} show that the matter has an elastic behavior until certain maximum stress. Above it, the magnetic stresses, quantified by the Maxwell tensor
${\cal M} \equiv B_iB_j/4\pi$, cannot be compensated by the elastic response (a more rigorous global condition is the von Mises criterion applied in \citealt{lander15}).
Crustal failures are treated in the most simplified manner as {\it star-quakes}. By evaluating the accumulated stress, \cite{pons11,perna11}  simulated the frequency and energetics of the internal magnetic rearrangements, which was proposed to be at the origin of magnetar outbursts. 
This model mimics earthquakes since, under terrestrial conditions, the low densities of the material allow for propagation of sudden fractures: the Earth mantle in this sense can be thought as brittle. However, materials subject to very slow shearing forces could behave differently and enter a plastic regime where, instead of sudden crustal failures, a slow plastic flow takes place. Despite the different dynamics, the energetic arguments relating the release of energy due to the accumulation of magnetic stresses are similar. Recent simulations \citep{lander19} show the features of such plastic flow under the assumption of Stokes flow, where a viscous term balances magnetic and elastic stresses.
They compare the crustal response under Ohmic and Hall evolution and find that there can be significant plastic-like motions in the external layers of the star.
Similar arguments have also been proposed to account for the deposition of heat by the visco-plastic flow and the propagation of thermo-plastic waves \citep{beloborodov14}. 
Depending on which hypotheses we make, the interpretation of the {\it velocities} in the advective term ($\vec{v} \times \vec{B}$) of the induction equation requires 
a proper physical and mathematical approach.

\subsection{Ambipolar diffusion in neutron star cores}
The number of works concerning mechanisms operating in NS cores is sensibly smaller, and most contain far less detail than the studies of the crust. 
Owing to its cubic dependence on $B$, ambipolar diffusion could be the dominant process driving the evolution of magnetars during the first $10^3-10^5$ yr, although there is some controversy.
In particular, we refer the reader interested in the role of chemical potential gradients, which is out of the scope of this review, to the literature. For example, \cite{goldreich92} or \cite{passamonti17b} derived 
an elliptic equation from the continuity and momentum equations to determine the small deviations from beta equilibrium. However, \cite{gusakov17}  question the validity of that
approach in stratified matter, and obtain a different equation from the momentum equation (implicitly assuming magnetostatic equilibrium), in which the small deviations of the chemical potentials from 
their equilibrium values do not depend on temperature and are determined by the Lorentz force.  With the same methodology, \cite{ofengeim18} calculate the
instantaneous particle velocities and other parameters of interest, determined by specifying the magnetic field configuration, and found that the evolution timescales could be shorter than expected.

The short way to incorporate ambipolar diffusion  is to generalize 
the form of the electric field by introducing the ``ambipolar velocity'' $\vec{v}_a$:
\begin{equation}\label{eq:efield_general}
\vec{E} =    \frac{\vec{j}}{\sigma} + \frac{1}{e n_e c}  \vec{j} \times \vec{B}  - \frac{\vec{v}_a}{c} \times \vec{B}~.
\end{equation}
The simplest case is realized in the regime where the system attains $\beta-$equilibrium faster than it evolves, and the ambipolar velocity is proportional to the Lorentz force 
\begin{equation}
\vec{v}_a = f_a  \vec{j}  \times \vec{B}~,
\end{equation}
where $f_a$ is a positive-defined drag coefficient.  For simplicity we only consider this case in the next sections.
We also note that, alternatively, the ambipolar term can be written as:
\begin{equation}
  - \frac{\vec{v}_a}{c} \times \vec{B} = \frac{f_a}{c} [B^2 \vec{j} - (\vec{j}\cdot\vec{B})\vec{B}] \equiv \frac{f_a}{c}B^2\vec{j}_\perp~,
\end{equation} 
where it explicitly takes the form of a resistive-like term, with a $B^2$-dependent coefficient, only acting on the currents perpendicular to the magnetic field ($\vec{j}_\perp$)
aligning the magnetic field with the current and bringing the system into a force-free configuration, characterized by definition by $\vec{j} \times \vec{B}=0$.
It is important to remark that the effect of this term is very sensible to the magnetic geometry, besides its strength: it has no consequences on the current flowing along magnetic field lines. 
This property has been used to introduce a formally similar term (differing only by a re-normalization factor $\propto 1/B^2$) in the so-called magneto-frictional method, 
used to obtain configurations of twisted force-free solar \citep{roumeliotis94} and NS \citep{vigano11} magnetospheres (see also Sect.~\ref{sec:extended_domains}). 

Most previous works studying ambipolar diffusion rely on timescale estimates, with few exceptions. Simulations are only available in
a simplified 1D approach \citep{2008A&A...487..789H,2010MNRAS.408.1730H} and very recently in 2D \citep{castillo17,passamonti17b,bransgrove18}, 
usually for constant coefficients.
However, in a realistic scenario, there is a further complication. 
The NS core cools down below the neutron-superfluid and proton-superconducting critical temperatures very fast, 
which has important implications, sometimes controversial.  \cite{goldreich92} argued that ambipolar diffusion would still be a significant process, but
\cite{glampedakis11b} studied in detail the ambipolar diffusion in superfluid and superconducting stars and concluded that its role on the magnetic field 
evolution would be negligible. Other recent works \citep{Graber2015,Elfritz2016} have also shown
that, without considering ambipolar diffusion, the magnetic flux expulsion from the NS core with superconducting protons is very slow.
In \cite{passamonti17a}  the various approximations employed to study the long-term evolution of the magnetic field in NS cores were revisited,
solving a recent controversy \citep{Graber2015,dommes}  on the correct form of the induction equation and the relevant evolution 
timescale in superconducting NS cores. 

\subsection{Mathematical structure of the generalized induction equation}\label{sec:induction_eq_char}

In order to understand the dynamical evolution of the system and to design a successful numerical algorithm, 
it is important to identify the mathematical character of the equations and the wave modes.
The magnitude of $\omega_B \tau_e$  defines the transition from a purely parabolic equation ($\omega_B \tau_e \ll 1$) to a hyperbolic regime ($\omega_B \tau_e \gg 1$). 
The Hall term introduces two wave modes into the system. \cite{huba03} has shown that,
in a constant density medium, the only modes of the Hall--MHD equation are the {\it whistler or helicon waves}. They are
transverse field perturbations propagating along the field lines. In presence of a charge density gradient, 
additional {\it Hall drift waves} appear. These are transverse modes that
propagate in the $\vec{B} \times \vec{\nabla} n_e$ direction. 
We also note that the presence of charge density gradients results in a Burgers-like
term \citep{Vai2000}. Furthermore, even in the constant density case but without planar symmetry, 
the evolution of the toroidal component also contains a quadratic term that resembles the Burgers equation \citep{PonsGeppert2007} with
a coefficient dependent on the distance to the axis. This term leads to the formation of discontinuous solutions (current sheets)  that require proper treatment.
 It is fundamental for a numerical Hall--MHD code to reproduce these modes and features, which are easily testable, as illustrated in Sect.~\ref{sec:tests}.

In \cite{vigano19} they give a complete description of the characteristic structure of the induction equation, including the Ohmic, Hall and ambipolar terms, in a flat spacetime, $e^\nu=e^\lambda=1$. 
By assuming a generic perturbation over a fixed background field $\vec{B}_o$:
\begin{equation}
\vec{B} = \vec{B}_o + \vec{B}_1 \, e^{i(\vec{k}\cdot \vec{x} - \omega t)}~,
\end{equation}
with a wavelength $\vec{k}$ much shorter than any other typical length of the system (typical variation scales of the Ohmic, ambipolar and Hall pre-coefficients), 
the eigenvalues are given by
\begin{equation}\label{eigenvalues}
i \frac{\omega^{\pm}}{k^2} = \eta + \frac{c f_a}{8 \pi} (B_{o}^2 + B_{ok}^2 ) \pm \frac{1}{8 \pi}\sqrt{\left({c f_a B_{op}^2}\right)^2  - 4 \left(\frac{c B_{ok}}{e n_e}\right)^{2}}~,
\end{equation}
where $B_{ok} = \hat{k}\cdot\vec{B}_o$, and $B_{op} = |\vec{B}_o - \vec{B}_{ok }\hat{k}|$. This relation explicitly confirms that the Hall term is the only 
one that could be associated with waves (take the limit  $\eta=f_a=0$), while the Ohmic and ambipolar terms are intrinsically dissipative.

\section{Magnetic field evolution in the interior of neutron stars: numerical methods}\label{sec:magnetic_methods}

In this section, we go through the most relevant aspects of numerical methods.
The first important choice is the formalism to be adopted. There are two options: i) to work directly with the magnetic field components, which does not require any further mathematical manipulation but implies to care about how to preserve the divergence-free condition, and ii) exploiting the solenoidal constraint to work with only two functions representing the two 
true degrees of freedom instead of three components: the so-called poloidal-toroidal decomposition (see Appendix~\ref{app:formalism}).
Finite-difference schemes have been developed for both formalisms, while spectral methods more often built on the poloidal-toroidal decomposition.
We begin with an overview of spectral methods, before turning into some key aspects of finite-difference schemes.

\subsection{Spectral methods with the toroidal-poloidal decomposition}\label{methods_spectral}

Using the notation of \cite{geppert91}, the basic idea  is to expand the poloidal ($\Phi$) and toroidal ($\Psi$) scalar functions in a series of spherical harmonics
\bear
\Phi = \frac{1}{r} \sum_{n,m} \Phi_{nm}(r,t) Y_{nm}(\theta,\varphi)~, 
\nonumber \\ 
\Psi = \frac{1}{r} \sum_{n,m} \Psi_{nm}(r,t) Y_{nm}(\theta,\varphi)~,
\label{expans} 
\ear
where $n=1,\ldots,n_{\rm max}$ and  $m=-n,\ldots,+n$.

Assuming a radial dependent diffusivity, $\eta=\eta(r)$, it can be shown that the Ohmic term for each multipole effectively decouples, and the 
set of coupled evolution equations for the radial parts ($\Phi_{nm}$ and $\Psi_{nm}$)  can be readily obtained \citep{geppert91}:
\bear
\frac{\partial \Phi_{nm}(r)}{\partial t} &=& e^{\nu} \eta (r)  \left[
\mathrm{e}^{-2\lambda} \frac{\partial^2 \Phi_{nm}}{\partial r^2} 
+ e^{-2\lambda} \left( \frac{d \nu}{d r} - \frac{d \lambda}{d r} \right)  \frac{\partial \Phi_{nm}}{\partial r}  - \frac{n(n+1)}{r^2}\Phi_{nm}
\right] + D_{nm}
\label{Phi_diff}
\nonumber \\
\frac{\partial \Psi_{nm}(r)}{\partial t}&=&
e^{-\lambda} \frac{\partial}{\partial r} 
\left( \eta(r) e^{-\lambda} \frac{\partial  (e^{\nu} \Psi_k)}{\partial r}\right)
-   \eta(r) \frac{n(n+1)}{r^2} e^{\nu} \Psi_{nm} + C_{nm}~.
\label{Psi_diff}
\ear
where we use $D_{nm}$ and $C_{nm}$ as a shorthand for the nonlinear Hall terms (the full expressions can also be found in \citealt{geppert91}). 
These include sums over running indices and coupling constants related to Clebsch--Gordan coefficients (the sum rules to combine angular momentum operators are used to determine
which multipoles are coupled to each other). All these coefficients can be evaluated once
at the beginning of the evolution and stored in a memory-saving form since only specific combinations of indices are non-zero.

In the most general case, however, the magnetic diffusivity also depends on the angular coordinates, for example through the temperature dependence of $\eta$ when the temperature is non-uniform.
In this case we can also expand the magnetic diffusivity in spherical harmonics 
\begin{equation}
\eta = \sum_{n,m} \eta_{nm}(r,t) Y_{nm}(\theta,\varphi)~, 
\end{equation}
where the sum must include the monopole term, $n=0,\ldots,n_{\rm max}$.
These new terms couple different multipoles of the same component (poloidal
or toroidal). 
The inclusion of additional terms in the electric field (e.g. ambipolar diffusion) would introduce even more complicated non-linear couplings (the theory has not yet been developed). 
In general, we end up with a system of the order of $\approx2n_{\rm max}^2$, strongly coupled, differential equations. The choice now is whether using a different spectral decomposition in the radial direction (usually Chebyshev polynomials) or employing a hybrid method, applying standard finite-difference techniques in the radial direction to solve the system of equations. 

The first multi-dimensional (2D) simulations of the evolution of the crustal magnetic field assumed a constant density shell  \citep{HR2002}  and were later extended
to include density gradients \citep{HR2004}. They used an adapted version of the spherical harmonic code described
in \cite{Hollerbach2000}, including modes up to $l =100$, and 25 Chebychev polynomials in the radial direction, but they were restricted to 
$\omega_B \tau_e<200$  by numerical issues.
In \cite{PonsGeppert2007,pons09}, they used a hybrid code (spectral in angles but finite-differences in the radial direction)  to perform 2D simulations in realistic profiles of NSs
over relevant timescales (typically, Myr). This approach allowed us to reach higher values of the magnetization parameter ($\omega_B \tau_e \approx 10^3$), and to 
study the Hall instability \citep{Pons2010}. 
The same approach is used in the 3D simulations of \cite{wood15,gourgouliatos16}, which were limited to magnetization parameters of the order of
$\simeq 100$. The main problem arises from the presence of non-linear Burgers-like terms, which naturally lead to discontinuities (see \S~\ref{sec:induction_eq_char}), which are notoriously poorly handled by spectral codes.
For this reason, subsequent works aiming at extending the simulations to more general cases have been gradually shifting towards the use of finite-difference schemes.

\subsection{Finite-difference and finite-volume schemes}

To study the interesting magnetar scenario in detail, the numerical codes must be able to go a bit further.
In \cite{vigano12a}, a novel approach making use of the well-know High-Resolution Shock-Capturing (HRSC) techniques \citep{toro97}, designed to handle shocks in hydrodynamics and MHD, was proposed. These techniques have been successfully applied to a range of problems, from a simple 1D Burgers equation to complex ideal MHD problems \citep{anton06,giacomazzo07,cerdaduran08}, 
avoiding the appearance of spurious oscillations near discontinuities. 
We refer to \cite{marti2015} for a general review on grid-based methods and to \cite{balsara17} for a review on finite-volume methods, applied to other astrophysical scenarios.
Let us review some of the main characteristics of these methods, of particular interest in our problem.

\subsubsection{Conservation form and staggered grids}

In hydrodynamics and MHD, the system of partial differential equations (PDEs) involve the divergence operator acting on vector or tensor fields. Thus, Gauss' theorem is usually 
employed in the design of the algorithms, exploiting the formulation of the equations in conservation form.
Analogously, for problems involving the induction equation, the presence of the curl operator makes it natural to apply Stokes' theorem to the equation.
Considering a numerical cell and its surface $\Sigma_\alpha$ normal to the $\alpha$ direction, delimited by the curve $C_\Sigma$, we have a discretized version of eq.~(\ref{induction0}):
\begin{eqnarray}
\frac{1}{c}\frac{\partial }{\partial t}\left[\int_{\Sigma_\alpha} \mathrm{e}^\nu B_\alpha d\Sigma_\alpha\right]  =  - \oint_{C_\Sigma} \vec{E} \cdot d\vec{l} ~.
\end{eqnarray}
The space-discretized evolution equation for the average of the magnetic field component normal to the surface over the cell surface  is then
\begin{eqnarray}\label{phialpha}
\frac{\partial \overline{\mathrm{e}^\nu B}_\alpha}{\partial t}  =  - \frac{c \sum_k E_k l_k}{\Sigma_\alpha} ~. \label{eq:induction_discretized}
\end{eqnarray}
Here, the circulation of the electric field is approximated by the sum $\sum_k E_k l_k$, where $E_k$ is the average value of the electric field over each 
cell of length $l_k$, and $k$ identifies each of the four edges of the face. 
For clarity, in this section, we omit relativistic metric factors that must be consistently 
incorporated in the definitions of lengths, areas, and volumes.

\begin{figure}[h]
 \centering
\includegraphics[width=.5\textwidth]{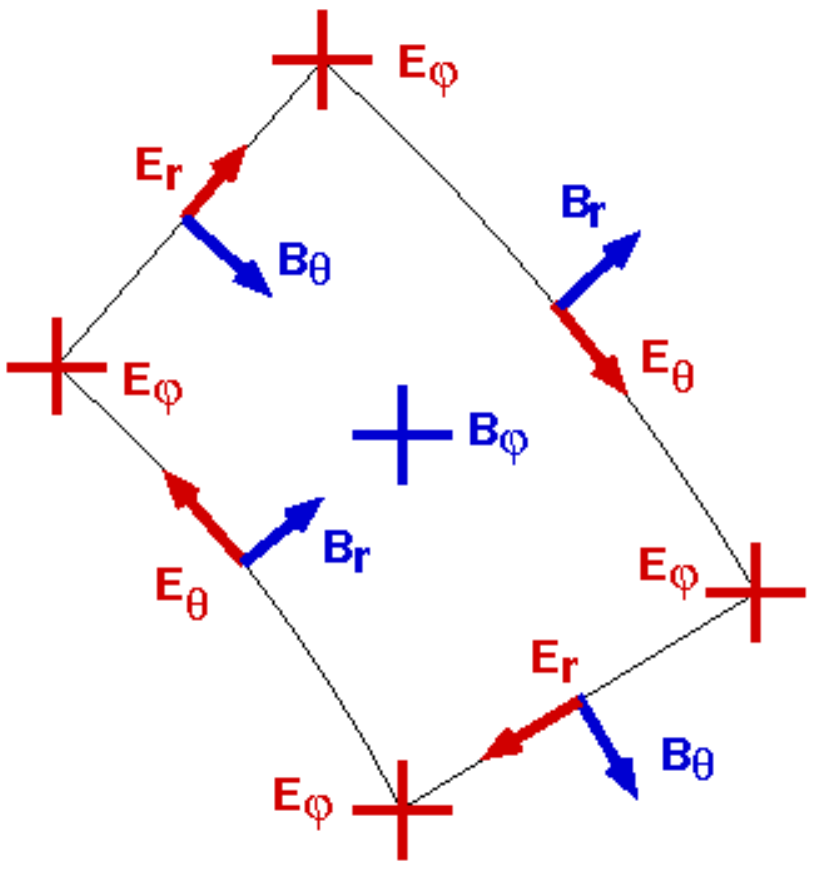}
\caption{Location of the variables on a staggered grid in spherical coordinates for the axisymmetric case.
Solid lines delimit the edges of the surface $\Sigma_\varphi$.} 
 \label{fig_staggered}
\end{figure}

The problem is then reduced to design an accurate and stable discretization method to calculate the $E_k$ components at each edge. 
A natural choice is to use staggered grids,
for which in each numerical cell the locations of the different field components are conveniently displaced, instead of being all located at the same position (typically, the center), 
as in standard centered schemes. 
In our case, we allocate the normal magnetic field components at each face center and electric field components along cell edges. 
Fig.~\ref{fig_staggered} shows an example of the location of the variables in a numerical cell in spherical coordinates $(r, \theta, \varphi)$, considering  axial symmetry (in the 
general 3D case, there would be a displacement of $B_\varphi, E_\theta, E_r$ in the direction orthogonal to the plane of the figure). 

Making use of Gauss' theorem, the numerical divergence can be evaluated, for each cell with volume $\Delta V$, as follows: 
\begin{equation}\label{divb_staggered}
  \vec{\nabla}\cdot\vec{B}=\frac{1}{\Delta V}\sum_\alpha{\overline{B}_\alpha \Sigma_\alpha}~.
\end{equation}
With this definition, the divergence-preserving character of the methods using the conservation form and advancing in time $\overline{B}_\alpha$ components, becomes evident: taking the time derivative of eq.~(\ref{divb_staggered}), and using eq.~(\ref{phialpha}), every edge contributes twice with a different sign and cancels out. 
By construction, the divergence condition is preserved to machine error for any divergence-free initial data. 
Examples of applications of such methods can be found, among many others, in  \cite{toth00,vigano12a,balsara15}.

\subsubsection{Evaluation of the current and the electric field}
\begin{figure}
 \centering
 \includegraphics[width=.45\textwidth]{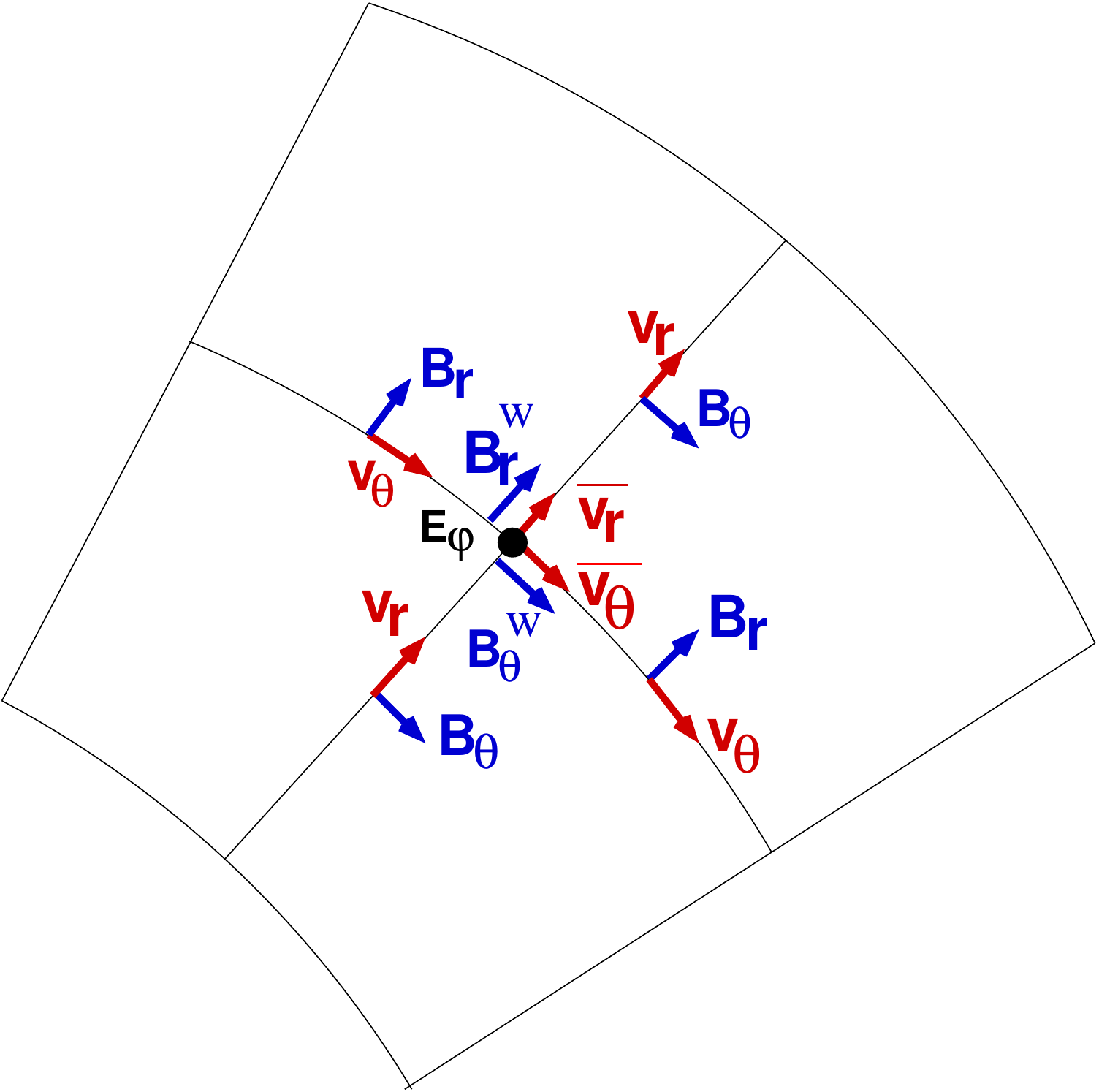}
\caption{Illustration of the procedure to calculate the electric field in a staggered grid: location of the 
components of velocity (red arrows) and magnetic field (blue) involved in the definition of contribution to  $E_\varphi$ (black dot) from the Hall term.}
 \label{fig_ewind}
\end{figure}

Let us consider a general electric field of the form:
\begin{eqnarray}\label{efield_v}
\vec{E} = \frac{\vec{j}}{\sigma} - \frac{\vec{v}}{c} \times \vec{B} ~,
\end{eqnarray}
where nonlinear (Hall and/or ambipolar) dependences on the magnetic field are implicitly contained in the expression of $\vec{v}$.

By considering the allocation of the components in the staggered grid (Fig.~\ref{fig_staggered}), the components of the current density can be naturally defined along the edges of the cells, in the same positions as the electric field components, exploiting the discretized version of the Stokes' theorem applied to $\vec{j}\propto  \vec{\nabla}\times(\mathrm{e}^\nu\vec{B})$. Therefore, the ohmic term in the electric field can be directly evaluated, but the other terms involving vector products require special care since they involve products of field components
that are not defined at the same place as the desired electric field component.
The simplest option is a direct interpolation of both $\vec{v}$ and $\vec{B}$ using the first neighbors, but this often results in numerical instabilities.

In the spirit of HRSC methods, we can instead think of the interpolated value of $\vec{v}$ as the advective velocity acting at that point (although it depends on $\vec{B}$ itself), and consistently take the {\it upwind} components $\overline{B}_\alpha^w$ of the magnetic field at each interface. For example, in the axisymmetric case and considering the evolution of the poloidal components ($B_r, B_\theta$), the contributions of $E_r$ and $E_\theta$ to the circulation cancel out and we only need to evaluate the contribution of $E_\varphi$, which is given by
\begin{eqnarray}\label{efield_wind}
{E_\varphi} =\frac{1}{\sigma}  {J_\varphi}  - \frac{1}{c} \left( \overline{v}_r {\overline{B}_\theta}^w -  \overline{v}_\theta {\overline{B}_r}^w \right) ~.
\end{eqnarray}
In Fig.~\ref{fig_ewind} we explicitly show the location of $E_\varphi$ (black point) and the location on the staggered grid of the quantities needed for its evaluation. 
First, $\overline{v}_r$ and $\overline{v}_\theta$ are calculated taking the average of the two closest neighbors; in the example, they point outward and to the right, respectively. Second, one considers the upwind values of  $\overline{B}_r^w$ and $\overline{B}_\theta^w$; in the example, they are taken from the bottom and left sides.

\subsubsection{Divergence cleaning methods in finite-difference schemes}

An algorithm built on a staggered grid can be designed to preserve the divergence constraint by construction, 
but the different allocation of variables makes its implementation relatively  complex, particularly in 3D problems and with the inclusion of 
quadratic and cubic terms in the electric field. Among alternative formulations that have recently gained popularity, and can also handle many MHD-like problems, a
relatively simple option is the family of {\it divergence-cleaning} schemes built on standard grids (all components of the fields are defined and evolved at every grid node).
A popular divergence-cleaning method \citep{dedner02}, extensively used in MHD, consists in the extension of the system of equations as follows: 
\begin{eqnarray}
  &&  \frac{1}{c} \frac{\partial \vec{B}}{\partial t} +  \vec{\nabla}\times (\mathrm{e}^\nu\vec{E}) + \vec{\nabla}\chi = 0 ~,\nonumber \\
  && \frac{\partial \chi}{\partial t} + c_h^2 \vec{\nabla}\cdot\vec{B} = -\gamma \chi ~,
\end{eqnarray}
where $\chi$ is a scalar field that allows the propagation and damping of divergence errors, and $c_h$ and $\gamma$ are two parameters to be tuned: $c_h$ is the propagation speed of the constraint-violating modes, which decay exponentially on a timescale $1/\gamma$. In principle, a large value of
$\gamma$ will damp and reduce divergence errors very quickly, but in practice 
the optimal cleaning is reached for {$c_h \approx \gamma \sim {\cal O}(1)$ because, if $\gamma$ is too large, the source term becomes stiff and more difficult to handle with explicit numerical schemes.

\subsubsection{Cell reconstruction and high-order accuracy}

The original upwind (Godunov's) method is well known for its ability to capture discontinuous solutions, but it is only first-order accurate: the variables are assumed to be constant on
each cell. This method can be easily extended to give second-order spatial accuracy on smooth solutions, but still avoiding non-physical oscillations near discontinuities, by using
a reconstruction procedure that improves the piecewise constant approximation. 

A very popular choice for the slopes of the linear reconstructed function is the {\it monotonized central-difference limiter}, proposed by \cite{vanleer77}. 
Given three consecutive points $x_{i-1},x_i,x_{i+1}$ on a numerical grid, and the numerical values of the function $f_{i-1},f_i,f_{i+1}$, the reconstructed function within the cell $i$ is given by $f(x)=f(x_i)+ \alpha (x-x_i)$, where the slope is
$$\alpha = {\rm minmod}\left( \frac{f_{i+1}-f_{i-1}}{x_{i+1}-x_{i-1}},2\frac{f_{i+1}-f_{i}}{x_{i+1}-x_{i}},
2\frac{f_{i}-f_{i-1}}{x_{i}-x_{i-1}}\right).$$
The ${\rm minmod}$ function of three arguments is defined by
\begin{equation}
{\rm minmod}(a,b,c) = \left\{ 
\begin{array}{cc}
{\rm min}(a,b,c) & {\rm if} ~a,b,c>0 ;\\
{\rm max}(a,b,c) & {\rm if} ~a,b,c<0 ;\\
0 & {\rm otherwise} .
\end{array}
\right.\nonumber
\end{equation}
Other popular higher order reconstructions, are PPM \citep{colella84}, PHM \citep{PHM}, MP5 \citep{suresh97},  the FDOC families \citep{bona09}, or the Weighted-Essentially-Non-Oscillatory (WENO) reconstructions \citep{jiang96,shu98,yamaleev09,balsara17}.
In \cite{vigano19} they presented and thoroughly tested a two-step method consisting of the reconstruction with WENO methods of a combination 
of fluxes and fields at each node, known as flux-splitting \citep{shu98}.
This reconstruction scheme does not require the characteristic decomposition of the system of equations (i.e., the full spectrum of characteristic velocities) and,  at the lowest order
of reconstruction, their flux formula reduces to the popular and robust Local-Lax--Friedrichs flux \citep{toro97}.

\subsection{Courant condition and time advance}

In explicit  algorithms to solve PDEs involving propagating waves, the time step is limited by the Courant condition, which essentially states that  waves cannot travel more than 
one cell length on each time step, avoiding numerical instabilities.
Since we want to evolve our system on long (Ohmic) timescales, the Courant condition makes the simulation computationally expensive for Hall-dominated 
regimes, $\omega_B^e \tau_e \gg 1$.
For each cell, we can estimate the Courant time related to the Hall term by 
\begin{equation}\label{estimate_courant_hall}
dt^h \approx  \frac{4\pi e n_c L ~\Delta l }{c B}~,  
\end{equation}
where $L$ is a typical distance in which the magnetic field varies (e.g., the curvature radius of the lines), 
$\Delta l$ is the minimum length of the cell edges in any direction. In the case of a spectral code, $\Delta l \simeq L_{dom}/n_{max}$, i.e., the ratio between the length  of the dominion and the maximum number of multipoles calculated.
  
The Courant condition related to the ambipolar diffusion term is
\begin{equation}\label{estimate_courant_ambip}
dt^a  \approx  \frac{4\pi  L ~\Delta l }{c f_a B^2}~,  
\end{equation}
which becomes more restrictive than the Hall term when $e n_c f_a B \gg 1$. The Courant condition is then
\begin{equation}
\Delta t = k_c \mbox{min} \left[ dt^h , dt^a \right]~,
\end{equation}
where $k_c$ is a factor $<1$ and the minimum is calculated among all the numerical cells.
For test-bed problems in Cartesian coordinates, taking $k_c=0.1-0.3$ is usually sufficient. 
In realistic models, however, numerical instabilities caused by the quadratic dispersion relation of the whistler waves arise. 
It becomes particularly problematic with spherical coordinates unless we use a very restrictive $k_c\approx 10^{-3}$. 

Recent work \citep{gonzalez18} includes other stabilizing techniques introduced in \cite{osullivan06} for the time advance of the non-linear terms. These techniques, namely the Super Time-Stepping and the Hall Diffusion Schemes, allow us to maintain stability and efficiently speed up the time evolution when the ambipolar or the Hall term dominates. 
Another common technique is the use of high-order dissipation (also called hyper-resistivity; \citealt{huba03}), or  a  predictor-corrector step advancing alternatively 
different field components.

\cite{vigano12a}  used a particularly simple method that significantly improves the stability of the scheme in spherical coordinates.
Their procedure to advance the solution 
from $t_n$ to $t_{n+1}=t_n+\Delta t$  can be summarized as follows:
\begin{itemize}
	\item
	starting from $\vec{B}^n$, all currents and electric field components are calculated \\ 
	$\vec{B}^n\rightarrow \vec{J}^n\rightarrow \vec{E}^n$;
	\item
	the toroidal field $\vec{B}_t^{n}$ is updated: $\vec{E}^n \rightarrow \vec{B}_t^{n+1}$;
	\item
	the new values $\vec{B}_t^{n+1}$ are used to calculate the modified current components and the toroidal part of the electric field $\vec{E}_t$: 
	$\vec{B}_t^{n+1} \rightarrow \vec{J}_p^\star \rightarrow \vec{E}_t^\star$;
	\item
	finally, we use the values of $\vec{E}_t^\star$ to update the poloidal components 
	$\vec{E}_t^\star \rightarrow \vec{B}_p^{n+1}$.
\end{itemize}
In \cite{Toth2008}, the authors discussed that such a two-stage formulation is equivalent to introduce a fourth-order hyper-resistivity. Since the toroidal component is advanced first,
it follows that the hyper-resistive correction only acts on the evolution of the poloidal components. In \cite{vigano12a} it was also shown that the additional correction given by $\vec{E}_t^\star$ contains higher-order spatial derivatives and scales with $(\Delta t)^2$, which is characteristic of hyper-resistive terms. They found a significant improvement in the stability of the method when comparing a fully explicit algorithm with the two-steps method, allowing to work with $k_c \approx 10^{-2}-10^{-1}$.

In the finite-difference schemes of \cite{vigano19}, the authors used a fourth-order Runge--Kutta scheme and found that the instabilities are especially significant when using fifth-order-accurate methods for the flux reconstruction (i.e.\ WENO5), which
needed to be combined with the application of artificial Kreiss--Oliger  dissipation  along each coordinate direction \citep{Calabrese2004}.   
A sixth-order derivative dissipation operator has a similar stabilizing effect, filtering the high-frequency modes which can not be accurately resolved 
by the numerical grid, at the cost of a potential loss of accuracy \citep{vigano19} . For this reason, they recommend using third-order schemes, that do not require any additional artificial Kreiss--Oliger dissipation. The typical Courant factors used were again quite low, $k_c \approx 10^{-2}-10^{-1}$.

The most advanced 3D code currently available \citep{wood15,gourgouliatos15a,gourgouliatos15b,gourgouliatos16,gourgouliatos18} uses spherical harmonic expansions of the magnetic potential functions for the angular directions (see Appendix \ref{app:formalism}), and a discretized grid in the radial one. The linear Ohmic terms are evaluated using a Crank--Nicolson scheme, while for the non-linear Hall terms an Adams-Bashforth scheme is used. The code is parallelized by considering spherical shells and uses the infrastructure of the PARODY code \citep{dormy98,aubert08}. Further details are available in \cite{gourgouliatos16}.

\section{Numerical tests}\label{sec:tests}

In order to calibrate the performance of  numerical methods or algorithms, it is crucial to provide analytical solutions against which the numerical results can be confronted. Unfortunately, there are not many such solutions in the 3D case with arbitrary coefficients in the generalized Ohm's law. For reference, we collect in this section a number of testbed cases with analytical solutions (most of them used in previous works \citealt{vigano12a,vigano19}), which probe different terms the induction equation. The successful completion of this battery of tests should be a good indicator of the performance of the codes. For the smooth tests below, \S~\ref{sec:whistler},\ref{sec:hall_drift},\ref{sec:pure_ohmic_mode},\ref{sec:ambipolar_test}, one can also check the convergence order of the numerical scheme, by computing the dependence of the relative errors (assessed for instance by a L2-norm) on the resolution used. The remaining two tests, where discontinuities form, are instead useful to test the robustness of the code, because near discontinuous solutions the convergence reduces to first order, regardless of the scheme. In all the following tests, we work in the Newtonian limit, $e^\nu=e^\lambda=1$.

\begin{figure}[t!]
	\centering
	\includegraphics[width=.8\textwidth]{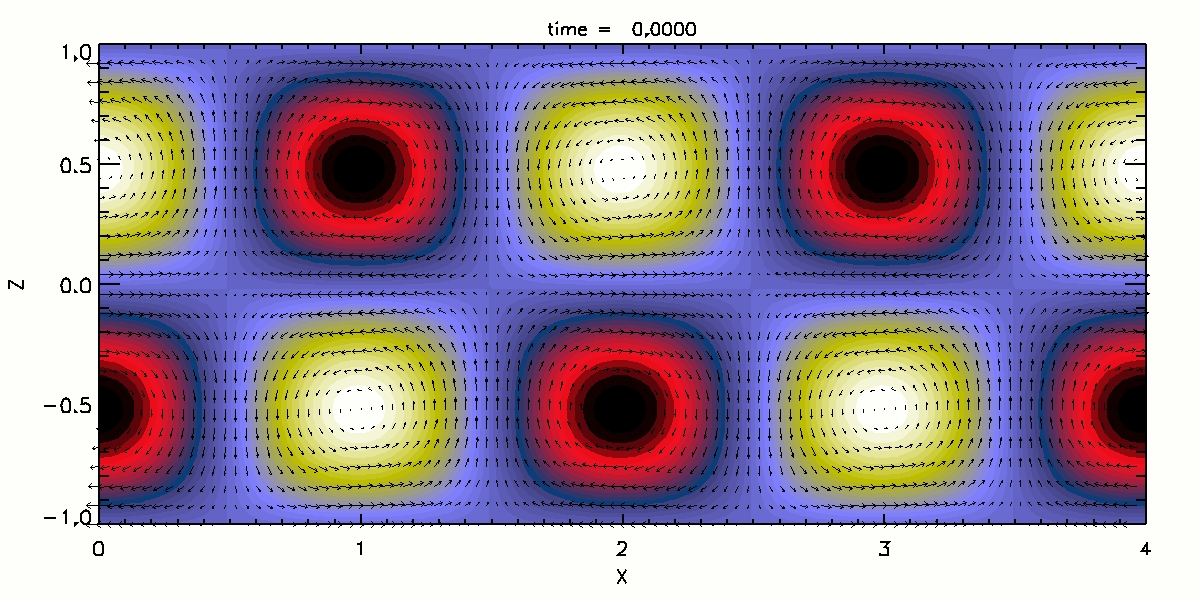}
	\includegraphics[width=.8\textwidth]{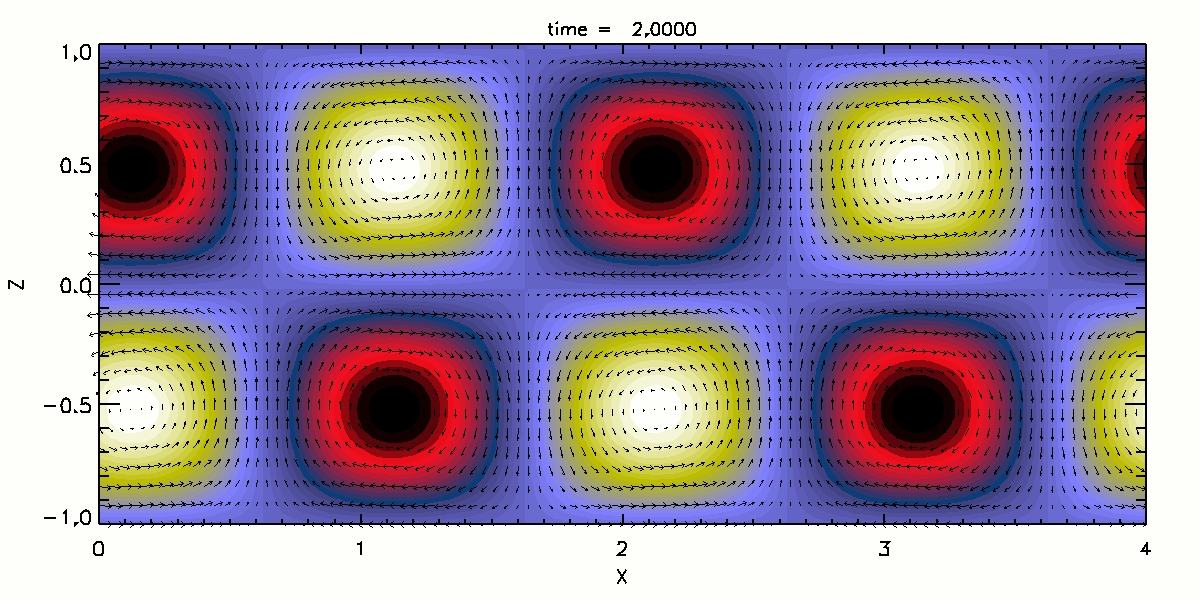}
	\caption{Evolution of the initial configuration defined by eqs.~(\ref{inimod1}) with $B_0=10^3~B_1$ and $k_x L=\pi$ from $t=0$ to $t=2$.  Arrows show the perturbed $B_x$ (subtracting $B_0$), and $B_z$-components, while the color scale represents the $B_y$ component (red/black positive, yellow/white negative). The travel time to cross over the whole domain ($4L$) is $t=2\sqrt{2}/\pi \tau_0 = 0.9003$, in units of $\tau_0$. (For video see supplementary material)} 
	\label{fig:whistler}
\end{figure}

\subsection{Whistler waves}\label{sec:whistler}

We begin by considering the case when only the Hall term is present in the induction equation.
In a constant density medium, the only modes of the Hall-MHD equation are the {\it whistler} or {\it helicon waves} (see \S~\ref{sec:induction_eq_char}), which consist in transverse field perturbations propagating along the magnetic field lines (notably known also in the terrestrial ionosphere \citealt{helliwell65,nunn74}). 
The first test we discuss is to follow the correct propagation of whistler waves. Consider a two-dimensional slab, extending from $z=-L$ to $z=+L$ in the vertical direction, with periodic boundary conditions in the $x$-direction, 
and assume that all variables are independent of the $y$-coordinate. 
For the following initial magnetic field:
\begin{eqnarray}
\label{inimod1}
B_x&=& B_0 + B_1\cos(k_x z)\cos(k_x x)~, \nonumber \\
B_y&=& \sqrt{2}B_1 \sin(k_x z)\cos(k_x x)~, \\
B_z&=& B_1\sin(k_x z)\sin(k_x x)~, \nonumber
\end{eqnarray}
where $k_x = n \pi /L$, $n=1,2,...$, and $B_1\ll B_0$, the linear regime admits a pure wave solution confined in the vertical direction and traveling in the $x$-direction, that is, the same 
eq. (\ref{inimod1}) replacing $x$ by $(x - v_w t)$, where the speed 
\begin{equation}\label{vel_whistler}
v_w=-\frac{c}{4\pi e n_0}\sqrt{2}~k_x B_0=-\sqrt{2}\frac{L^2 k_x}{\tau_0}~.
\end{equation}
Here we have defined the reference Hall timescale as
\begin{equation}\label{tau0_wave}
\tau_0=\frac{4\pi e n_0 L^2}{c B_0}~.
\end{equation}
As an example, in Fig.~\ref{fig:whistler} and attached movie - online version only - , we report the evolution of this initial 
configuration with $B_0=10^3~B_1$, and $k_x L=\pi$, from $t=0$ to $t=2$ (in units of $\tau_0$), 
in a $200 \times 50$ Cartesian grid.  
The perturbations travel through the horizontal domain twice, with negligible dissipation or dispersion.
\cite{vigano12a} ran the test for hundreds of Hall timescales without any indication of instabilities, 
even though electrical resistivity is set to zero.  By varying the values $k_x$ and $B_0$, one can confirm that the velocity of the perturbations in the simulation scales 
linearly with both parameters. 
An additional twist is to consider the same problem in a 2D or 3D box, but with an arbitrary rotation of the coordinates, in order to test the correct propagation in a more general direction, not aligned to any axis.

\begin{figure}[t]
	\centering
	\includegraphics[width=.45\textwidth]{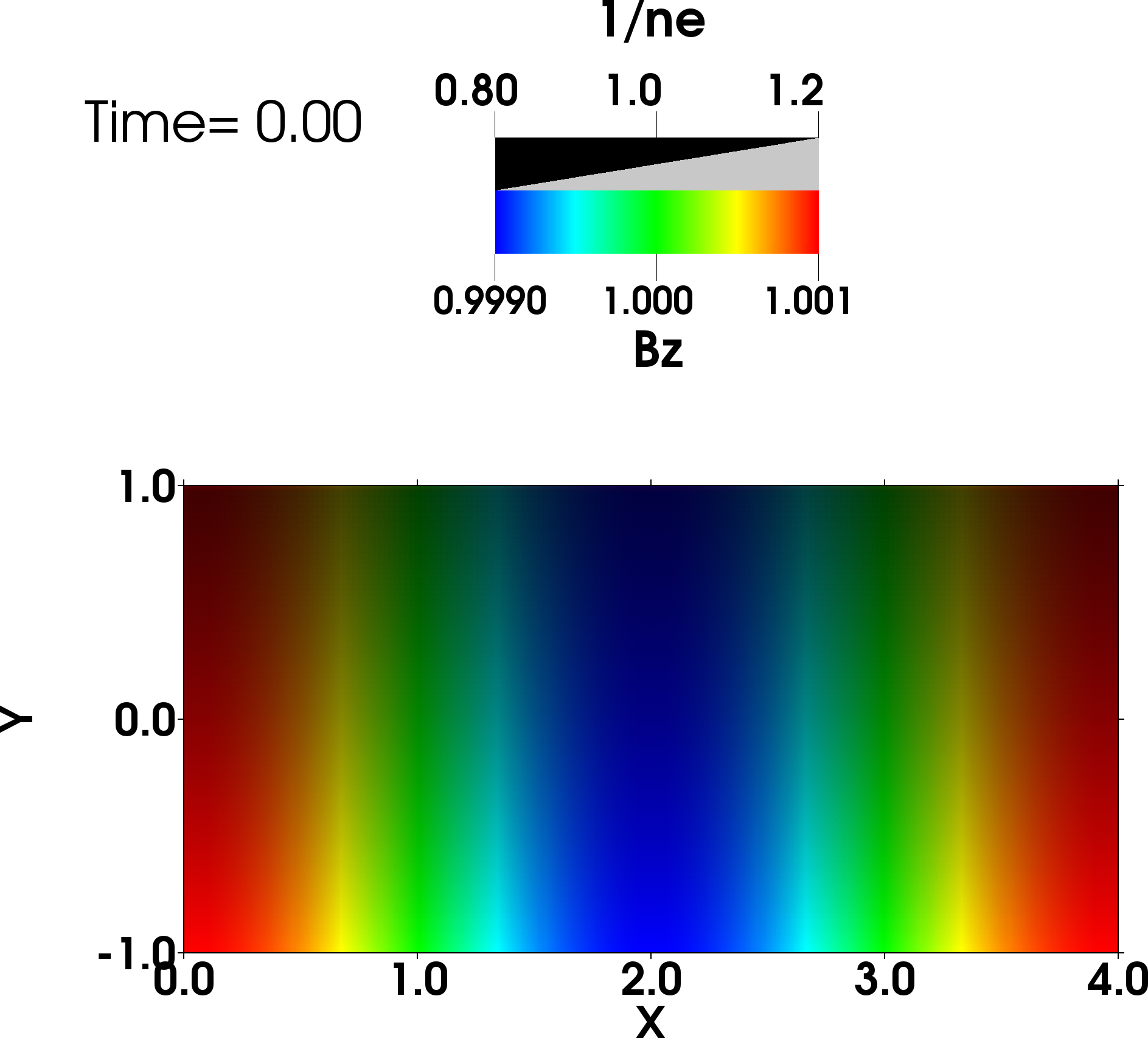}
	\includegraphics[width=.45\textwidth]{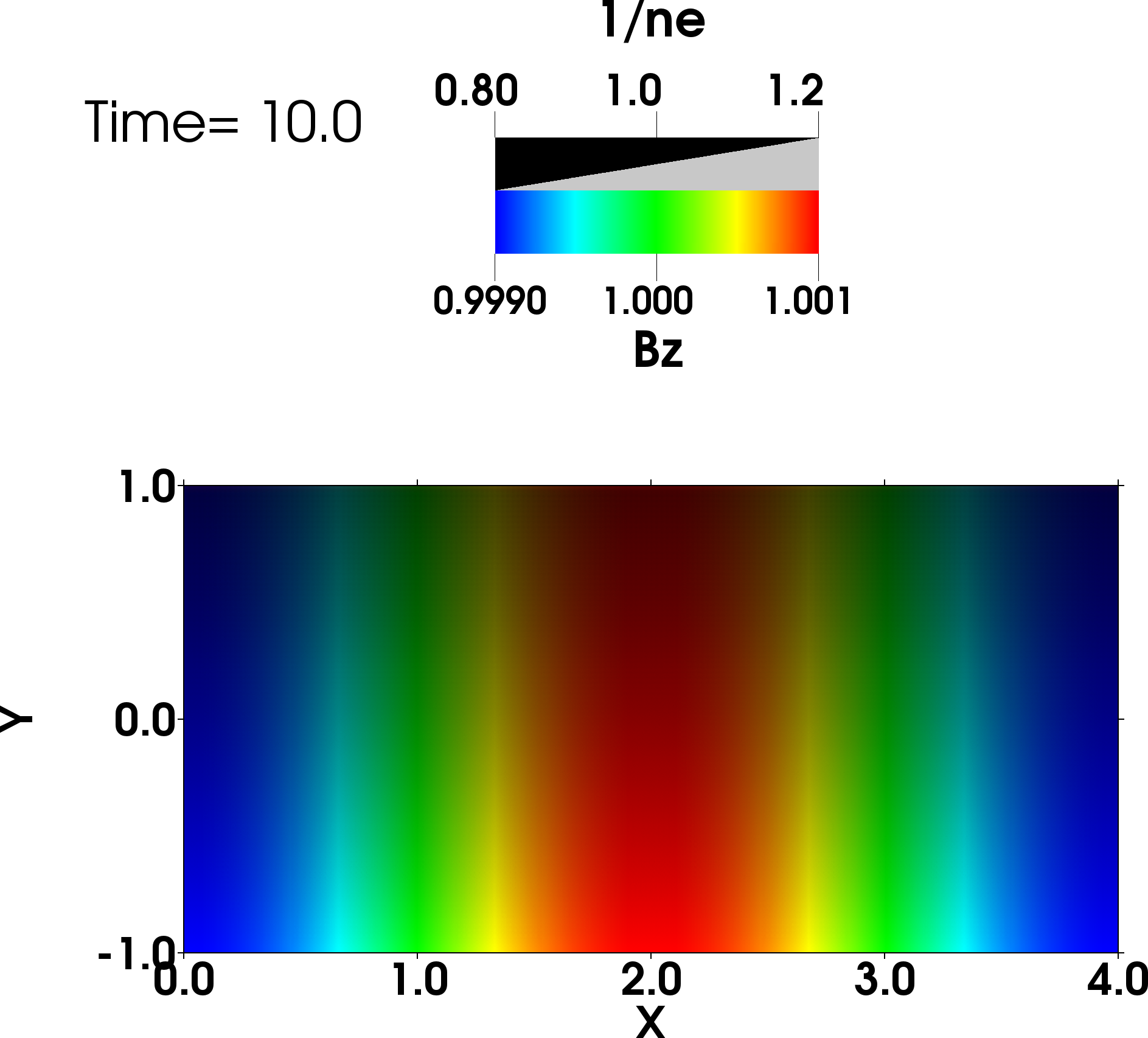}\\
	\vspace{0.6cm}
	\includegraphics[width=.45\textwidth]{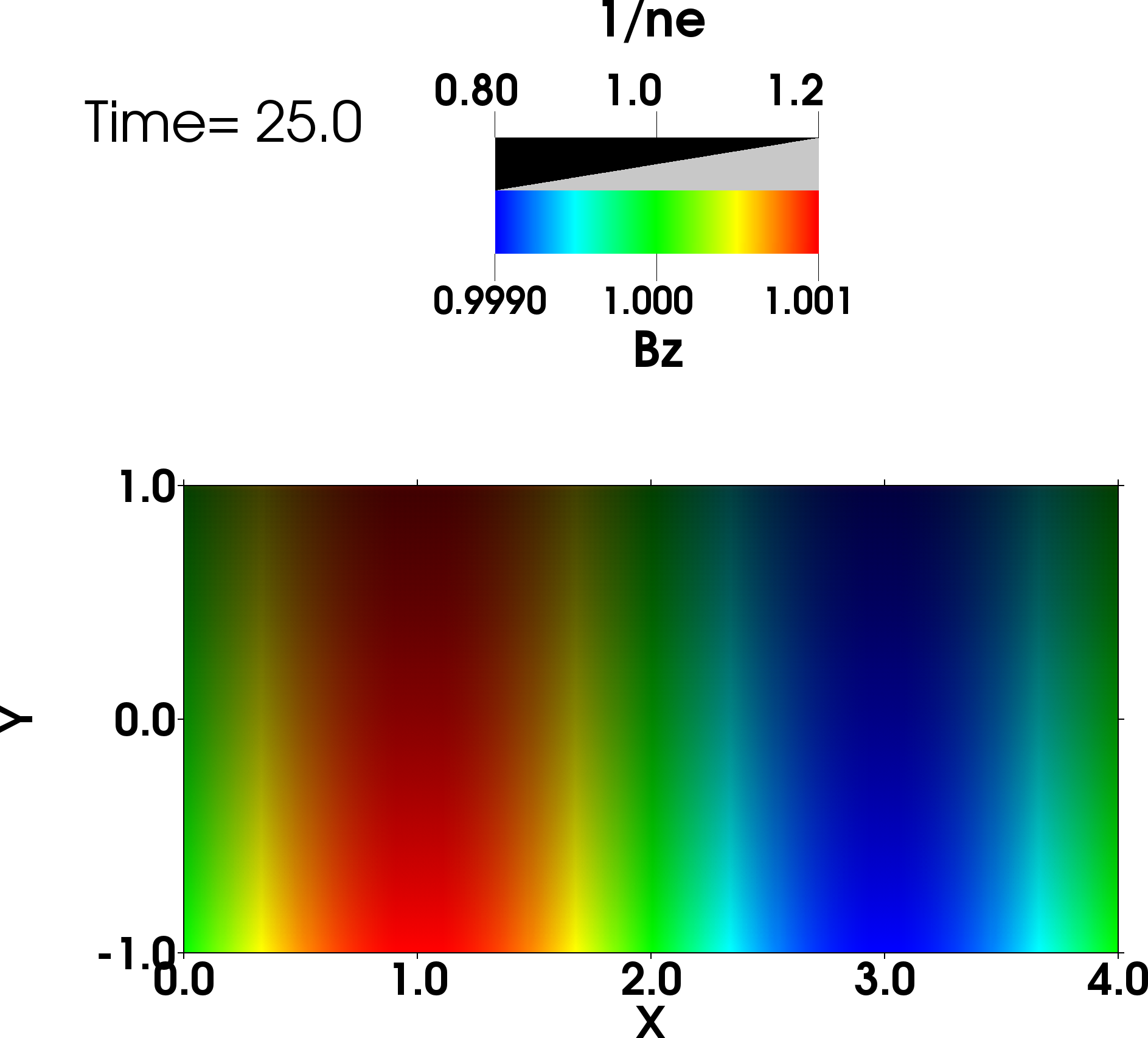}
	\includegraphics[width=.45\textwidth]{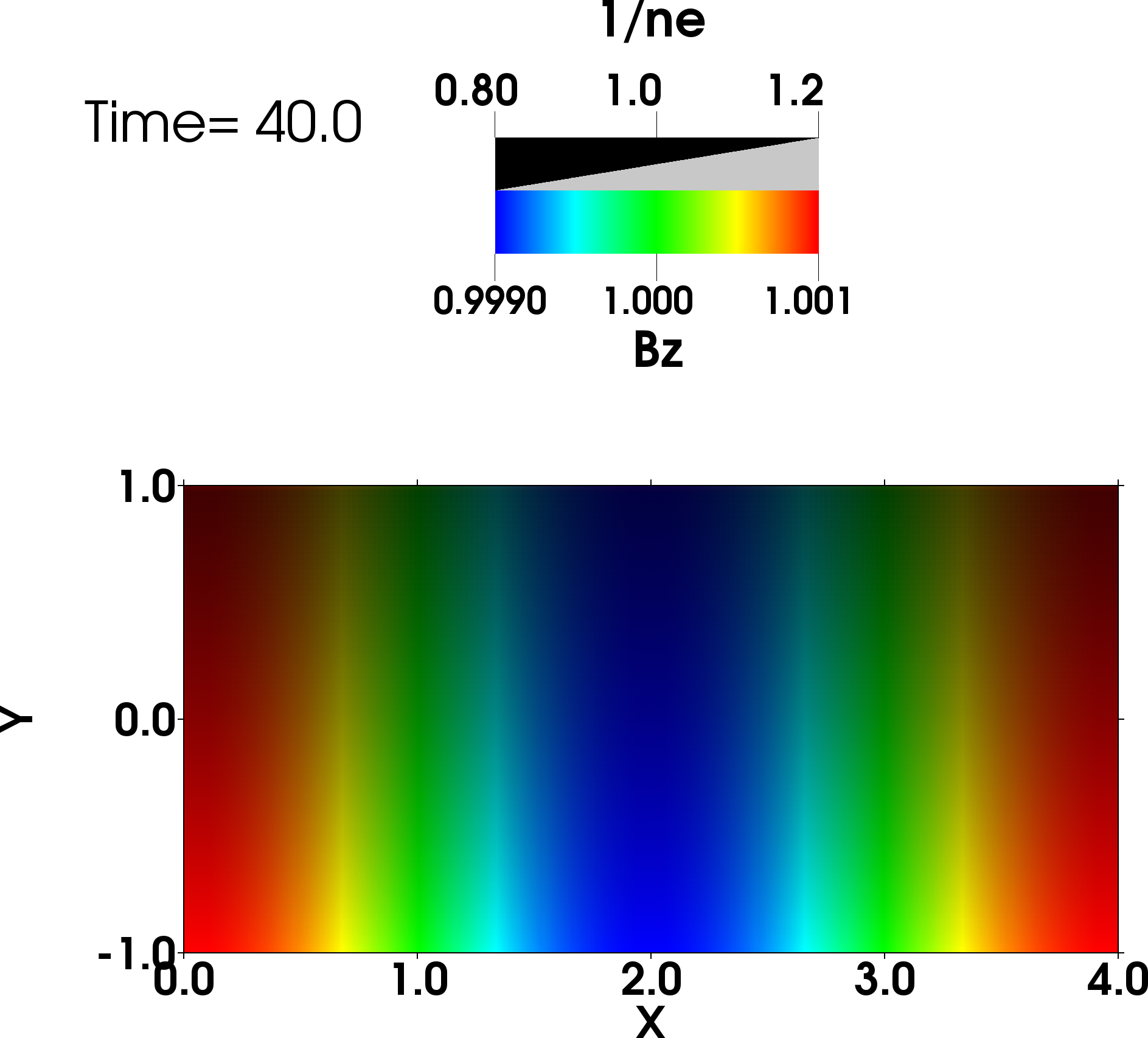}
	\caption{Hall drift wave: evolution of the initial configuration defined by eq.~(\ref{inimod2}) with $B_0=10^3~B_1$, $L=1$ and $k_x L=\pi/2$  at $t/\tau_0=0,10,25,40$ (from left top to right bottom), corresponding to 0, 0.5, 1.25, and 2 crossing times, respectively. Blue-to-red color scales indicate the values of $B_z$, while the black shadowing increases with the value of $1/n_e(y)$. Simulation run with the Simflowny-based code \citep{vigano19}, with a resolution of $200\times100$ points. (For video see supplementary material)} 
	\label{fig:drift}
\end{figure}

\subsection{Hall drift waves}\label{sec:hall_drift}

In the second test of the Hall term, we remove the assumption of a constant charge density background. In presence 
of a charge density gradient, additional transverse modes appear, the so-called {\it Hall drift waves}, 
which propagate in the $\vec{B} \times \vec{\nabla} n_e$ direction. 
Let us consider the same domain as in the previous test, but with a stratified background in the $y$-direction with  $n_e$ given by
\begin{equation}\label{n_wave}
n_e(y)= \frac{n_0}{1+\beta_L y}~,
\end{equation}
where $n_0$ is a reference density, with an associated Hall timescale $\tau_0$ defined in eq.~(\ref{tau0_wave}), and $\beta_L$ is a parameter with dimensions of inverse length. 
We apply periodic boundary conditions in the $x$-direction, while in the $y$-direction, an infinite domain can be simulated by copying the values of the magnetic field in the uppermost and lowermost cells ($y=\pm L$) into their first neighbor ghost cells.


For the following initial configuration:
\begin{eqnarray}\label{inimod2}
B_x&=& 0~, \nonumber \\
B_y&=& 0~, \nonumber \\
B_z&=& B_0 + B_1\cos(k_x x) ~,
\end{eqnarray}
and small perturbations ($B_1\ll B_0$), the solution at early times consists in pure Hall drift waves traveling in the $x$-direction with speed
\begin{equation}\label{vel_hd}
v_{hd}=\frac{\beta_L L^2}{\tau_0}~.
\end{equation}
The solution in the linear regime can be obtained by replacing $x$ by $(x - v_{hd} t)$ in eq. (\ref{inimod2}).
For the particular model shown in Fig.~\ref{fig:drift}, with $B_0=10^3~B_1, k_x=\pi/2, L=1$ and $\beta_L=0.2$, we have a horizontal 
drift velocity of $v_{hd}=0.2 L/\tau_0$, corresponding to a crossing time of 20 $\tau_0$. 
The figure shows the initial configuration of $B_z$ (top left) and the evolution of the perturbation after 0.5, 1.25 and 2 crossing times, respectively (top right, bottom left and bottom right). The shadow increases with the value of $1/n(y)$. For the Hall drift modes, the propagation velocity scales linearly with both $B_0$ and the gradient of $n_e^{-1}$, but it is independent of the wavenumber of the perturbation. All these properties are correctly reproduced. After many cycles (the number depending on the $B_1/B_0$ ratio), deviations from the purely advected, smooth solution begin to be visible. This is an expected non-linear effect that we discuss next.

\subsection{The nonlinear regime and Burgers flows}\label{sec_burgers}

\begin{figure}
	\centering
	\includegraphics[width=.75\textwidth]{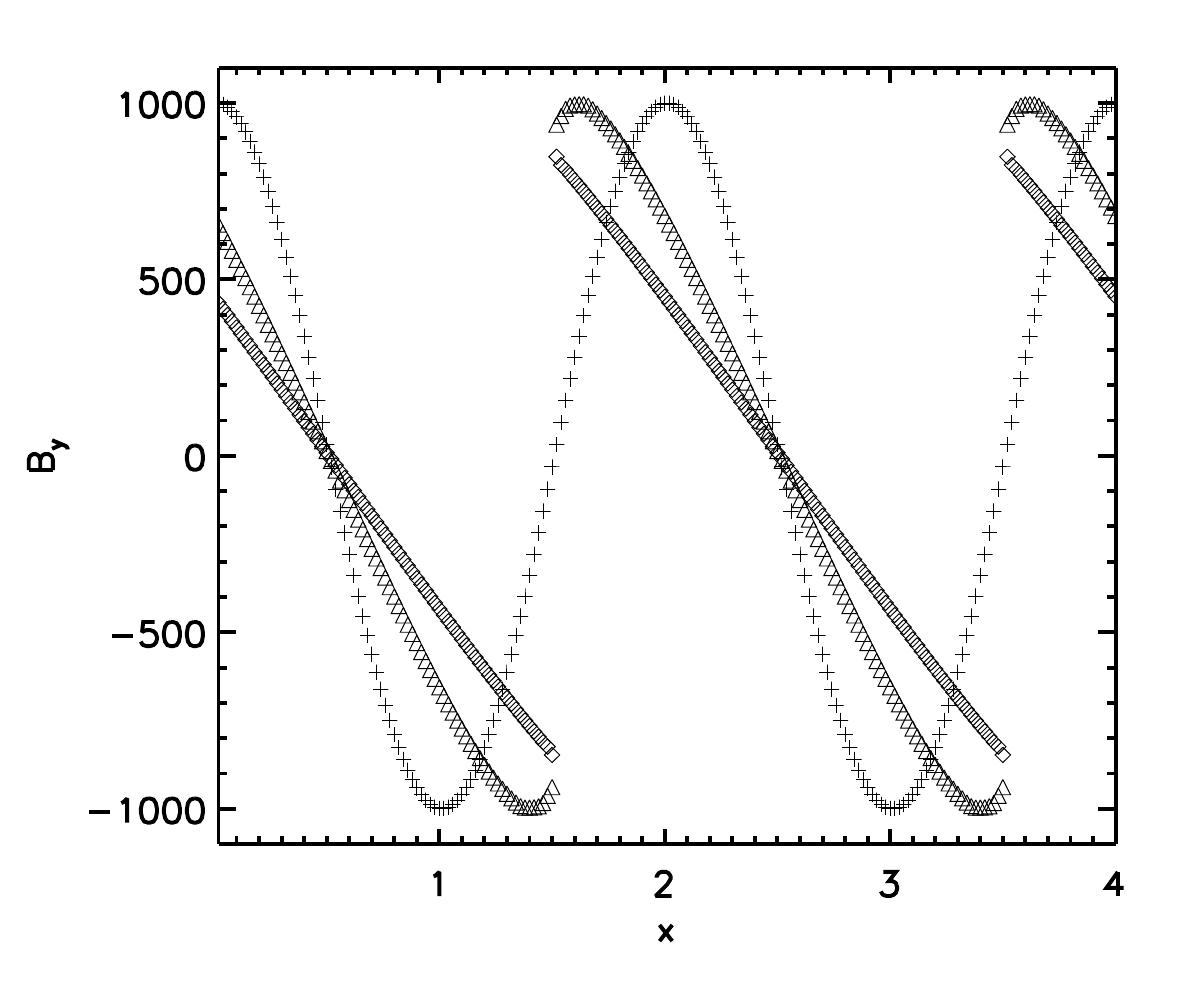}
	\caption{Horizontal section of the evolution of the initial configuration defined by eq.~(\ref{inimod3}) with $B_0=10^3$ and $k_x L=\pi$ at $t=0$  (crosses), $t=2 \tau_0$ (triangles)
	and $t=4 \tau_0$ (diamonds). The shock forms at $t=2 \tau_0$. The classical sawtooth shape developed during the evolution of the Burgers equation is evident. Figure courtesy of \cite{vigano12a}.} 
	\label{fig:burg1}
\end{figure}

With the two previous tests, we can check if a numerical code can reproduce the propagation of the fundamental modes at the correct speeds. However, these are valid solutions only in the linear regime. Let us consider more carefully the evolution of the $B_y$ component in a medium stratified in the $z$-direction. Assuming that $B_x=B_z=0$, 
 the governing equation reduces to:
\begin{equation}
\frac{\partial B_y}{\partial t} + g(z) {B_y}\frac{\partial B_y}{\partial x}=0 ~.
\label{eq_burg0}
\end{equation}
This is a version of the Burgers equation (which solution is well known) in the $x$-direction with a coefficient that depends on the $z$ coordinate:
\begin{equation}
g(z)=-\frac{d}{dz}\left(\frac{c}{4\pi e n_e}\right)~.
\end{equation}
If we consider the following initial configuration:
\begin{eqnarray}
B_x&=& 0~, \nonumber \\
B_y&=& B_0 \cos(k_x x) ~,  \\
B_z&=& 0~,\nonumber
\label{inimod3}
\end{eqnarray}
on a stratified background with $n_e(z)= \frac{n_0}{1+\beta_L z}$, we have  $g(z)=-\beta_L L^2/\tau_0 B_0$ and we 
can directly compare to the solution of the Burgers equation in one dimension, which evolves to form discontinuities from smooth initial data.

To handle this problem, we can make use of well-known HRSC numerical techniques to design a particular treatment to the quadratic term in $B_y$
\footnote{In axial symmetry, we find an analogous equation for the $\varphi$-component}. 
A key issue is to consider the Burgers-like term in conservation form:
\begin{equation}
\frac{\partial B_y}{\partial t} + \frac{\partial \hat{F}}{\partial x}=0~,
\end{equation}
where $\hat{F}=g(z) B_y^2/2$, which can now be treated with an upwind conservative method \citep{vigano12a}.
In this case, the wave velocity determining the upwind direction is given by $g(z)B_y$. 

Expressing the evolution equations in conservative form is crucial when solving problems with shocks or other discontinuities, since non-conservative methods may result in the incorrect propagation speed of discontinuous solutions \citep{toro09}. 
In Fig.~\ref{fig:burg1} we show snapshots of the evolution of the initial conditions (\ref{inimod3}) with $k_x L=\pi$, $B_0=10^3$, and $\beta_L L=0.2$, taken from \cite{vigano12a}. It follows the typical Burgers evolution. The wave breaking and the formation of a shock at $t=2 \tau_0$ is clearly captured. We remark again that this test is done with zero physical resistivity, i.e., in the limit $\omega_B \tau \rightarrow \infty$, which is not reachable by spectral methods or centered-difference schemes in non-conservative form. In \cite{vigano19}, the reader can find 
more details about the solutions obtained with different reconstruction schemes. 

\begin{figure*}[t]
	\centering
	\includegraphics[width=.45\textwidth]{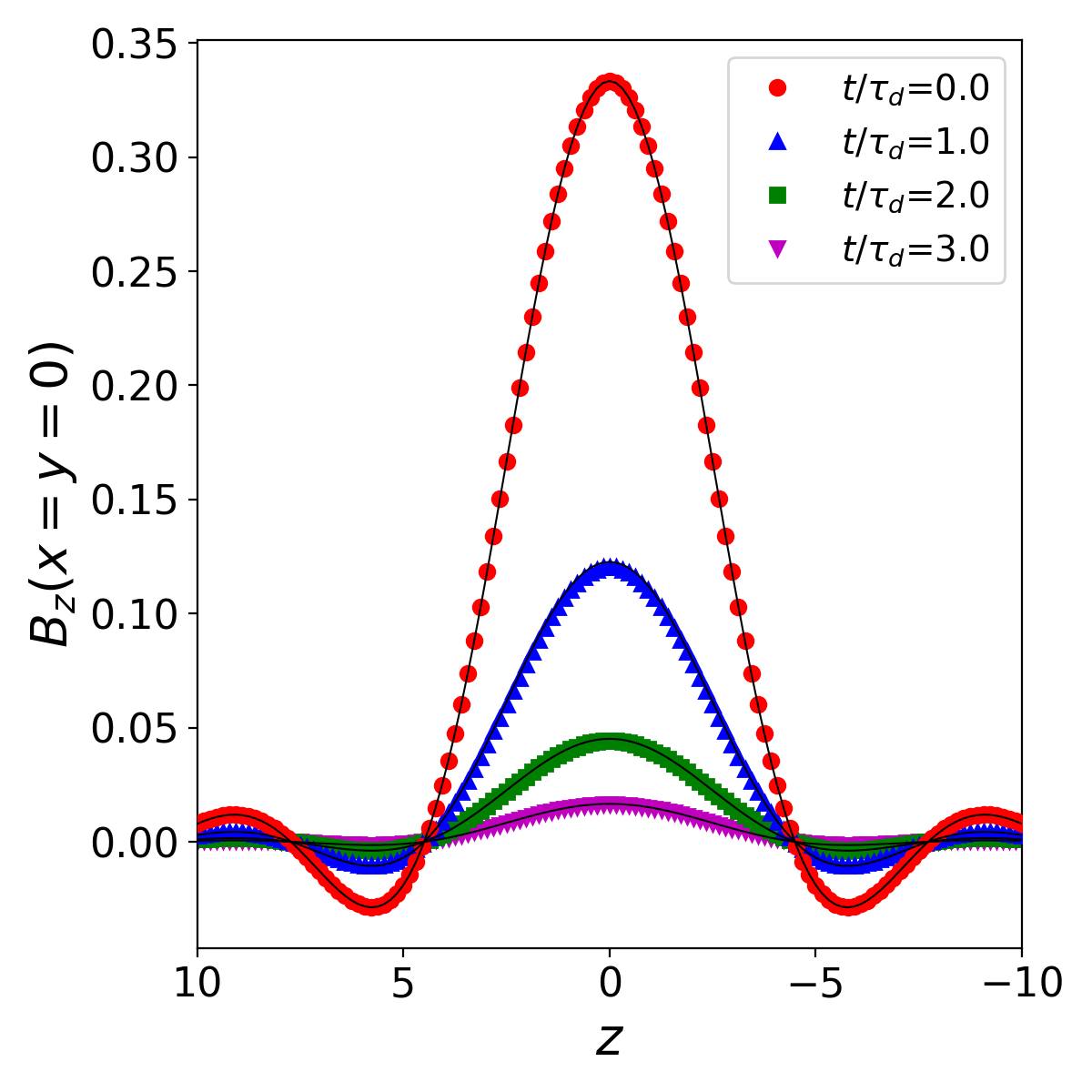}
	\includegraphics[width=.45\textwidth]{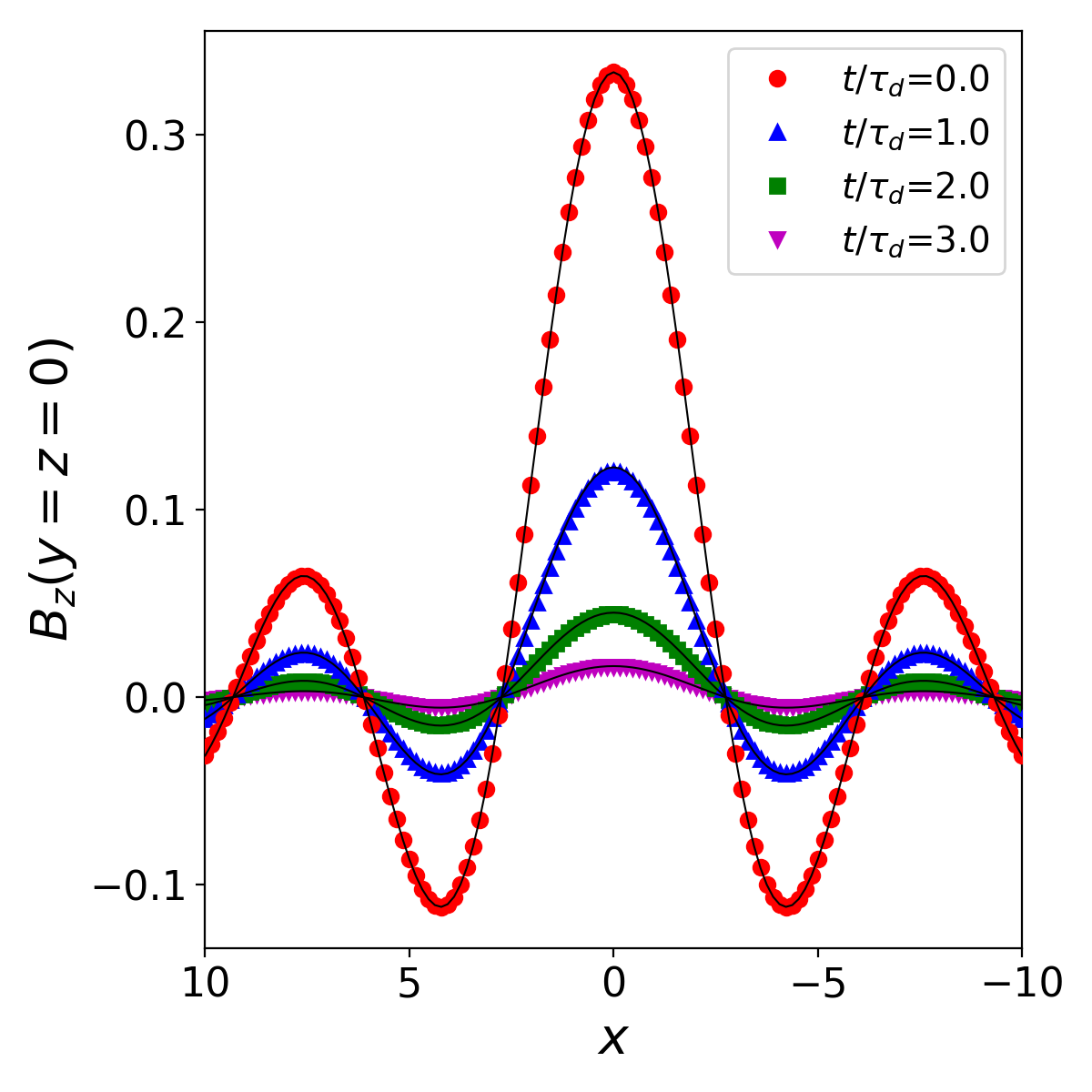}
	\includegraphics[width=.45\textwidth]{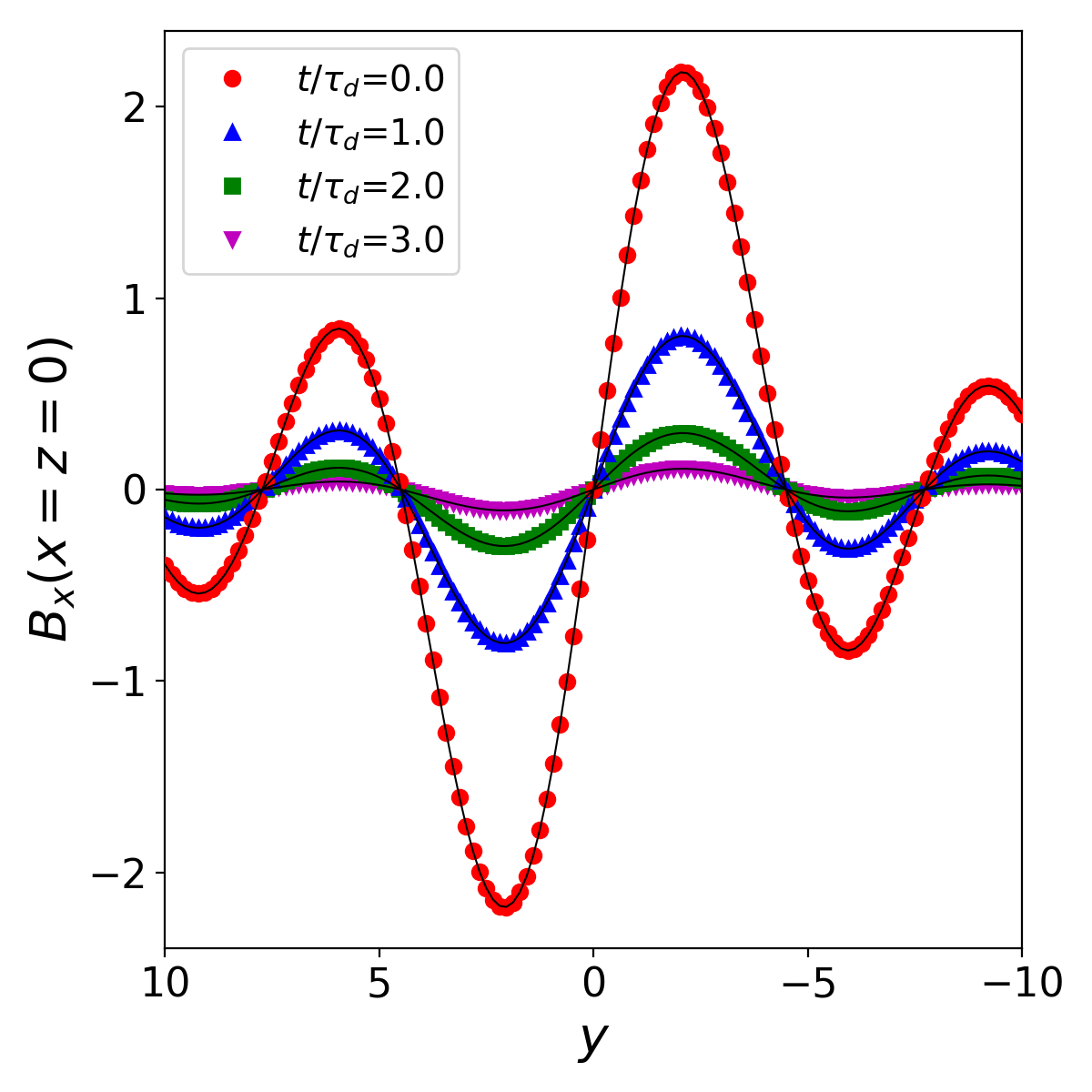}
	\caption{Evolution of the purely Ohmic modes, eqs.~(\ref{eq:mf_ohmic1})-(\ref{eq:mf_ohmic3}), with $\alpha =1$, at $t=0, 1, 2,$ and $3$ diffusion times ($\tau_d$).
		The simulation has been run in a $[-10,10]^3$ cubic domain, with a resolution of $128^3$ equally-spaced Cartesian grid points with the Simflowny-based code \citep{vigano19}. 
		In the figure we compare the analytical (black lines) and numerical (color symbols) profiles of $B_z(x=0,y=0,z)$ (i.e., $B_r(r,\theta=0)$, top left), $B_z(x,y=0,z=0)$ (i.e., $-B_\theta(r,\theta=\pi/2)$, top right), and $B_x(x=0,y,z=0)$ (i.e., $B_\varphi(r,\theta=\pi/2, \varphi=\pi/2)$, bottom).}
	\label{fig:bessel_decay}
\end{figure*}

\subsection{Ohmic dissipation: self-similar axisymmetric force-free solutions}\label{sec:pure_ohmic_mode}

In spherical geometry, one of the few existing analytical solutions is the evolution of pure Ohmic dissipation modes.
Considering the limit $\omega_B \tau \rightarrow 0$, and a constant $\eta$, the induction equation reads:
\begin{equation}
\frac{\partial \vec{B}}{\partial t}=-\eta\vec{\nabla}\times(\vec{\nabla}\times\vec{B})~.
\end{equation}
It is straightforward to show that a force free  magnetic field satisfying $\curlB=\alpha\vec{B}$, with constant $\alpha$, is an Ohmic eigenmode, since the induction equation is reduced to
\begin{equation}\label{eq:induction_ohmic_mode}
\frac{\partial \vec{B}}{\partial t} = - \eta \alpha^2\vec{B}~.
\end{equation}
Therefore, each component of the magnetic field decays exponentially with the diffusion timescale $\tau_d=(\eta \alpha^2)^{-1}$. 
We note that the evolution of each component is completely decoupled in this case.

In spherical coordinates, the solutions of eq.~(\ref{eq:induction_ohmic_mode}) are described by factorized functions, which radial parts involve the spherical Bessel functions.
The regularity condition at the center selects only one branch of the spherical Bessel functions (of the first kind), which, for the $(l,m)=(1,0)$ mode are 
\begin{eqnarray}
&& B_r=\frac{B_0R}{r}\cos\theta \left(\frac{\sin x}{x^2} - \frac{\cos x}{x}\right) ~, \label{eq:mf_ohmic1}\\
&& B_\theta=\frac{B_0R}{2r}\sin\theta\left(\frac{\sin x}{x^2}-\frac{\cos x}{x}-\sin x\right)~, \label{eq:mf_ohmic2}\\
&& B_\varphi=\frac{k B_0R}{2}\sin\theta \left(\frac{\sin x}{x^2}-\frac{\cos x}{x}\right) ~, \label{eq:mf_ohmic3}
\end{eqnarray}
where $x=\alpha r$ and $k=\pm 1$.
With this initial condition, we follow the evolution of the modes during several $\tau_d$, until the magnetic field is almost completely dissipated.  As boundary conditions, 
we impose the analytical solutions for $B_\theta$ and $B_\varphi$. 
Fig.~\ref{fig:bessel_decay} compares the numerical (crosses) and analytical (solid lines) solutions of $B_r$ and $B_\varphi$ at different times, for a model with $\alpha=1$, and a $[-10,10]^3$ cubic domain, run with the Simflowny-based code \citep{vigano19}.

\subsection{Ambipolar diffusion: the Barenblatt-Pattle solution}\label{sec:ambipolar_test}

\begin{figure}[ht] 
	\centering
	\includegraphics[width=0.8\linewidth]{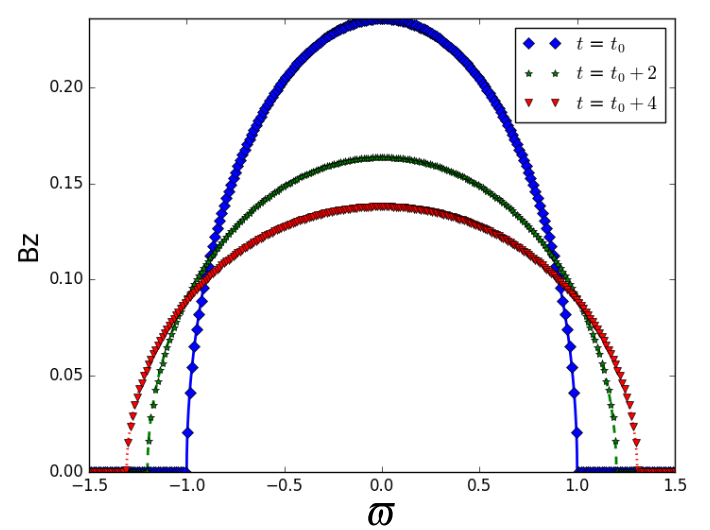} 
	\caption{Barenplatt-Pattle solution for three different times in a Cartesian grid with a resolution of $256^2$. Solid lines correspond to the analytical solutions and symbols represent the numerical solutions. Figure courtesy of \cite{vigano19}.} 
	\label{fig:ambipolar} 
\end{figure}

To test the ambipolar term, we now consider the case of a constant $f_a$ and set to zero the Hall and Ohmic coefficients. 
In axial symmetry and cylindrical coordinates, there exists an analytical solution corresponding to the diffusion of an infinitely long magnetic flux. Let us consider the 
evolution of the only component of the magnetic field 
$\vec{B}= B_z \hat{z}$, in the direction of the fluxtube, which only depends on the cylindrical radial coordinate ($\varpi$). 

The currents are perpendicular to the magnetic field so that $(\vec{j}\times\vec{B})\times\vec{B}=-B^2\vec{j}$, and the induction equation with the ambipolar term is reduced to
\begin{eqnarray}
\derparn{t}{B_z} = f_a \frac{1}{\varpi}\frac{\partial}{\partial\varpi}\left[ \varpi\left(B_z^2 \frac{\partial B_z}{\partial\varpi}\right) \right].
\end{eqnarray}
This form is analogous to the non-linear diffusion equation 
\begin{equation}
\derparn{t}{u} = \vec{\nabla}\cdot(mu^{m-1}\vec{\nabla} u)~,
\end{equation}
where $m$ is a power index. The analytical 2D solutions proposed by Barenblatt and Pattle \citep{barenblatt52,pattle59} consist of a delta function of integral $\Gamma$ at the origin, which diffuses outwards with finite velocity. We note that the diffusion front is clearly defined, contrarily to the infinite front speed of a linear diffusion problem. This analytic solution can be explicitly written as follows:
\begin{equation}
u(\varpi,t)=\max\left\{0, t^{-\alpha}\left[\Gamma - \frac{\alpha(m-1)}{2dm}\frac{\varpi^2}{t^{\frac{2\alpha}{m}}}\right]^{\frac{1}{m-1}}\right\}~,
\end{equation}
where $d$ is the dimension of the problem, and $\alpha=(m-1+2/d)^{-1}$. The initial pulse spreads with a front located at a distance $\varpi_f$ from the origin, given by
\begin{equation}
\varpi_f(t)=\left(\frac{2\Gamma dm}{\alpha(m-1)}\right)~t^{\alpha/d} ~.
\end{equation}
In \cite{vigano19}, they studied the evolution of the model with $d=2$, $m=3$, $\alpha=1/3$, and $f_a=3$, which gives the explicit solution 
\begin{equation}
B_z(\varpi,t) = t^{-(1/3)}\left[\Gamma - \frac{1}{18}\frac{\varpi^2}{t^{1/3}}\right]^{1/2}~.
\end{equation}
In Fig.~\ref{fig:ambipolar} we show three snapshots of the evolution, starting with $t_0=1$, and $\Gamma=1/18$. The front propagates according to $\varpi_f(t)=t^{1/6}$.
The numerical results correctly reproduce the expected shape of the expanding flux tube and the propagation speed of the front. The sharp discontinuity in the slope of $B_z$ near the front end was found to be well-reproduced even for low resolutions.

\subsection{Evolution of a purely toroidal magnetic field}\label{tor_sec}

Finally, to conclude our proposed series of tests and examples, we consider the evolution of a pure toroidal magnetic field confined into a spherical shell, $R_{\rm core}<r<R$, under the combined action
of both Ohmic dissipation and the Hall term.  
This case does not have an analytical solution, but we believe it is an important (yet relatively simple) test that can highlight some relevant issues.
For simplicity, we impose as boundary conditions that all components of the magnetic field vanish at both boundaries. 

We consider the realistic NS background profile of Fig.~\ref{fig:ns_profile}, with $R_{\rm core}=10.8~$km, $R=11.6~$km, and we set a constant temperature of $10^8$ K, which corresponds to a density-dependent magnetic 
diffusivity in the range $\eta\sim 0.01-10 $ km$^2$/Myr. Our initial magnetic field is given by the following expression:
\begin{equation}\label{btorq}
B_\varphi=-B_0\frac{(R-r)^2(r-R_{\rm core})^2\sin\theta\cos\theta}{r}~,
\end{equation}
where $B_0$ is a normalization factor adjusted to fix the initial maximum value that the toroidal magnetic field reaches across the star (denoted by $B_t^0$). 
\begin{figure}[t]
	\centering
	\includegraphics[width=.32\textwidth]{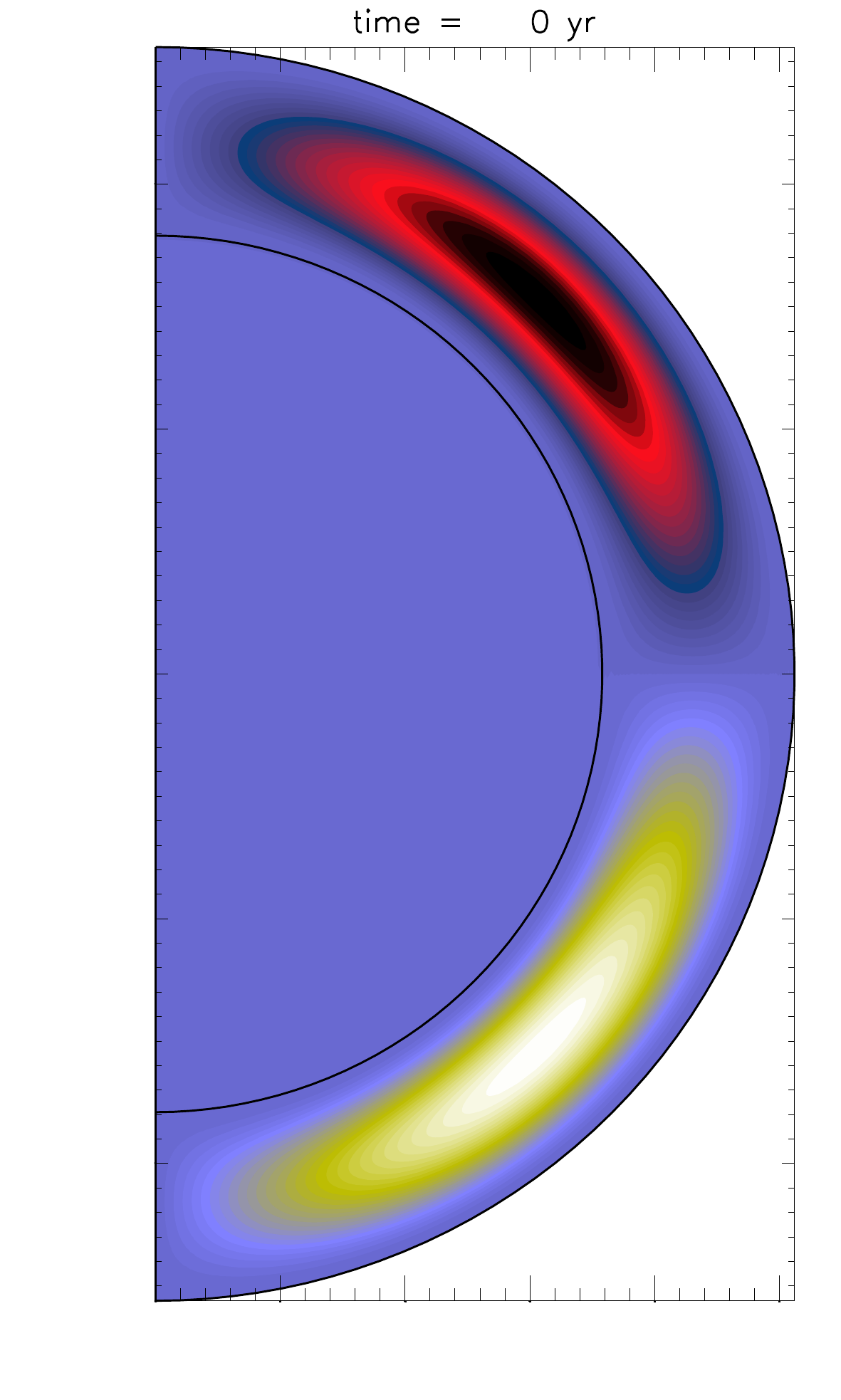}
	\includegraphics[width=.32\textwidth]{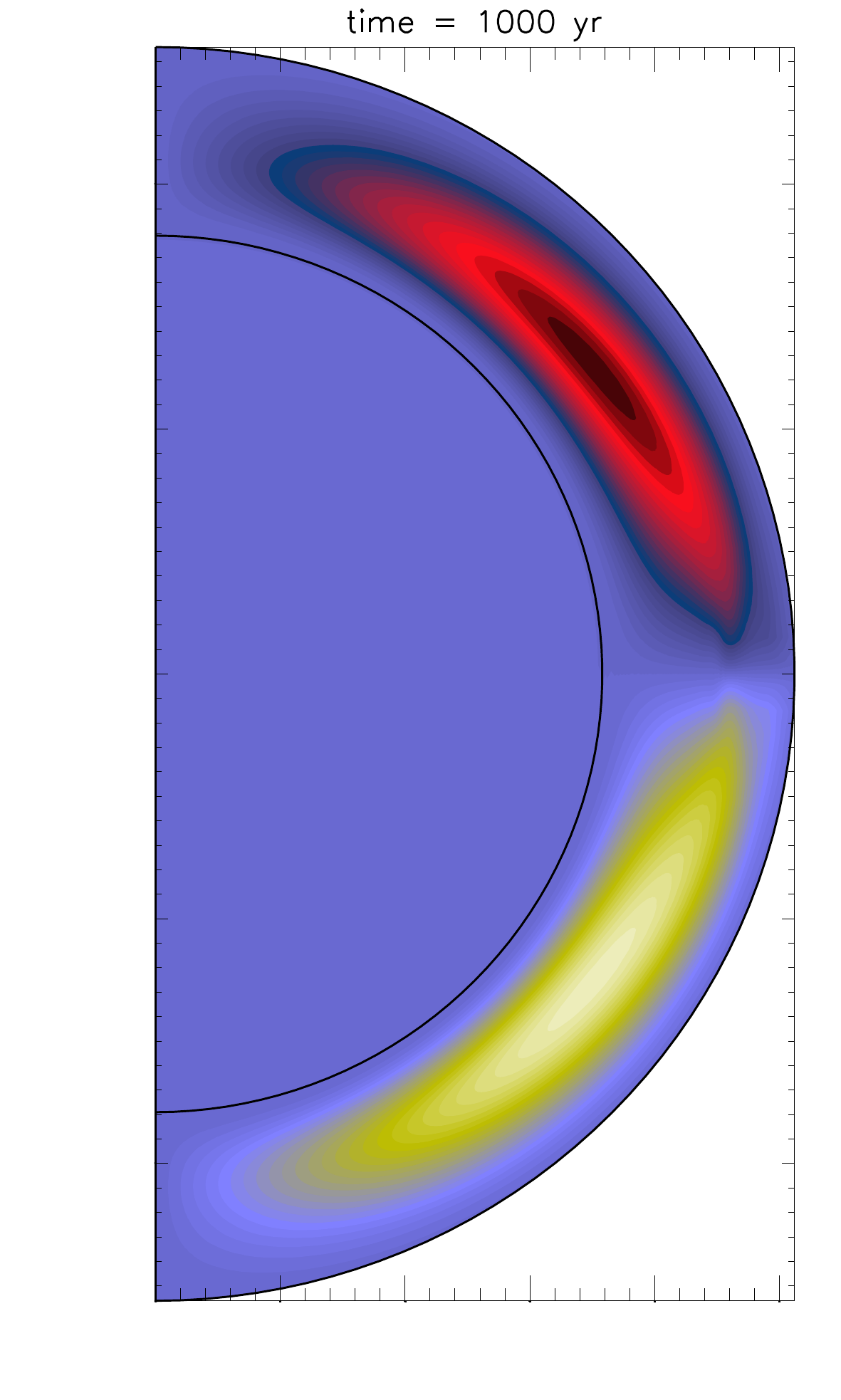}
	\includegraphics[width=.32\textwidth]{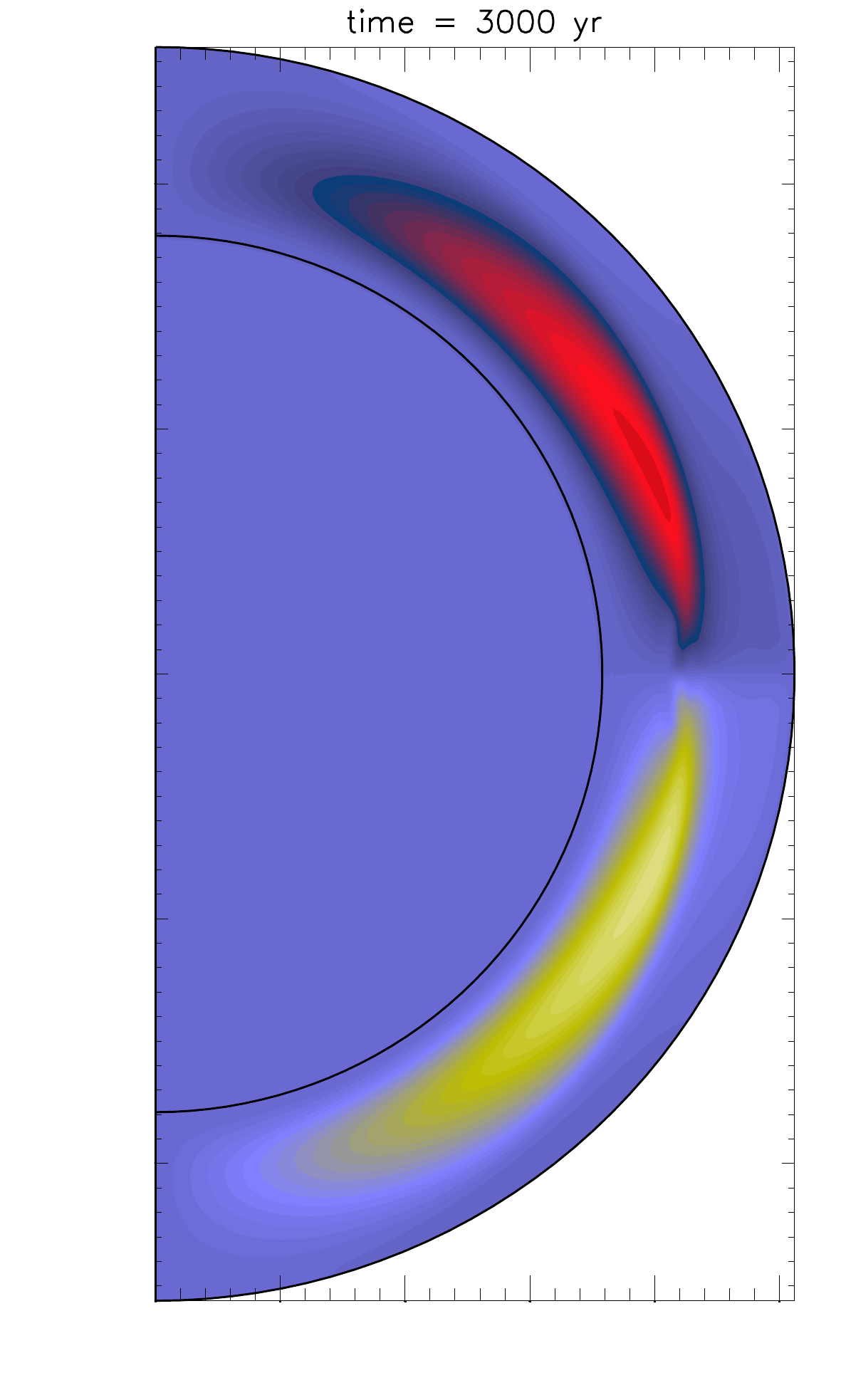}
	\includegraphics[width=.32\textwidth]{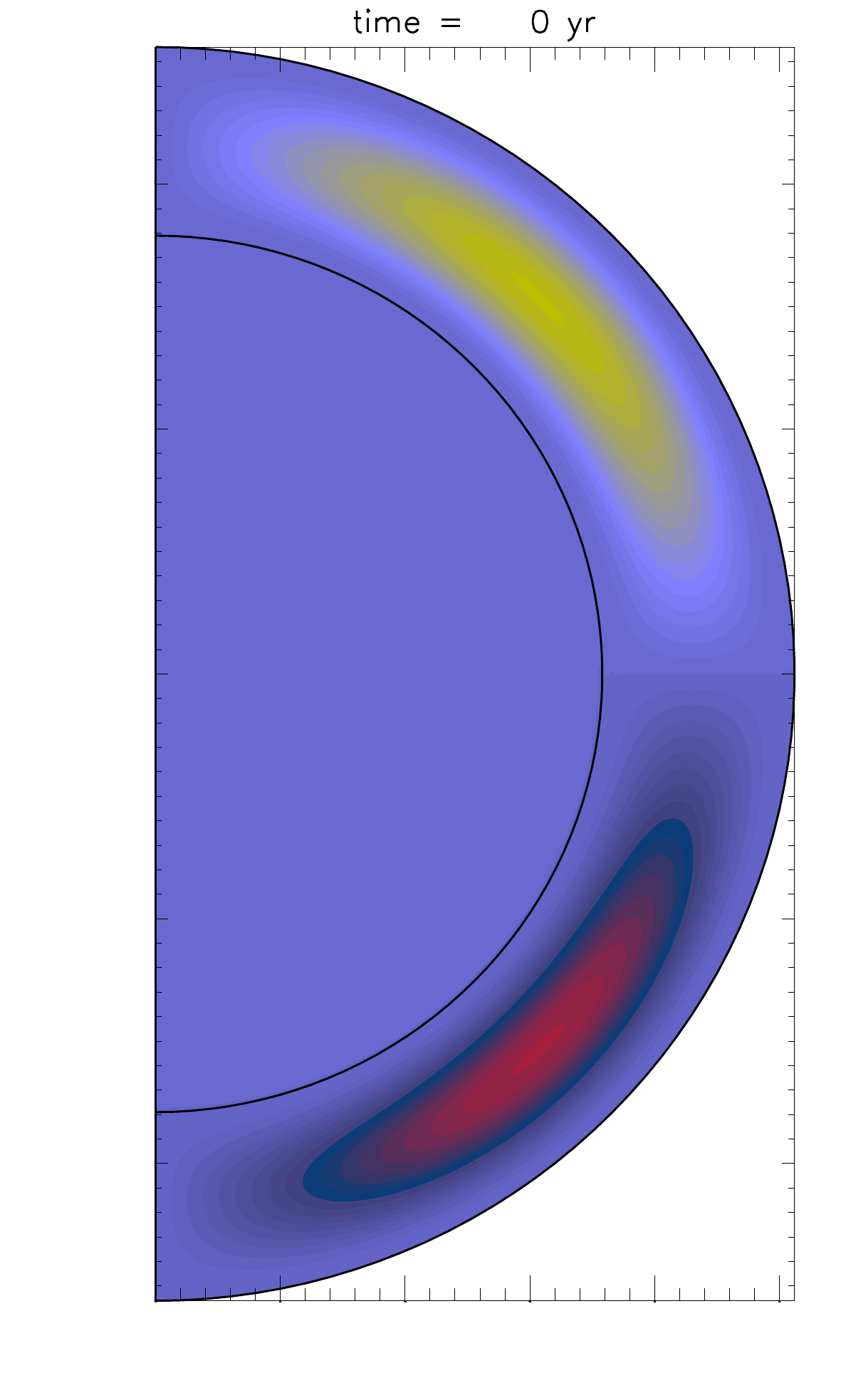}
	\includegraphics[width=.32\textwidth]{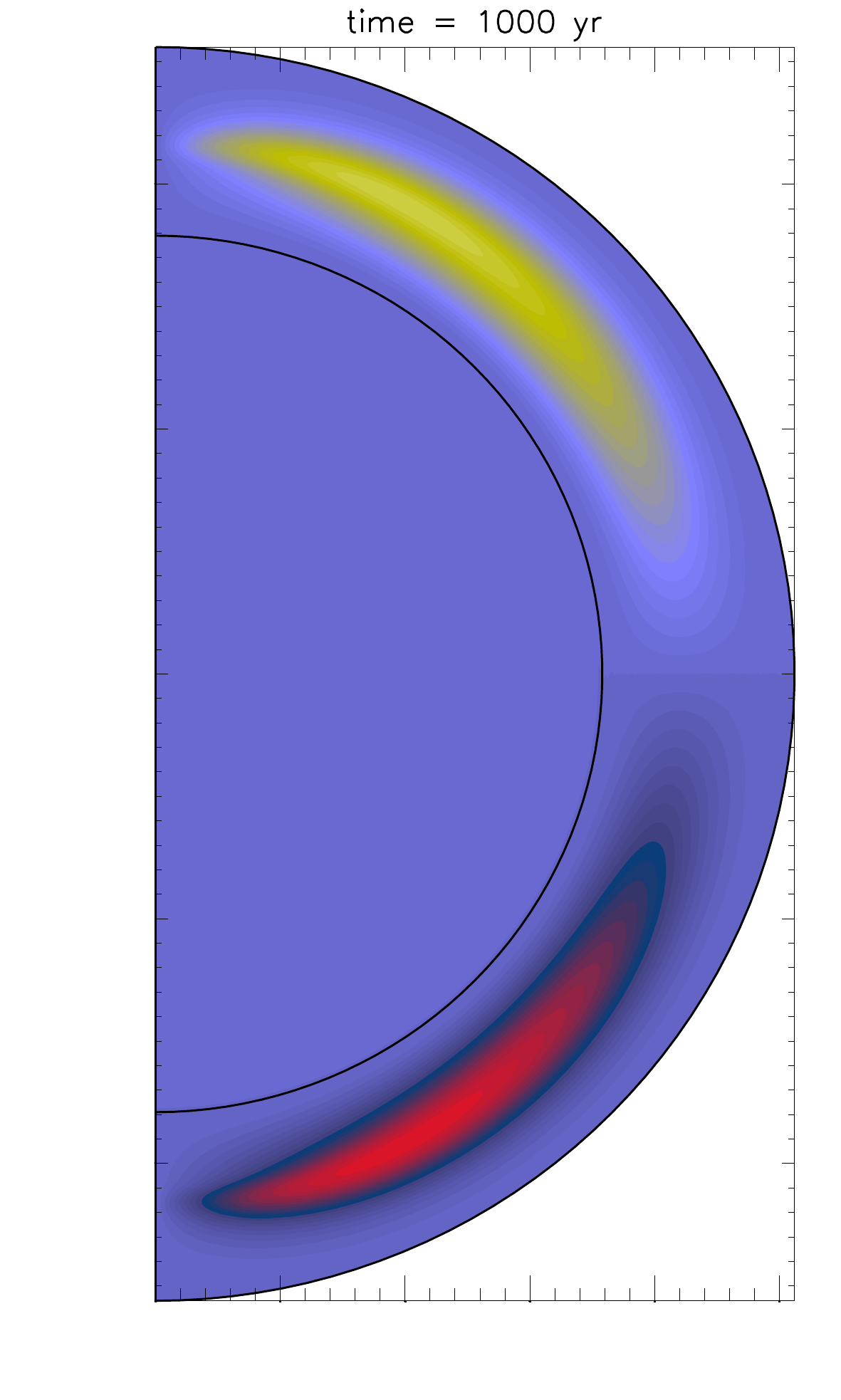}
	\includegraphics[width=.32\textwidth]{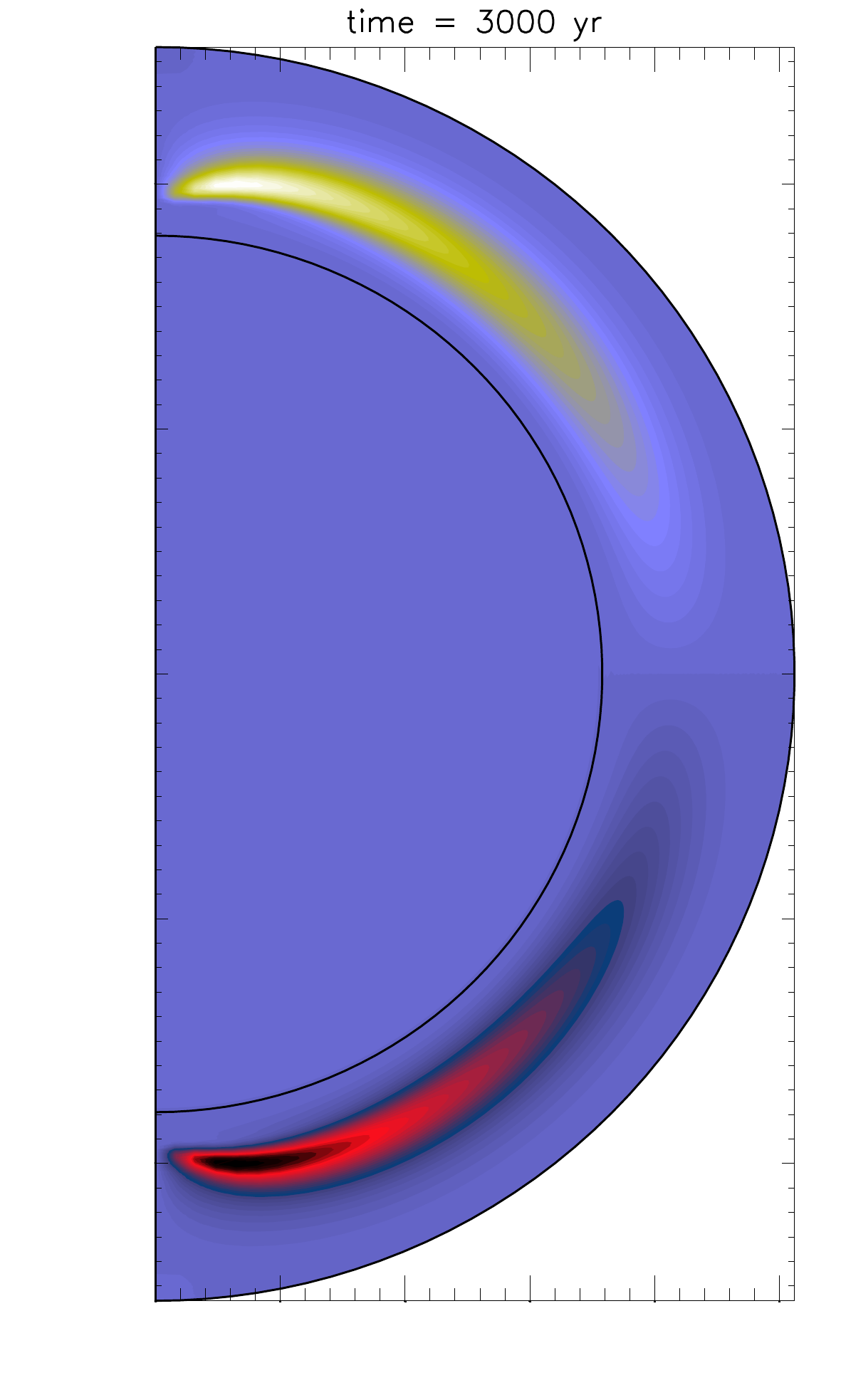}
	\caption{Evolution of a quadrupolar toroidal magnetic field, with a maximum initial value of $3 \times 10^{15}$ G confined into the NS crust. We show snapshots $t= 0$, 1000, and 3000 yr. Color contours show the toroidal magnetic field strength, the reddish corresponding to negative $B_\varphi$ and the yellowish to positive $B_\varphi$. The only difference between the top and bottom panels is the reverse sign of the initial field. In the figure, the size of the crust has been amplified by a factor of 4 for better visualization.} 
	\label{fig:btor_evol}
\end{figure}

According to the Hall induction equation, any initial toroidal configuration must remain purely 
toroidal during the evolution, but its shape and location (and the associated currents) vary with time.
As discussed in the literature \citep{HR2002,PonsGeppert2007,vigano12a}, the evolution has two characteristics: (i) a vertical drift, northward or 
southward depending on the sign of $B_\varphi$; (ii) 
a drift towards the interior of the star due to the existence of a charge density gradient. In the top panels of 
Figure \ref{fig:btor_evol} we show three snapshots of the evolution of an ultra-strong toroidal field ($B_t^0=3\times10^{15}$ G), 
such that the first effect (drift towards the equator of both rings) occurs faster. In this model the maximum value of $\omega_B\tau_e$ is  $\approx 300$, although it varies throughout the crust. 
After 1000 yr, a radial current sheet  (i.e., a sharp discontinuity of the toroidal magnetic field in the meridional direction) is created in the equator and Ohmic dissipation is locally enhanced. 
We also notice a global drift towards the interior: compare the distance to the surface of the models at $t=1000$ and $t=3000$ yr.  The bottom panels show the evolution of the initial model with the reverse sign. We observe how the drift proceeds in the opposite direction, creating a strong toroidal ring around the axes, near the poles. 
This simple model is useful to understand the evolution when the initial toroidal field is dominant, even in the presence of a weaker poloidal field. As \cite{geppert14} 
have shown, starting with a very large fraction  ($\gtrsim 99\%$) of magnetic energy stored in the toroidal component is a potential way to create local 
magnetic spots near  the poles, where the lines are more concentrated and the magnetic field intensity can be one or two orders of magnitude higher than the average value. 

In addition, because of the formation of the current sheet in the equator or the localized rings in the poles,  this model is also useful to check several issues concerning energy conservation, numerical viscosity, and current sheet formation, amply discussed in Sect.~5.1 of \cite{vigano12a}. 
A necessary test for any numerical code is to check the instantaneous (local and global) energy balance.
To remark one of the
key points, let us recall the magnetic energy balance equation:
\begin{equation}\label{en_cons}
\frac{\partial}{\partial t}\left(e^{\nu} \frac{B^2}{8\pi}\right)= -  e^{2\nu}  Q_j - \vec{\nabla}\cdot ( e^{2\nu}  \vec{S} ) ~,
\end{equation}
where $Q_j=4\pi\eta j^2/c^2$ is the Joule dissipation rate and $\vec{S}=c\vec{E}\times\vec{B}/4\pi$ is the Poynting flux. 
During the evolution, the magnetic energy in a cell can only vary due to local Ohmic dissipation and by the interchange
between neighbor cells (Poynting flux).
Integrating eq.~(\ref{en_cons}) over the volume of the numerical domain, we obtain the following energy balance equation:
\begin{equation}\label{integrated_balance}
\frac{\partial}{\partial t}{\cal E}_b + {\cal Q}_{tot} + {\cal S}_{tot}=0~,
\end{equation}
where ${\cal E}_b=\int_V (e^{\nu} B^2/8\pi) dV$ is the total magnetic energy , ${\cal Q}_{tot}=\int_V e^{2\nu} Q_j dV$ the total Joule dissipation rate, and 
${\cal S}_{tot}=\oint_{\partial V} e^{2\nu} \vec{S}\cdot \vec{d}{\Sigma}$ the Poynting flux through the boundaries.
Numerical instabilities usually show up as a strong violation of energy 
conservation and careful monitoring of the energy balance is a powerful diagnostic.

This was one of the simplest possible initial configurations, yet capturing interesting physics. When we introduce an initial poloidal component, higher multipoles,
stratified microphysical (Ohmic/ambipolar) coefficients, etc., it becomes non-trivial to design benchmark tests. In \S~\ref{sec:examples}, we give a summary 
of previous attempts to gradually approach realistic scenarios.

\section{Magnetosphere-interior coupling and rotational evolution}
\label{sec:magnetosphere}

An open issue in realistic simulations of the magnetic field evolution of NSs concerns the correct implementation of boundary conditions at the star surface. 
In the external region of a NS, the mass density is over twenty orders of magnitude smaller than in the outer crust, where numerical grids usually end. 
One needs to match two regions, with radically different physical conditions and timescales, through the thin ($\approx 100$ m) layer between them. The usual procedure is to assume that, on the slow secular evolution timescales, the exterior is immediately readjusted (light crossing timescale) to the stationary solution imposed by the surface values of magnetic fields and currents. In other words, for long timescales, the magnetosphere can be seen as a perfect conductor where currents quickly respond to cancel electromagnetic forces out. Thus, the interior evolution provides the surface values of the field that determine the external configuration. However, in a numerical code the interior 
also needs, at each time step, a recipe for the outer boundary condition to proceed with the evolution, so both problems are interlinked and must be consistently treated.

Under the assumption that the dynamics of the magnetosphere is dominated by the electro-magnetic field, and the plasma pressure as well as its inertia are negligible, a reasonable 
approximation is to consider that the  large-scale  structure  of  the  magnetosphere  is  given by force-free configurations, in which the electric and magnetic forces 
on the plasma balance each other. 
For magnetar conditions, one can safely neglect the effects of rotation in the magnetospheric region near the star. Under this approximation, which we follow hereafter, the electric force is neglected ($\vec{E} = - \vec{v} \times \vec{B}$, with $v \ll c$). Thus, the force-free condition reduces to $\vec{j}\times \vec{B}=0$: the electric currents flow parallel to the magnetic field lines that they sustain (since $\vec{j} \propto \nabla\times\vec{B}$, a force-free magnetic field is a Beltrami vector field).
Within the family of possible solutions, the most trivial (and popular) one is the current-free, or potential solution, $\vec{j}= 0$, which also holds in vacuum. Matching the interior magnetic
field to a magnetospheric potential field is equivalent to physically avoid that the current escapes (enters) from (into) the star, although the non-vanishing Poynting flux across the boundary allows the two regions to interchange magnetic energy (but not magnetic helicity). 

While the potential solution is acceptable as a first approximation, to advance toward more realistic models, we need more general solutions. 
As a matter of fact, electrical currents can stably flow in the closed magnetic field line region, similar to the Solar coronal loops. 
The current system lasts on relatively long timescales, from months to decades \citep{beloborodov09},  presumably sustained by the interior dynamics. 
There is indirect observational evidence of such currents in some magnetars, where the presence of a plasma much denser than the Goldreich-Julian value has been inferred.
Soft X-ray photons emitted from the star surface are up-scattered to higher energy \citep{lyutikov06,rea08,beloborodov13} through resonant Compton processes, resulting in the observed spectra.  Equilibrium solutions of force-free twisted magnetospheres in the magnetar context were considered by several recent works \citep{fuji14,glampedakis14,pili15,akgun16,kojima17}.
However, the  evolution of the interior sometimes leads to solutions that cannot be smoothly connected to a force-free solution. This implies 
discontinuities in the tangential components at the surface, corresponding to current sheets, that may cause numerical instabilities.

While rotation has negligible effects on the magnetic evolution, the opposite is not true: the spin period evolves due to electromagnetic torques determined by the magnetospheric configuration.  Compared to the magnetic and thermal evolution,  the equations describing the rotational evolution are simpler, but they predict the observable 
timing properties of isolated NSs. 
In the remaining of this section we review the methodology to prescribe boundary conditions to the magnetic field when one solves the induction equation with different types of code, commenting on some problems that arise at the practical level, and we provide the recipe for the rotational evolution.

\subsection{Potential boundary conditions}\label{app:vacuum_spectral} 

\paragraph{Spectral methods.}
Using the same notation as in Sect.~\ref{methods_spectral} for the poloidal/toroidal decomposition,
the requirement that all components of the magnetic field be continuous (no current sheets at the surface) implies that
the scalar potentials $\Phi_{nm}$ and $\Psi_{nm}$, and their derivatives $\frac{\partial \Phi_{nm}}{\partial r}$, 
are continuous through the outer boundary. Therefore, the $\nabla \times \vec{B}= 0$ condition translates into
\beq
\Psi_{nm}=0 ~,
\eeq
and the following differential equation for each radial function $\Phi_{nm}(r)$
\beq
(1-z) \frac{\partial^2 \Phi_{nm}}{\partial r^2} 
+ \frac{z}{r}\frac{\partial \Phi_{nm}}{\partial r}  
-\frac{n(n+1)}{r^2} \Phi_{nm} = 0~,
\label{ode2}
\eeq
where we assume the metric~(\ref{Schw}),
and $z\equiv \frac{2 G M}{c^2 r}$. We note that there is no $m-$dependence in the equation, so that the solution depends only on $n$ and we will omit the $m$ subindex hereafter.
 
In general, the family of solutions of Eq.~(\ref{ode2}) for any value of $n$
can be expressed in terms of generalized hypergeometric functions ($F([],[],z)$), 
also known as Barnes' extended hypergeometric functions, as follows:
\bear
\Phi_n
=  C_n ~r^{-n} ~F([n,n+2], [2+2n], z) + D_n ~r^{n+1} ~ F([1-n,-1-n], [-2n], z)~, 
\label{outerPhi}
\ear
where $C_n$ and $D_n$ are arbitrary integration constants that correspond to the weight of each magnetic multipole $n$. Note that regularity at $r=\infty$ requires $D_n=0$ for each $n$. 
For any given value of $n$, one can also express the solution in closed analytical form. The explicit expressions for $n=1$ and $n=2$ are
\bear
\Phi_1 &=& C_1 r^2 \left[ \ln(1-z) + z + \frac{z^2}{2}\right] ~,
\\
\Phi_2 &=& C_2 r^3 \left[ (4-3z) \ln(1-z) + 4z - {z^2} - \frac{z^3}{6}\right]~. 
\ear
If we consider the Newtonian limit ($z\rightarrow0$), Eq.~(\ref{ode2}) simplifies to:
\beq  
 \frac{\partial^2 \Phi_{n}}{\partial r^2}   - \frac{n(n+1)}{r^2}\Phi_{n} = 0~.
\eeq 
The only physical solution (regular at infinity) of this equation is $\Phi_n=C_nr^{-n}$. Therefore, the requirement  
of continuity across the surface results in  
\beq 
\left. \frac{\partial \Phi_{n}}{\partial r}\right |_{r=R} = -\frac{n}{R}\Phi_n ~.
\label{OBC_Phi} 
\eeq 
In the relativistic case, we can implement Eq.~(\ref{outerPhi}) directly, or the most practical form, analogous to the Newtonian case:
\beq
\left.  \frac{\partial \Phi_{n}}{\partial r}\right |_{r=R} = -\frac{n}{R} f_n \Phi_n  ~,
\label{relBC}  
\eeq
where the $f_n$'s are relativistic corrections that only depend on the value of $z$ at the star surface,
$z(r=R)$ (in the Newtonian limit all $f_n=1$), and can be evaluated numerically 
only once  with the help of any algebraic manipulator and 
stored\footnote{See \citealt{radler2001} for  an alternative form to evaluate $f_n$ based on the expansion in a series of powers of $1/r$}.

\paragraph{Finite-difference schemes.} If we do not use a spectral method, we must apply boundary conditions to magnetic field components instead of the individual multipoles. In general, we need to provide the field components, in one or more ghost cells outside the physical grid, in terms of the components at the last grid point. However, we still can make use of the previous form of the boundary conditions, as a relation between the poloidal radial function and its derivative for each multipole,.
Let us explain an accurate and elegant procedure to impose the current-free constraint in the axisymmetric case.

From Eq.~(\ref{relBC}) and the expression of the field components in terms of the poloidal and toroidal functions, one can easily show that the potential solution expansion in terms of Legendre polynomials ($P_l$) reads:
\begin{eqnarray}
  && B_r      =   \sum_l b_l (l+1)P_l(\cos\theta) \left(\frac{R}{r}\right)^{-(l+2)}~,  \\
  && B_\theta = - \mathrm{e}^{-\lambda}  \sum_l f_l b_l \frac{dP_l(\cos\theta)}{d\theta} \left(\frac{R}{r}\right)^{-(l+2)}~, 
  \label{bvacuum}
\end{eqnarray}
where we denote by $b_l$ the weights of the multipoles
\begin{equation}
 b_l  = \frac{2l+1}{2(l+1)} \int_{0}^{\pi} B_r(R,\theta) P_l(\cos\theta) d\theta  ~. \label{eq:bl}
\end{equation} 
At a practical level, one can proceed as follows: 
\begin{itemize}
\item
First, at each time step, obtain the $b_l$ coefficients  from the Legendre decomposition of the radial component of the magnetic field over the surface $B_r(r=R,\theta)$. In a discretised scheme, values of $b_l$ can be calculated up to a maximum multipole $l_{\max}=n_\theta/2$, where $n_\theta$ is the number of angular points of the grid.
\item
Second, from the $b_l$'s, reconstruct the values of $B_r$ and $B_\theta$ in the external ghost cells, as required by the method, by using Eq.~(\ref{bvacuum}).
\item
Finally, simply set $B_\varphi =0$ for any cell $r \ge R$.
\end{itemize}

This method is very accurate for smooth functions $B_r$. In the case of sharp features in $B_r$, which may be created by the Hall term, 
the largest multipoles acquire a non-negligible weight, and, since $l_{\max}$ is limited,  fake oscillations in the reconstructed $B_\theta$ may appear (Gibbs phenomenon). An alternative method to impose potential boundary conditions is based on the Green's representation formula, a formalism often used in electrostatic problems
able to correctly handle the angular discontinuities in the normal components. 
Details about the derivation of the Green's integral relation between $B_r$ and $B_\theta$ at the surface are given in Appendix B.

Note that, in 3D, applying the potential boundary conditions is a challenge for parallelization. The easiest solution is to parallelize the dominion by spherical shells, so that the integration over the star's surface, needed by either the spherical harmonic expansion or the Green's method, is in charge of a single processor (as in \citealt{wood15} and following works). If, on the other hand, the parallelization is done by geometrically optimized patches (cubic in the simplest case, \citealt{vigano19}), then the star's surface would be covered by different processors. In this case, the calculation at each point depends on calculations done by other processors, thus enlarging the needed stencil. This results in an excessive intercommunication load and prevents optimal scaling.

\subsection{Force--free boundary conditions}\label{sec:forcefree}

The construction of relativistic, axisymmetric, force-free magnetospheres for (non-rotating) magnetars is a well studied problem (see, e.g., \citealt{kojima17} and references therein).
In \cite{akgun18b} the authors explored a method to impose such boundary conditions by solving the Grad-Shafranov equation, at each time step, to match the internal evolution of the star. Let us review their approach. 
Considering axial symmetry, the magnetic field can be written as follows: 
	\bear
	\vec{B} = \frac{(\partial P/\partial \theta)}{r^2 \sin\theta} {\hat{r}} - e^{-\lambda}  \frac{(\partial P/\partial r)}{r \sin\theta} {\hat\theta} + e^{\lambda} \frac{T}{r\sin\theta} {\hat\varphi} ~,
	\label{mag_PT}
	\ear
where $P$ and $T$ are functions defining the poloidal and toroidal components, respectively (see more details in Appendix~\ref{app:formalism}).

The force-free condition ($\vec{j} \times \vec{B} = 0$) implies that the electrical currents flow along magnetic surfaces, which are defined by constant $P$.  
Thus, the mathematical requirement of a vanishing azimuthal component of the local Lorentz force implies that the poloidal and toroidal functions must be functions of one another, say $T = T(P)$,
that is, the poloidal and toroidal functions $P$ and $T$ are constant on the same magnetic surfaces\footnote{The magnetic flux
through the area enclosed by the corresponding magnetic surface is $2\pi P$, and the current through the same area is $c T/2$. }.

From the definition of the current, one can arrive at the so-called Grad--Shafranov equation
\bear
 \frac{\partial}{\partial r}\left( e^{- 2 \lambda}  \frac{\partial P}{\partial r}\ \right) +  \frac{\sin\theta}{r^2}  \frac{\partial}{\partial \theta}\left( \frac{1}{\sin\theta}  \frac{\partial P}{\partial \theta}\ \right) =
 - e^{2 \lambda} T(P) T'(P)   \ ,
\label{GS_eq}
\ear
where $T'(P) = dT/dP$. The current-free limit (potential solution)  is simply recovered by taking the right hand side equal to zero.

In principle, there is an infinite family of external force-free solutions for a given radial magnetic field at the surface, because of the freedom to choose the functional form of $T(P)$.
The main problem of this approach is how to continuously match the arbitrary field configuration, resulting from the evolution in the crust, while enforcing the force-free solution outside. 
 In the crust, any line bundle marked by a given magnetic flux $P$ has in general different values of $T$ because, internally, the force-free condition does not hold. 
As discussed in \cite{akgun18b}, there is an intrinsic inconsistency in the possibly multi-valued function $T(P)$, if we strictly take it from the values at the surface ($r=R$).
They address this problem by symmetrizing the numerical function $T(P)$, which is physically equivalent to allow to propagate through the surface only the modes compatible 
with solutions of the Grad-Shafranov equation. This is motivated by the results from MHD simulations of the propagation of internal torsional oscillations \citep{gabler14}, who
found that antisymmetric modes cannot propagate into the magnetosphere and are reflected back into the interior.

In Fig.~\ref{fig:twisted_taner}, we show the evolution of a magnetospheric configuration physically connected to the interior. The initial model consists of both 
poloidal and toroidal dipolar components, with the latter extending beyond the surface. 
As the internal magnetic field evolves, the external magnetic field is consistently twisted, by the injection of magnetic helicity (i.e., currents) in the magnetosphere. 
Force-free solutions are calculated at each time step until a critical point, 
where numerical solutions cannot be found anymore. At this point,  the magnetosphere is expected to become unstable, resulting in a global reconfiguration 
with the opening of the twisted field lines and magnetic reconnection. 
This mechanism was studied in more detail in force-free electrodynamics simulations in 2D and 3D, and both the Newtonian and general relativistic cases \citep{parfrey13,carrasco19}.

\begin{figure}[t]
	\centering
	\includegraphics[width=.9\textwidth]{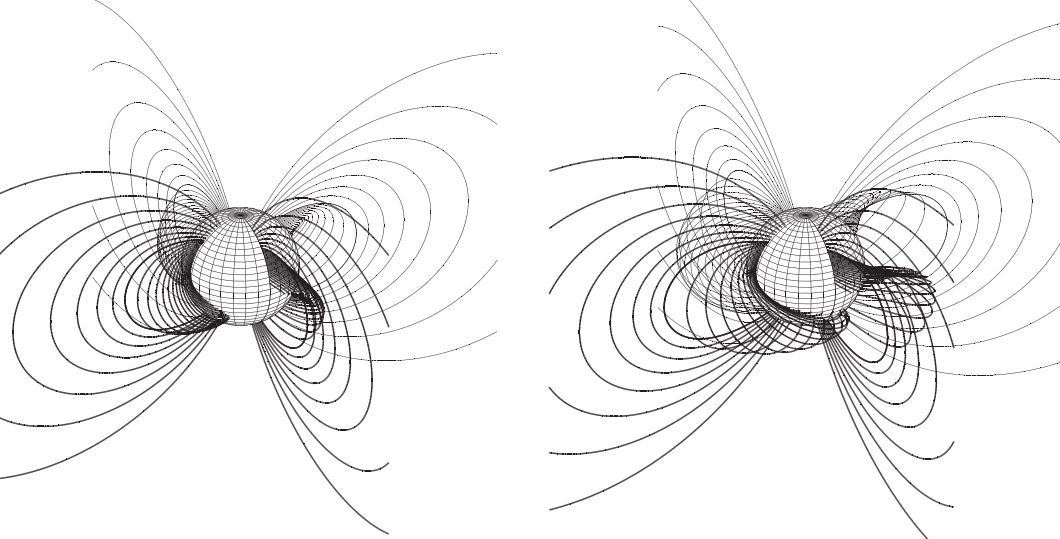}
	\caption{Evolution of a twisted magnetosphere with real-time coupling with the interior. The left and right panels show snapshots at $t=0$ and $t \sim 1.58$ kyr, the critical time when the magnetosphere has stored the maximum possible twist. Figure courtesy of \cite{akgun17}.}
	\label{fig:twisted_taner}
\end{figure}

\subsection{Extended domains}\label{sec:extended_domains}

An alternative to imposing a precise mathematical boundary condition at the surface is to consider an extended domain, where we evolve at the same time all components of the field, but with physical coefficients that enforce the solution to meet the required conditions. Instead of imposing a boundary condition at the last numerical cell, this approach considers a generalized induction equation, where, at the surface, there is a sharp transition in the values of the pre-coefficients describing the physics ($\eta, 1/n_e, f_a$). In the numerical GRMHD context, this approach has been successfully used to describe at the same time the resistive and ideal MHD inside and outside a NS \citep{palenzuela13}. 

The idea is that, since the magnetospheric timescales are many orders of magnitude shorter than the interior, the long-term evolution of the magnetosphere can be seen as a series of equilibrium states, attained immediately after every time step of the interior. Therefore, one can activate an artificial term that dynamically leads to the force-free solution. This approach is similar to the {\it magneto-frictional method} \citep{yang86,roumeliotis94}, as known in solar physics. The modified induction equation employed in the exterior of the star has a mathematical structure equivalent to an ambipolar term, which forces currents to gradually align to magnetic field lines, without having to solve the elliptical Grad-Shafranov equation at every time step (which is numerically expensive). This also allows us to account for the transfer of helicity and provides a mechanism to continuously feed currents that twist the magnetosphere. The caveat is that the ambipolar coefficient must be fine-tuned to prevent the exterior dynamics from being neither too fast (it would excessively limit the time step), nor too slow (it would not manage to relax to a force-free configuration and would cause non-negligible, unphysical feedback on the interior).

In our context, such a strategy has only been explored preliminarily in the 3D Cartesian parallelized code used in \cite{vigano19}. This code was built by using {\em Simflowny} \citep{arbona13,arbona18}, a versatile platform able to automatically generate parallelized codes for partial differential equations. It employs the {\it adaptive mesh refinement} libraries of {\em SAMRAI} \citep{hornung02}, and a graphical user interface that easily allows us to implement equations and to choose among different time and space discretization schemes.
The code has not been applied yet to realistic simulations and is very different from previous codes.
The Cartesian grid, with parallelization by regular cubic patches, imposes numerical challenges. 
One problem associated to Cartesian grids is that the geometry is not adapted to the physically preferred radial direction, along which gradients are usually much larger. 
Thus, one cannot improve the resolution in the radial direction alone, causing a rapid increase in the computational cost, compared to spherical coordinates-based codes ($\propto N^3$ instead of $N_r$). Furthermore, the Cartesian discretization implies the appearance of numerical noise at the (physically spherical) surface and crust/core interfaces (where the pre-coefficients in the induction equation show a sharp transition as mentioned above). Finally, one has also to take care that the outer domain (placed far enough from the surface by using different mesh refinements) does not introduce noise and allow a regular solution at infinity. These challenges need to be tackled soon to make this alternative method, still at its infancy, numerically feasible.

\subsection{Evolution of spin period and obliquity}

As NSs age, they spin down because of angular momentum losses due to magnetospheric torques \citep{spitkovsky06,beskin13,philippov14}. This mechanism is effectively ruled by the dipolar component since other multipoles decay faster with the distance to the star and are negligible. Thus, the magnetic field evolution not only affects the surface temperature but also determines the rotational properties of the star.
In the general case, the equations describing the coupled evolution of the spin period $P$ and the angle between the magnetic dipolar moment and the rotation axis, $\chi$, are \citep{philippov14}:
\begin{eqnarray}
&& \dot{P} = \beta \frac{B_p^2}{P} (\kappa_0 + \kappa_1 \sin^2{\chi})~,\label{eq:spindown} \\
&& \dot{\chi} = - \kappa_2 \beta \frac{B_p^2}{P^2} \sin{\chi} \cos{\chi}~, \label{eq:alignment} 
\end{eqnarray}
where we have defined the auxiliary quantity
\begin{eqnarray}
\beta \equiv \frac{\pi^2 R^6}{I c^3} ~, 
\end{eqnarray}
$I$ is the moment of inertia of the part of the star co-rotating with the magnetosphere, and $B_p$ is the value of the {\em dipolar component of the magnetic field at the 
 magnetic pole}. The latter is a function of time and can be provided by the simulations of the internal magnetic field.
 
 The coefficients $\kappa_0$, $\kappa_1$, $\kappa_2$ depend on the magnetosphere geometry and its physical conditions and determine the magnetospheric torque.
For the classical vacuum dipole formula $\kappa_0=0$, $\kappa_1=\kappa_2=2/3$, while for a realistic plasma-filled magnetosphere, $\kappa_0 \approx \kappa_1 \approx 1$, with the last coefficient varying between 0 and 1, depending
on the assumptions. This coefficient can be fitted from results from 3D simulations for force-free and resistive magnetospheres \citep{philippov14}, who found that the  alignment of the rotation and magnetic axis in pulsars with vacuum magnetospheres proceeds much faster (exponential, with characteristic time $\tau_0 = \frac{P_0}{2 \beta B_0}$) than for realistic plasma-filled magnetospheres (a power-law). 

A very important remark is that using the classical dipole formula, with $\kappa_0=0$, may mislead to the wrong conclusion that an aligned rotator does not exert any torque, thus stopping the spin-down of the star \citep{johnston17}. This is not physically correct, and 
realistic models predict $\kappa_0\approx1$, which at most results in about a factor two correction \citep{spitkovsky06,philippov14}. Therefore, alignment cannot completely stop the period evolution, which can only happen if the magnetic field becomes negligible.

Besides the magnetic field strength and the inclination angle, $I$ could also change with time. If there is a superfluid component (e.g., neutrons in the core or the inner crust), it is generally rotationally decoupled from the rest of the star. Therefore, it does not contribute to $I$, which only accounts for the matter rigidly co-rotating with the magnetosphere. Then, there are two possible effects. First, the volume of the superfluid component can change with time, since the phase transition depends on density and temperature 
(as the star cools down superfluid components occupy a larger volume, thus $\beta$ slowly increases). Second, normal and superfluid components can suddenly and temporarily couple during glitches, modifyingf $I$. 
This can be formally considered with a  two-fluid description or neglected, by assuming a constant $I$ corresponding to the rigid co-rotation of the whole star. 
For realistic stars, the moment of inertia is $I \sim 1.5 \times 10^{45}$~g~cm$^2$, with a $50\%$ uncertainty. This gives $\beta \sim 6 \times 10^{-40}$ s G$^{-2}$. We note that either angle variations (alignment) or moment of inertia variations, result in corrections to the torque by a factor $\lesssim 2$, while magnetic field decay can result in torque variations of several orders of magnitude, and consequently in large and relatively fast variations of $P$ and $\dot{P}$.

Finally, we would like to address why the effects of rotation have negligible feedback on the magneto-thermal evolution.  Magnetospheres of spinning isolated NSs have been studied in the last 50 years, analytically (starting from the seminal work of \citealt{goldreich69}) and numerically for some geometries (mostly inclined rotating dipoles). Examples of 2D and 3D numerical simulations can be found in \cite{spitkovsky06,contopoulos99}. First of all, the rotationally-induced electric field, $\vec{E}=-(\Omega  r\sin\theta/c)\hat{\phi}\times\vec{B}$, is negligible in the interior regions of the magnetosphere ($r\sin\theta \ll c/\Omega$), thus justifying the non-rotating force-free approximation above, as a boundary condition at the star surface. Secondly, rotation opens up a bundle of lines close to the magnetic poles, which are stretched and twisted and are supposedly responsible for the radiated emission. The surface polar cap containing the footprints of the open lines has an opening angle of only $\theta\sim \arcsin[(R\Omega/c)^{1/2}] \sim 0.8^\circ P{\rm [s]}^{-1/2}$: a negligible fraction of the NS surface, especially for magnetars ($P\sim 1-10$ s). A recent work \citep{karageorgopoulos19}, based on the minimization of the Joule dissipation rate, has proposed how the rotationally-induced polar currents close within the crust. The dissipated power by the Joule effect was estimated to be more than 10 orders of magnitude smaller than the rotational energy losses, which does not affect the cooling history.

\section{Magneto-thermal evolution of neutron stars}\label{sec:examples}


The first question to answer when one plans to simulate the evolution of magnetic fields in NSs is the choice of the initial model. Since we are mostly interested in understanding the evolution of highly magnetized NSs, we should use a physically motivated initial model. A first approach is to consider that the hot, liquid initial phase lasts long enough to establish an MHD equilibrium, and reduce the
pool of possible initial models to perfect MHD equilibria solutions, which have been calculated for different geometries \citep{colaiuda2008,ciolfi2013}. 
Unfortunately, the formation of a NS during a supernova explosion is a very complex process, and the origin of strong magnetic fields in NSs and their presumed topology remains unclear.
One of the most promising mechanisms at work is the magnetorotational instability, which, in the presence of differential rotation, can amplify exponentially fast a weak initial magnetic field in a proto-NS to a dynamically relevant strength \citep{guilet2017}. A recent simulation showed an effective dynamo \citep{mosta15}, able to create an amplified large-scale toroidal magnetic field. 
A different mechanism is based on the interplay between compression and convection in the hot-bubble region between the proto-NS and the stalled shock \citep{ober2015}.  
Another relatively recent idea gaining popularity is the formation during a NS-NS merger \citep{ciolfi2019}.
In any case, all the viable mechanisms involve some degree of turbulence, so that the outcome is plausibly different  from a perfectly ordered dipole.  
With all these caveats in mind, one must choose to start the simulations, preferably with a few free parameters, to establish some qualitative trends in the long term evolution. For this reason, most of the existing works simply take a dipolar configuration, or at most some combination of dipolar and quadrupolar poloidal and toroidal components. 

In the literature, there is a clear distinction between crustal-confined and core-threading fields. The former has been studied in-depth, with special focus on the Hall term. 
It assumes a type I superconducting core (not realistic) or, equivalently, that some other mechanism acts on very short timescales to expel most of the magnetic flux from the core.
Core-threading magnetic fields are less studied, due to uncertain core physics. 
A popular configuration is the twisted-torus (e.g., \citealt{ciolfi2013}): an MHD equilibrium solution with a large magnetic helicity where a dipole threads the core and the closed field lines within the star contain a toroidal field. We note the two fundamental differences between crustal-confined and core-threading configurations: first, the field curvature changes by one order of magnitude (roughly the size of the star versus the size of the crust); second, the location of most of the currents (core or crust) determines where Ohmic dissipation occurs, and the two regions have very different conductivities (see Fig.~\ref{fig:cond_el}).

Regarding the temperature evolution, the initial conditions are much easier: it is well known that a few hours or days after birth, most of the star is nearly isothermal, so it is a good approximation to assume a constant temperature, between $10^9$ to $10^{10}$ K. Moreover,  the particular choice of the initial temperature only affects the evolution in the first few days, which is completely irrelevant for following the NS evolution for thousands or millions of years. In any case, since the majority of existing works do not couple the magnetic field evolution to the temperature (most of them assume a constant temperature, or in some cases an independently prescribed function of time), we will begin by reviewing the main results of models with only magnetic evolution (2D and 3D), to conclude the section revising the only fully consistent magneto-thermal simulations available (in 2D).

\subsection{Magnetic field evolution in the neutron star crust}
\label{subsec:3.1}


\paragraph{2D simulations.}
Many works \citep{PonsGeppert2007,vigano12a,vigano13,gourgouliatos13,gourgouliatos14a,gourgouliatos14b} have agreed on the general picture of the Hall-driven dynamics of a crustal-confined field in axial symmetry.  For typical field strengths of $10^{14}$ G, and
starting from a predominantly poloidal dipolar field, we observe a stage dominated by the Hall drift (readjusting from initial conditions), which creates higher-order multipoles, followed by a quasi-stationary Ohmic stage. This structure, which has been called the Hall attractor \citep{gourgouliatos14b},
is characterized by a nearly constant angular velocity of the electron fluid ($\Omega \approx j/e n_e r$) along each poloidal field line, and proportional to the magnetic flux. 
This result holds even if the initial state is a high multipole, say $l$, with the system relaxing to a mixture of modes dominated by the $l$ and $l + 2$, but again 
with the electron angular velocity linearly related to the flux \citep{gourgouliatos14a}. 
It is also relevant to remark that the Hall drift may noticeably accelerate the dissipation of magnetic fields,  by continuously redistributing magnetic field energy towards smaller scales, where Ohmic dissipation is more effective.
In the supplementary material, we provide the animations of two models with an initial dipolar poloidal magnetic field with surface polar intensity $B_p=10^{14}$ G plus a toroidal field with a maximum intensity $B_{\rm tor}=10^{15}$ G. The models differ in the initial multipole of the toroidal field ($l=1$ or $l=2$).


\paragraph{3D simulations.}
Using a mixed spectral/finite-difference code, \cite{wood15,gourgouliatos16} presented the first 3D simulations of crustal-confined fields,  with an exterior boundary condition consisting of a general potential solution. The temperature was not included in the simulations, and the resistivity and density profiles were some radial dependent analytical functions, fitted to mimic a realistic model at $T=10^8$ K \citep{cumming04}. 
These 3D studies show new dynamics and the creation of km-size magnetic structures persistent over long timescales. Even using initial axisymmetric conditions, the Hall instability breaks the symmetry and new 3D modes quickly grow. These have lengthscales of the order of the crust thickness. 
A typical model is shown in Fig. \ref{fig:kostas_3d}. The surface field is highly irregular, with small regions in which the magnetic energy density exceeds by at least an order of magnitude the average surface value. By exploring many different initial models,  \cite{gourgouliatos16} found that magnetic instabilities can efficiently transfer energy to small scales, which in turn enhances Ohmic heating and powers the star persistent emission, confirming the 2D results. 

\begin{figure}
	\includegraphics[width=.49\textwidth]{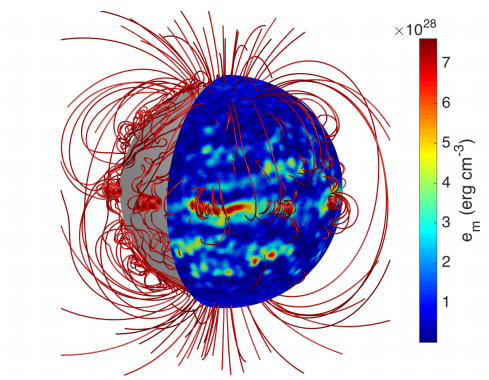}
	\includegraphics[width=.49\textwidth]{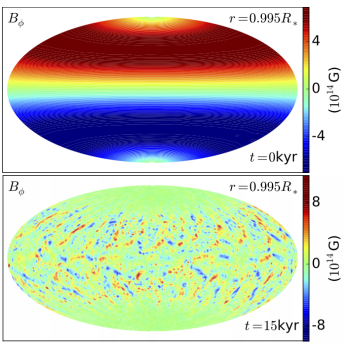}
	\caption{Left: Magnetic field lines and magnetic energy density maps on the star surface (in colors),  at $t=15$ kyr, for an initial model consisting of an $l=1$ poloidal field, and $l=2$ toroidal field, plus a small non-axisymmetric perturbation. Right: Contour plot of the azimuthal component of the magnetic field at $r = 0.995R_\star$, with $R_\star$ being the star radius, for the same model. Figures courtesy of \cite{gourgouliatos16}.}
	\label{fig:kostas_3d}
\end{figure}

More recently, \cite{gourgouliatos18} explored magnetic field configurations that lead to the formation of magnetic spots on the surface of NSs, extending previous 2D works \citep{geppert14}, as described in the final part of \S \ref{tor_sec}. They show how an ultra-strong toroidal component is essential for the generation of a single spot, possibly displaced from the dipole axis, which can survive on very long timescales.
We must note that boundary conditions arguably play a very important role to determine the scale of the, initially unstable, dominant modes since the thickness of the crust 
sets a preferred scale in crustal-confined models.

\subsection{Coupled magneto-thermal simulations}

\begin{figure}[t]
	\centering
	\includegraphics[width=.32\textwidth]{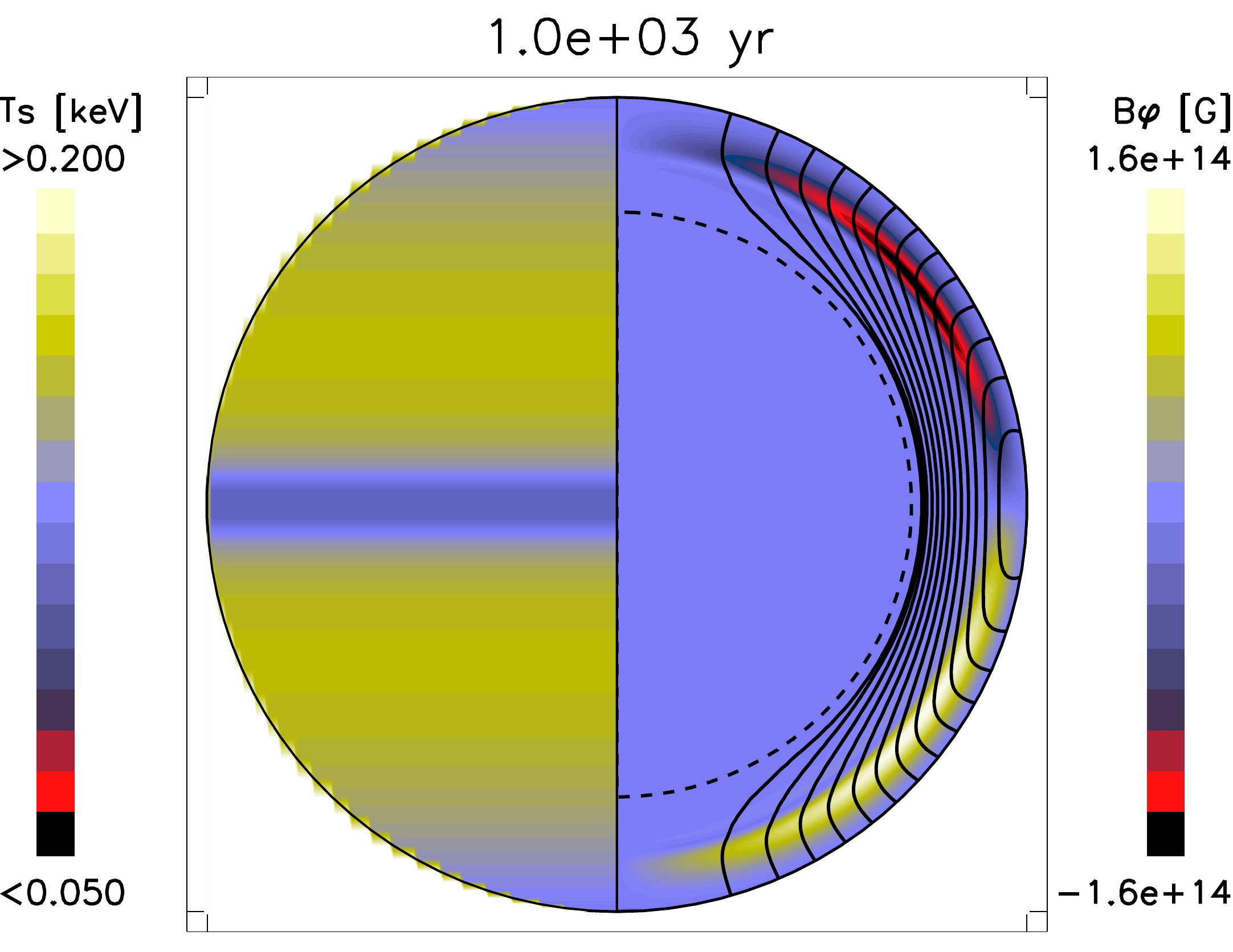}
	\includegraphics[width=.32\textwidth]{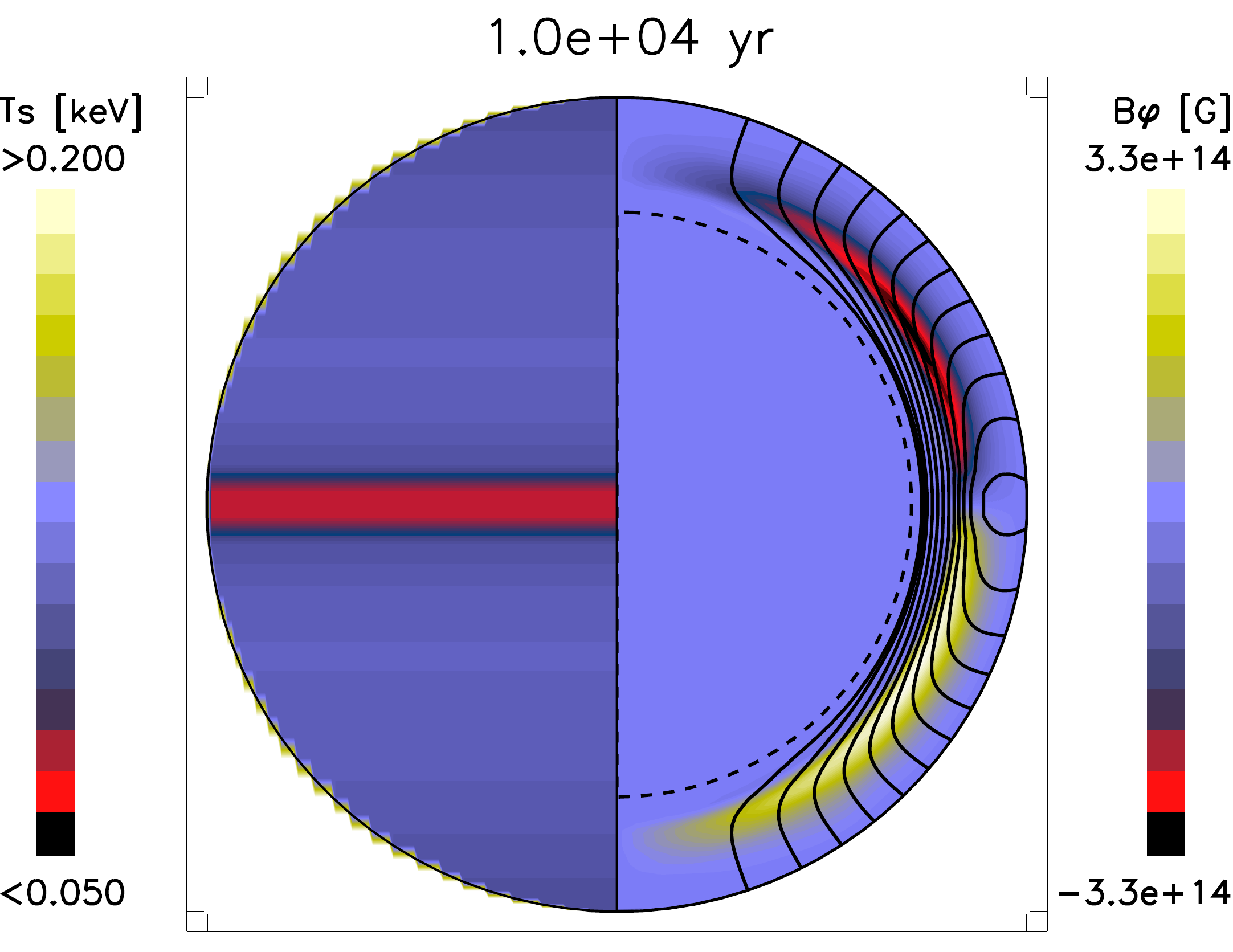}
	\includegraphics[width=.32\textwidth]{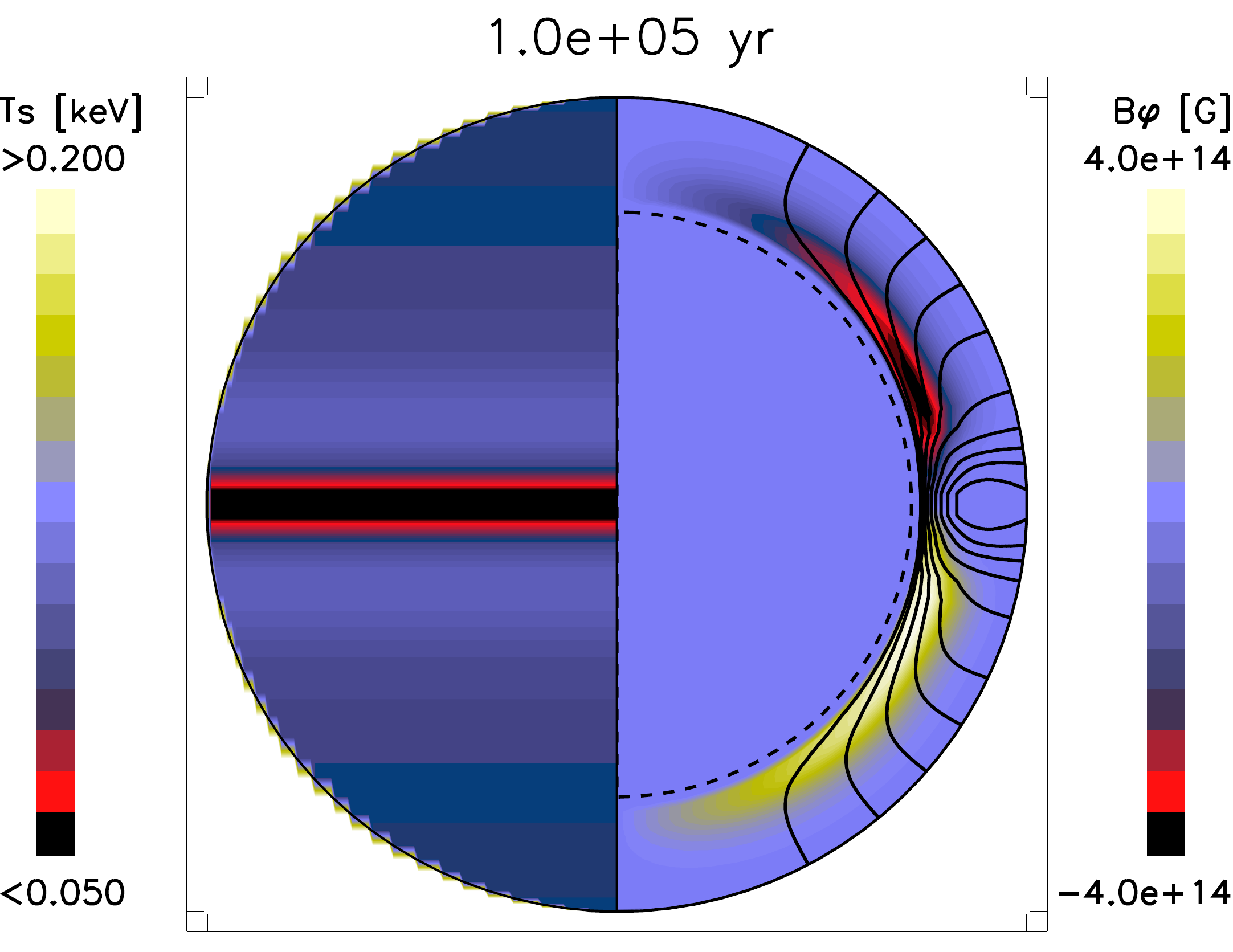}\\
	\vskip0.2cm
	\includegraphics[width=.32\textwidth]{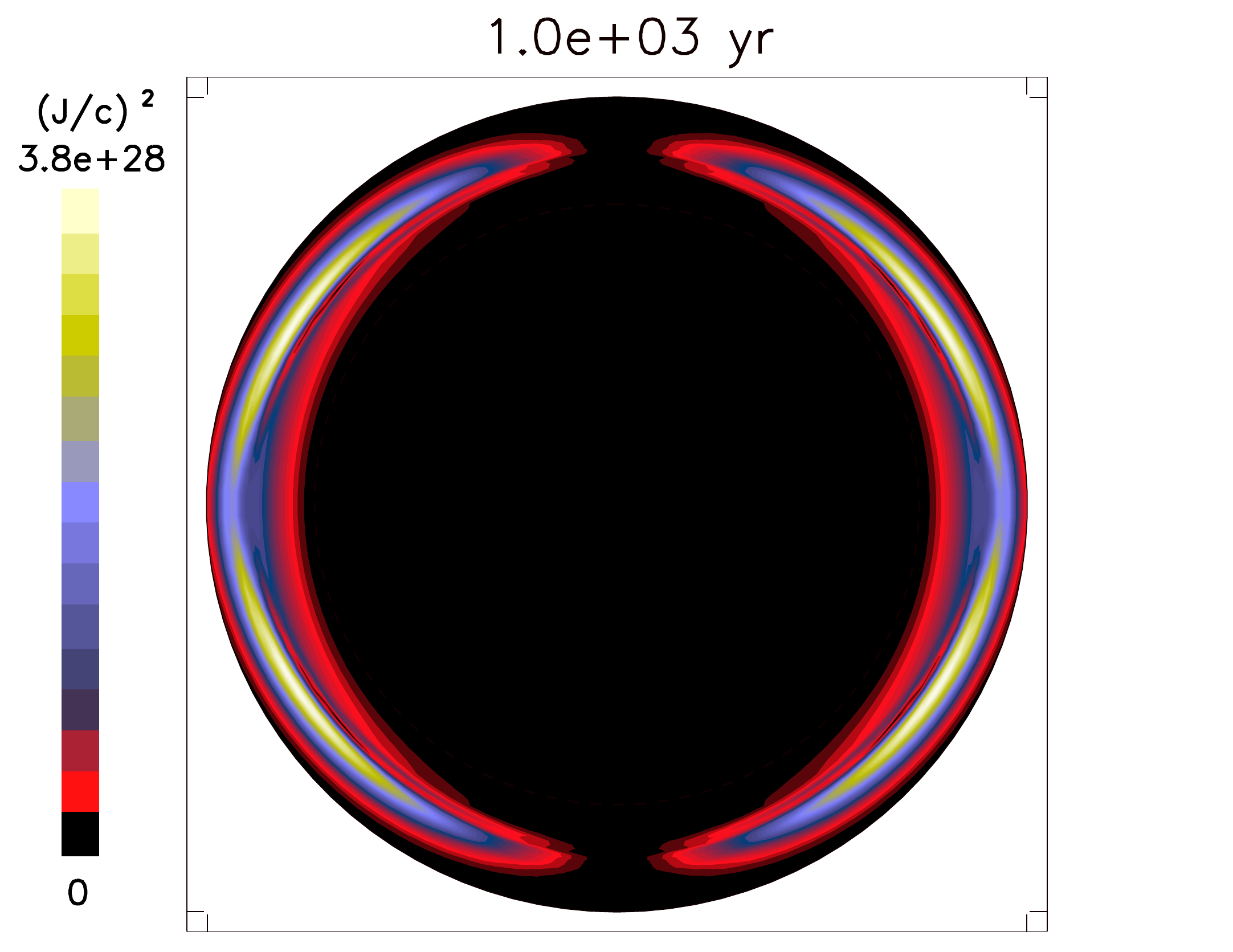}
	\includegraphics[width=.32\textwidth]{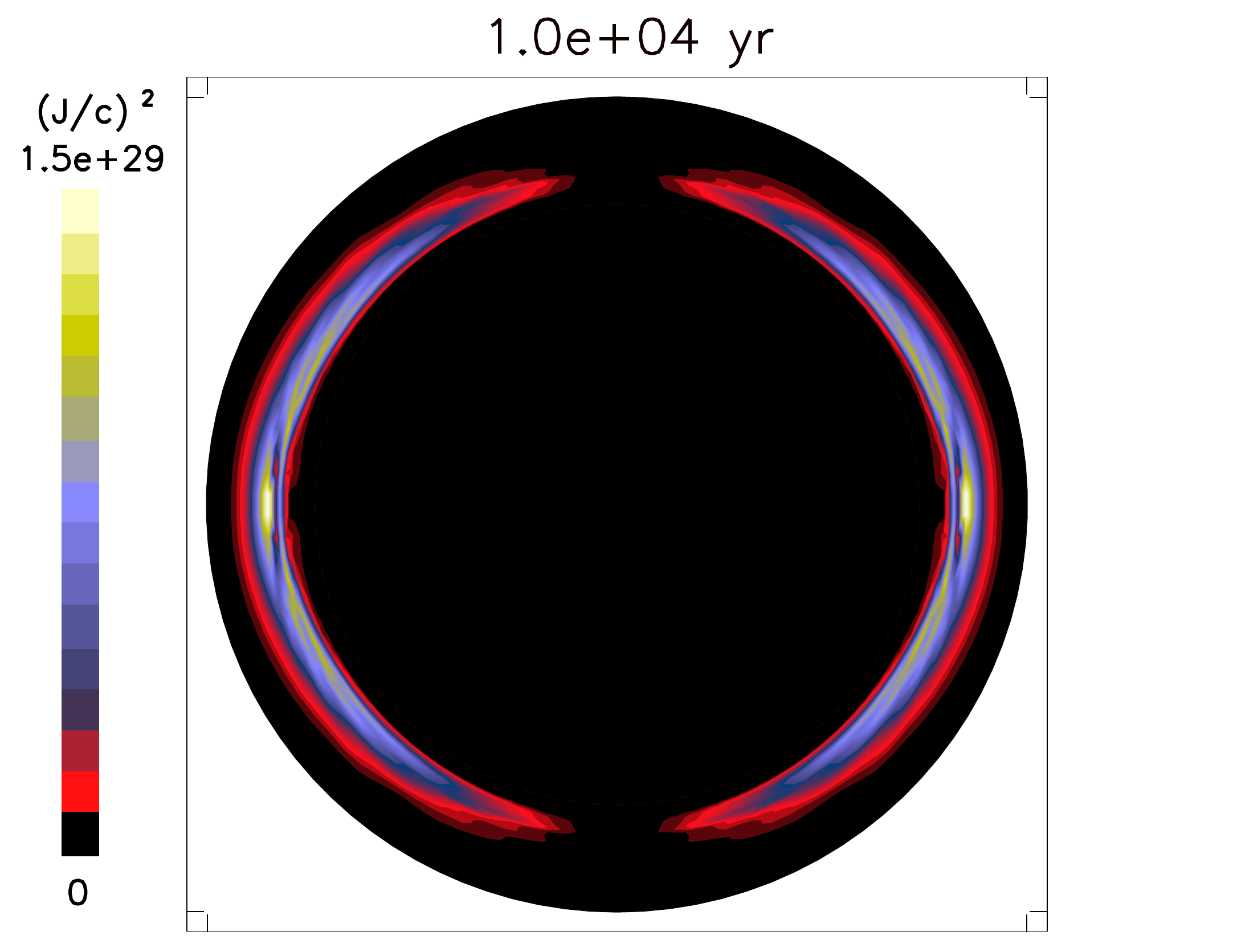}
	\includegraphics[width=.32\textwidth]{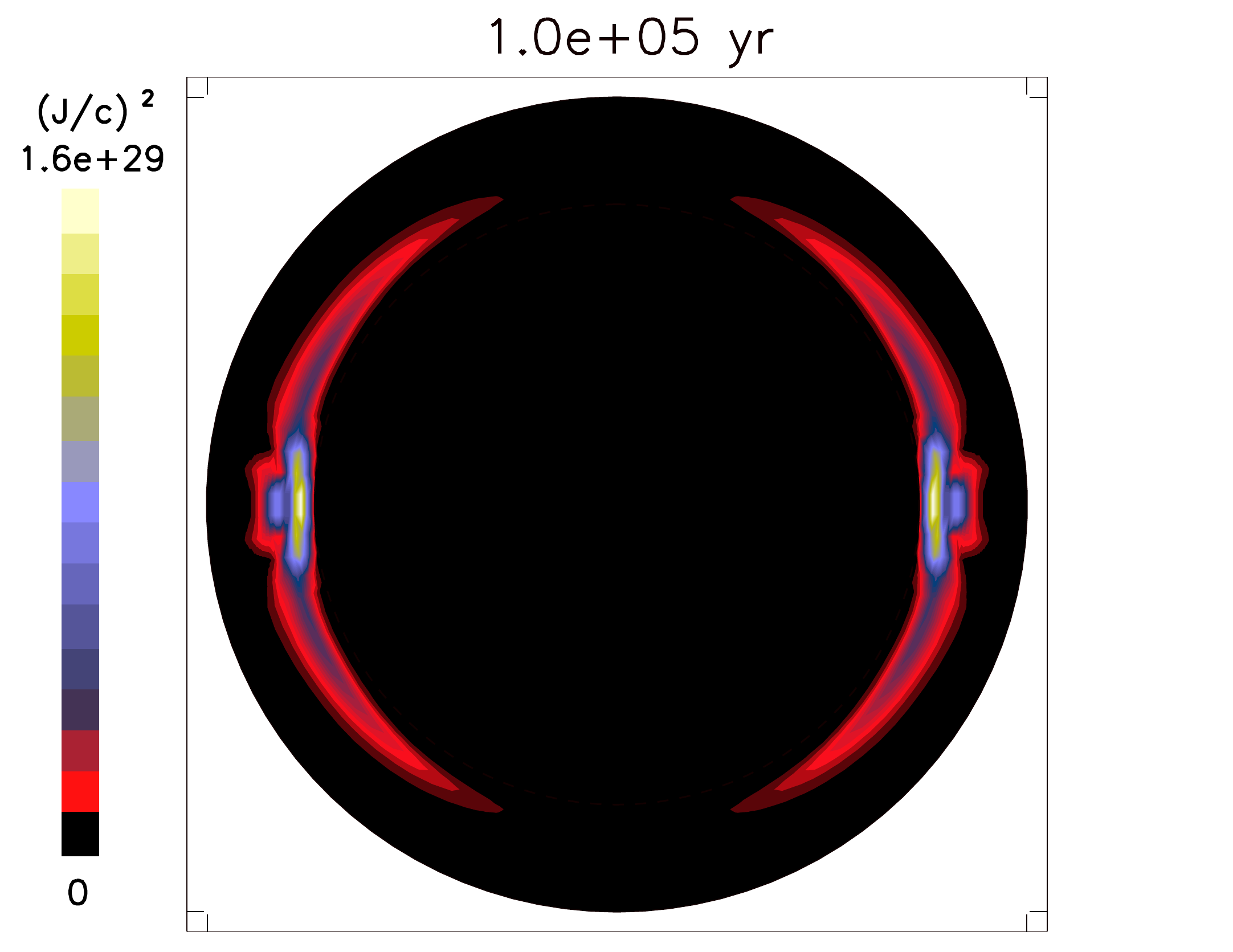}\\
	\vskip0.2cm
	\includegraphics[width=.32\textwidth]{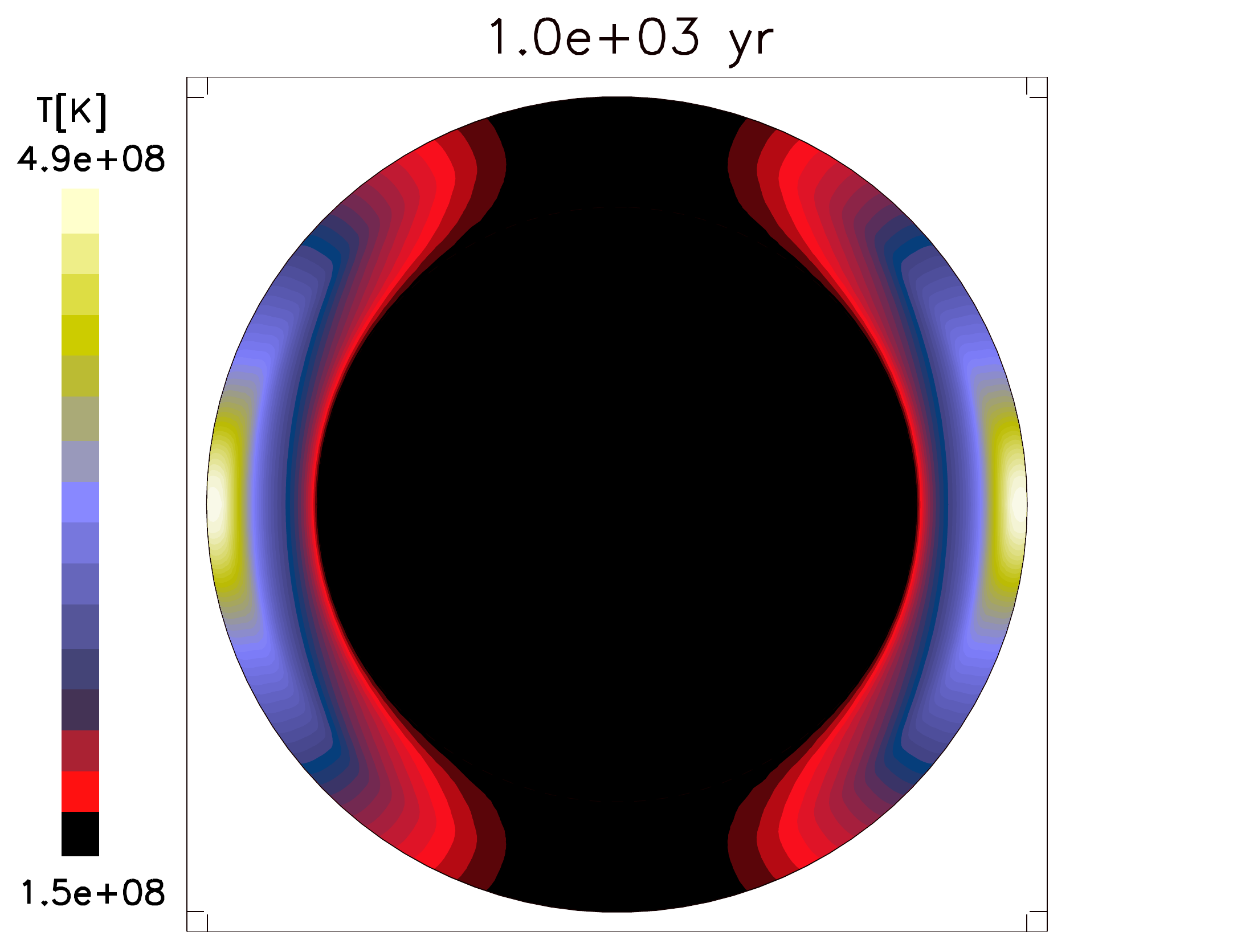}
	\includegraphics[width=.32\textwidth]{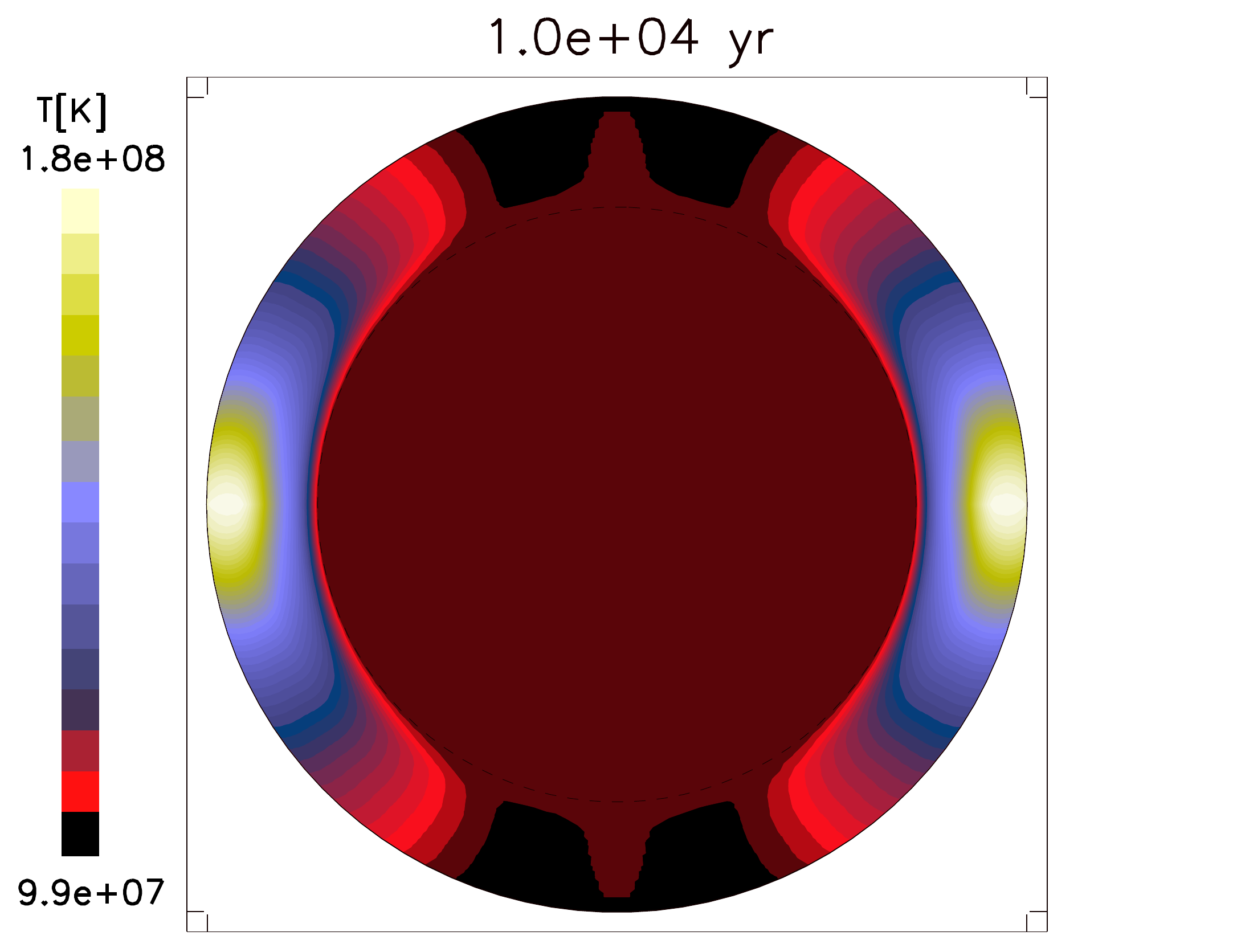}
	\includegraphics[width=.32\textwidth]{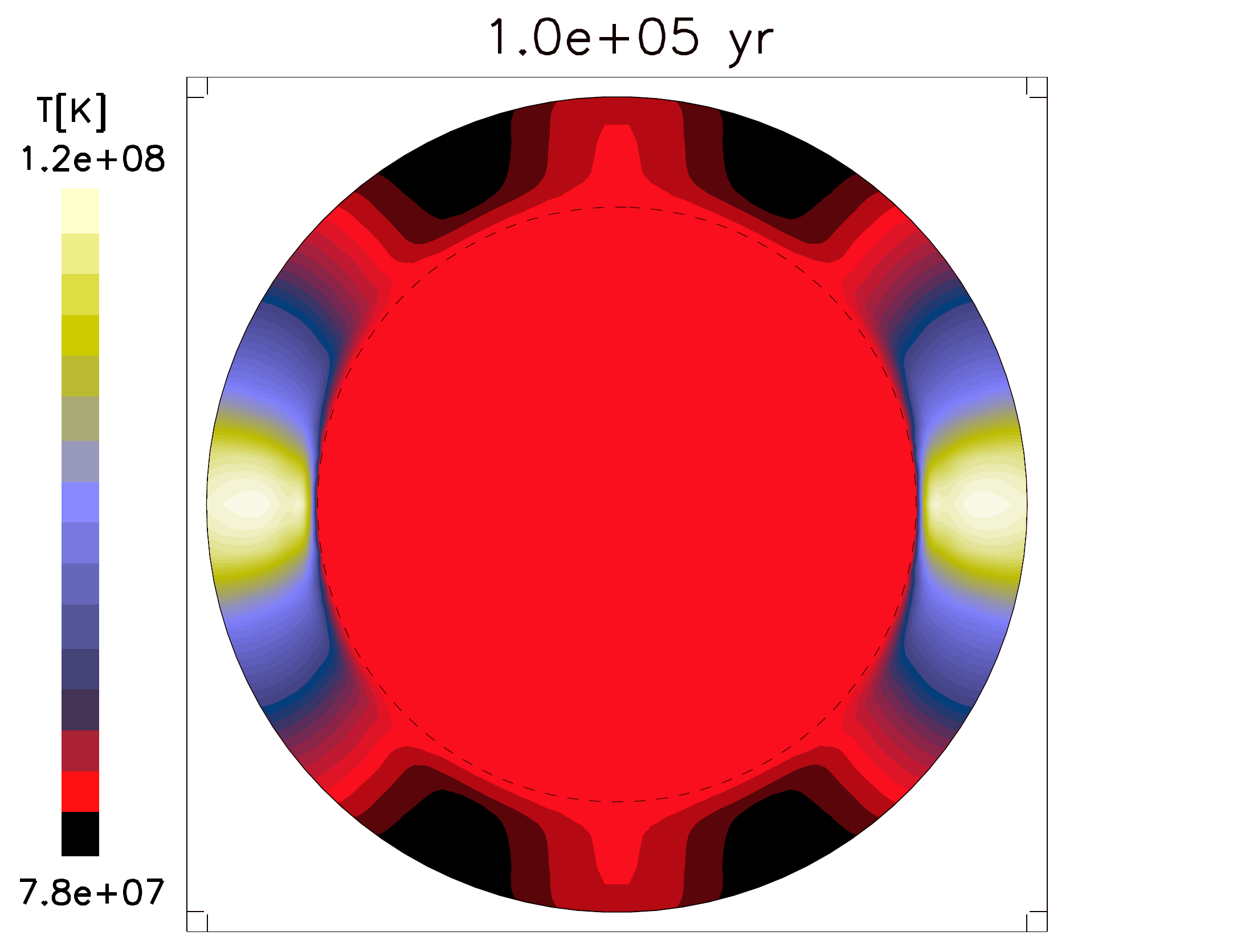}
	\caption{Snapshots of the magneto-thermal evolution of a NS model at $10^3, 10^4, 10^5$ yr, from left to right. Top panels: the left hemisphere shows in color scale the surface temperature, while the right hemisphere displays the magnetic configuration in the crust. Black lines are the projections of the poloidal field lines and the color scale indicates the toroidal magnetic field intensity (yellow: positive, red: negative). Middle panels: intensity of currents; the color scale indicates $J^2/c^2$, in units of $($G/km$)^2$. Bottom panels: temperature map inside the star. In all panels, the thickness of the crust has been enlarged by a factor of 4 for visualization purposes. Figure courtesy of \cite{vigano13}. Animations available in the supplementary material.}
	\label{fig:b14_evo}
\end{figure}

We now turn to the complete problem: solving the temperature evolution coupled to the induction equation with realistic microphysics. 
To our knowledge, the only existing  work studying the fully coupled magneto-thermal evolution of a realistic NS was \citep{vigano13}, where they presented the results of 2D simulations. 
This work also re-analysed in a consistent way the available data on isolated, thermally emitting NSs (a sample of 40 sources), and compared the theoretical models to the data, concluding that the evolutionary models can explain the phenomenological diversity of isolated NSs by only varying their initial magnetic field, NS mass, and envelope composition. 

As an  example, in Fig.~\ref{fig:b14_evo}  we show three snapshots of the evolution of a crustal confined model, initially an
$l=1$ poloidal field with $B_p=10^{14}$ G (labelled as model A14 in \citealt{vigano13}). Many of the general features described 
in previous more simple cases are also visible in this realistic model. Let us recap the most important details:
\begin{itemize}
\item
The first effect of the Hall term in the induction equation is to couple the poloidal and toroidal components so that, even if the latter is zero at the beginning, it is quickly created. After $\sim 10^3$ yr, a 
quadrupolar toroidal magnetic field with a maximum strength of the same order of the poloidal magnetic field has been created, with $B_\varphi$ being negative in the northern hemisphere and positive in the southern hemisphere.
\item 
Thereafter, under the effect of the Hall drift, the toroidal magnetic field rules the evolution, dragging the currents into the inner crust (see middle panels), and compressing the magnetic field lines.
The Hall term is thus responsible for the energy redistribution from the large scale dipole to small scales (higher-order multipoles are locally very strong), possibly creating current sheets in some situations (here, in the equator).
\item 
Where sufficiently small-scale components are present,  the locally enhanced ohmic dissipation balances the effect of the Hall drift and  
a quasi-stationary state (resembling the Hall attractor) is reached.
After $\sim 10^5$ yr, the toroidal magnetic field is mostly contained in the inner crust.  
\item
We note that, at this point, most of the current circulates close to the crust/core interface. Therefore, the dissipation of magnetic energy is regulated by the resistivity in this precise region. 
In the model, there was a highly resistive layer in the {\it nuclear pasta region} leading to a rapid decay of the magnetic field, which
has a direct imprint on the observable rotational properties of X-ray pulsars \citep{pons13}.
\item 
Joule heating modifies the map of the internal temperature. We can observe in the bottom panels of Fig.~\ref{fig:b14_evo} how, at $t=10^3$ yr, the equator is hotter than the poles by a factor of 3. This is caused by the insulating effect of the strong magnetic field discussed in \S \ref{sec:T_anis}.
The presence of strong tangential components ($B_\theta$ and $B_\varphi$) insulates the surface against the interior. In a dipolar geometry, the magnetic field is nearly radial at the poles, which remain thermally connected with the interior, while the equatorial region is insulated by tangential magnetic field lines. This has a two-fold effect: if the core is warmer than the crust, the polar regions will be warmer than the equator; however, if ohmic dissipation heats the equatorial regions, the situation is reverted. The temperature reflects the geometry of the poloidal magnetic field lines, which channel the heat flow.
\end{itemize}

\begin{figure}
	\centering
	\includegraphics[width=.8\textwidth]{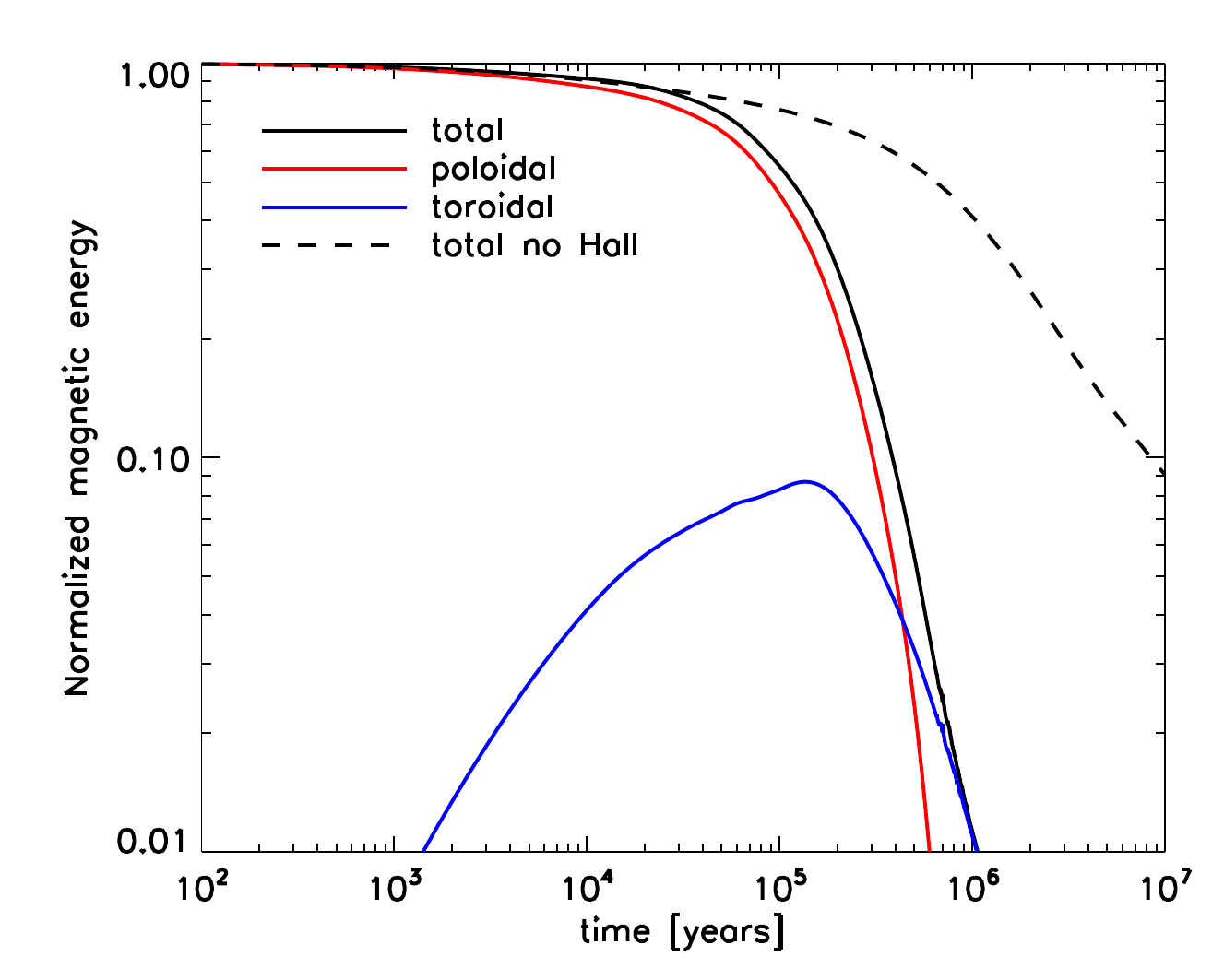}
	\caption{Magnetic energy in the crust (normalized to the initial value) as a function of time, for the sane model of Fig. \ref{fig:b14_evo}. The solid lines correspond respectively to the total magnetic energy (black), the energy in the poloidal component (red), and the energy in the toroidal component (blue). The dashed line shows the evolution of the same model when 
	the Hall term is deactivated (only Ohmic dissipation). Figure courtesy of \cite{vigano13}.}
	\label{fig:entransf}
\end{figure}

In order to show more clearly the enhanced dissipation caused by the combined action of Hall and Ohmic terms, in 
Fig.~\ref{fig:entransf} we show the evolution of the total magnetic energy stored in each component, comparing the evolution of the previous model 
with another model with the same initial data but switching off the Hall term (purely resistive case). In this case, there is no creation of a toroidal magnetic field or smaller scales. When the Hall term is included, $\sim 99 \%$ of the initial magnetic energy is dissipated in the first $\sim 10^6$ yr, compared to only the $60\%$ in the purely resistive case. At the same time, a $\sim 10\%$ of the initial energy is transferred to the toroidal component in $10^5$ yr, before it begins to decrease. Note that the poloidal magnetic field, after $10^5$ yr, is dissipated faster than the toroidal magnetic field. The poloidal magnetic field is supported by toroidal currents concentrated in the inner, equatorial regions of the crust. Here the resistivity is high for two reasons: the effect of the nuclear pasta phase, and the higher temperature (see right bottom panel of Fig. \ref{fig:b14_evo}). Conversely, the toroidal magnetic field is supported by larger loops of poloidal currents that circulate in higher latitude and outer regions, where the resistivity is lower. As a result, at late times most of the magnetic energy is stored in the toroidal magnetic field.
This example is very illustrative of the importance of knowing in detail the topology of the field and the location of currents at different stages.
We refer the interested reader to \cite{vigano13} for an extended analysis of different models, and how the initial magnetic field configuration affects the evolution.  The qualitative behavior is similar to that shown in 
model A14, but subtle differences can arise when the strength or geometry of the initial field is modified.

\begin{figure}
	\centering
	\includegraphics[width=.9\textwidth]{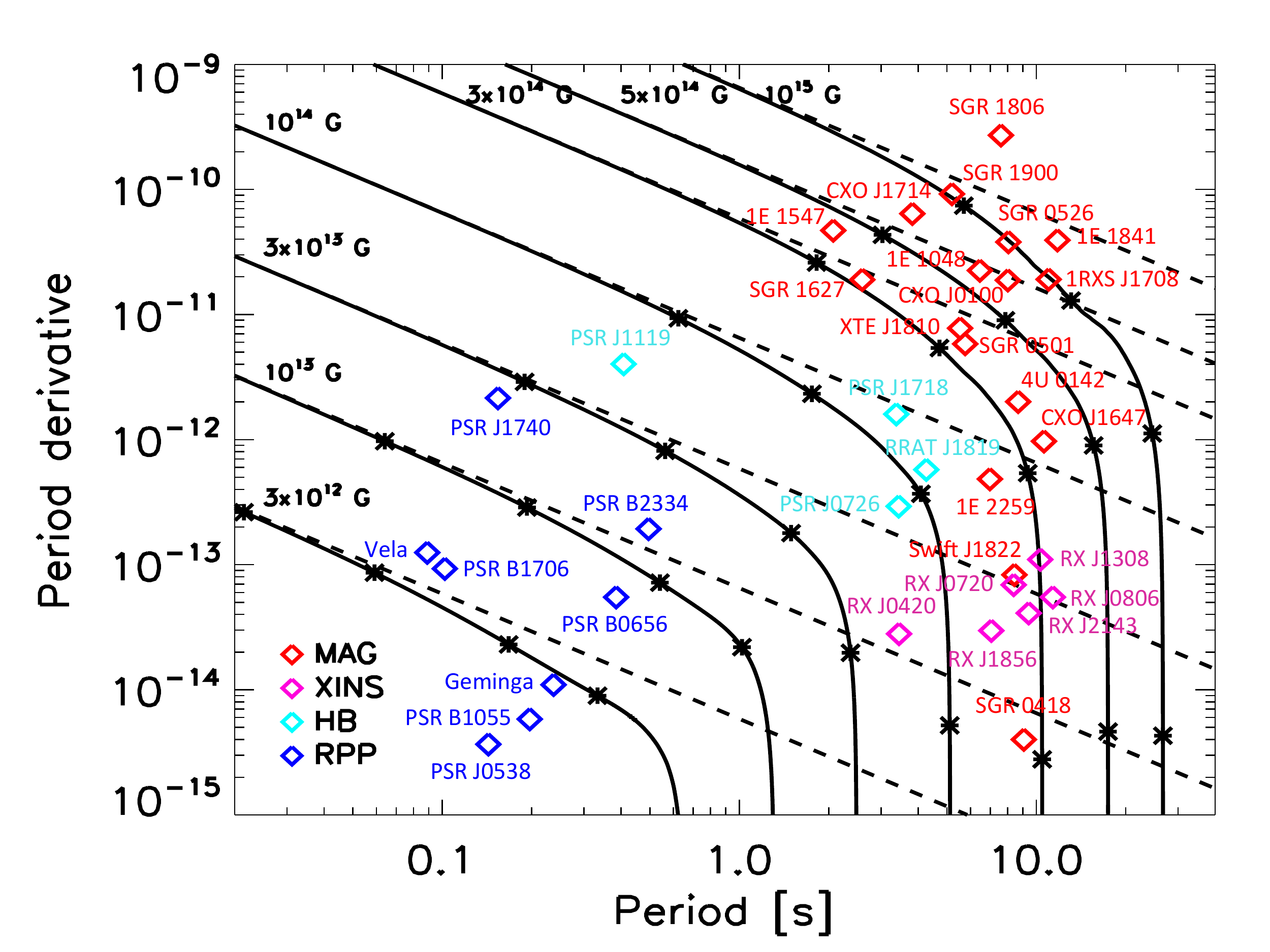}
	\caption{Evolutionary tracks in the $P-\dot{P}$ diagram of a typical NS with mass of $1.4 M_\odot$ and radius of 11.6 km, with different initial magnetic field strengths: $B_p^0= 3\times10^{12}, 10^{13}, 3\times10^{13}, 10^{14}, 3\times10^{14}, 10^{15}$ G, evolving under the action of the Hall drift and Ohmic dissipation. Asterisks indicate the points when the star reaches the age of $t=10^3,10^4,10^5,5\times10^5$ yr. Dashed lines show the tracks followed without considering magnetic field decay. The figure includes the sample of X-ray pulsars
with thermal emission analysed in \cite{vigano13}, which contains magnetars (MAG), nearby X-ray isolated NSs (XINS), rotation powered pulsars (RPP), and high magnetic field pulsars (HB). Figure courtesy of \cite{vigano13}.} 
\label{fig:ppdot}
\end{figure}

Finally, we turn our attention to the rotational evolution of NSs. In Fig.~\ref{fig:ppdot} we show evolutionary tracks in the
$P-\dot{P}$ diagram for a typical NS of 1.4 $M_\odot$ with different initial values of the initial magnetic field strength. The magnetic field configuration employed is the
type A geometry (crustal confined) in \cite{vigano13}.
Dashed lines show the results for models assuming a constant magnetic field, which are straight lines in the diagram. The solid lines, which account for
realistic field evolution, show significant differences from the constant field models.
Initially, the tracks overlap ($B_p$ is almost constant during an initial epoch,
$t\lesssim 10^3-10^5$ yr, which depends on the initial $B_p^0$), but eventually, the field dissipates faster than the spin period evolution timescale
and the lines bend down, at nearly constant $P$. This effect has been proposed to be the main reason for the observed clustering of periods of isolated X-ray pulsars \citep{pons13}. 
The particular value of the limit period mainly depends on the initial magnetic field and the resistivity at the crust-core interface. 
The large differences between the spin period evolution for models assuming a constant magnetic field and more realistic models make evident that 
the coupling between temperature, magnetic, and rotational evolution has to be considered.

\section{Future prospects}\label{sec:conclusions}

After having reviewed the status-of-the art of the field of the long-term magneto-thermal evolution of NSs, we highlight the three main areas that, in our opinion, need the focus of the researchers for the near- and mid-term future:

\begin{itemize}

\item Although the numerical solution of the 3D heat equation is a well-studied problem in the literature of numerical methods, in our particular context of NSs, the first paper implementing a full 3D temperature evolution with realistic microphysics is yet to come. The main reason for this lack of models is that the study of the 3D temperature evolution alone does not add much to the problem, and only its coupling with the magnetic field evolution is of great interest. From the observational point of view, the possible  existence of small-scale hotspots associated with the properties of the X-ray spectra is plausibly connected to the creation of small magnetic structures, and localized heat deposition. Thus, a necessary future step is to  implement consistent 3D temperature evolution in the few existing 3D magnetic field evolution codes.

\item
As discussed in Sect.~\ref{sec:magnetosphere}, another open issue is the correct implementation of realistic, more general, boundary conditions at the star surface. 
Going beyond the popular and simple potential/vacuum solution seems a necessity, that has only begun to be considered. 
The magnetar observational data favor the presence of twisted magnetospheres, that can influence  the interior dynamics in a significant manner. Among the different possible solutions, 
some mentioned in the text (solving elliptic equations, extended domains), it is unclear which has a better balance between physical motivation and computational cost.
This point is of particular relevance when connected to the first one: the creation of localized hot spots may be strongly dependent on the applied boundary condition, 
because currents passing through the envelope may have the key to understand the very high temperatures of magnetars.

\item
Finally, the core evolution is arguably the less explored part of the problem. Concerning the temperature, the core is almost isothermal to a very good approximation. 
But the complexities of the interaction between superfluid neutrons and superconducting protons result in uncertainties of many orders of magnitude in the transport 
coefficients that determine the magnetic field evolution. A full 3D study of the evolution of the field penetrating the core and including all relevant physics does not exist, 
and it should be a high priority task for the incoming years. 

\end{itemize}

All these efforts, in combination with the continuous upgrades of the microphysics ingredients, and the improving quality of the observational data 
with the new instruments, will allow us to decipher some of  the fascinating physical processes taking place in the interiors of NSs.

\begin{acknowledgements}
This work is partially supported by the PHAROS COST Action CA16214.
DV acknowledges the support from the Spanish Ministry of Economy, Industry and Competitiveness grants AYA2016-80289-P and AYA2017-82089-ERC (AEI/FEDER, UE). We thank K.~Gourgouliatos for fruitful discussions.
\end{acknowledgements}

\appendix
\normalsize

\section{Poloidal-toroidal decomposition of the magnetic field}\label{app:formalism}

Any three-dimensional, solenoidal vector field $\vec{B}$, can be expressed in terms of  its {\em poloidal} and {\em toroidal} components
\begin{equation} 
\vec{B}=\vec{B}_{\rm pol} + \vec{B}_{\rm tor}~.
\end{equation} 
In the literature, one can find different formalisms and notations to describe the two components. In this appendix we go through some of the ideas of the mathematical formalism and compare the most common notations.

Adopting the notation of \cite{geppert91}, the magnetic field can be written in terms of two scalar functions $\Phi (\vec{r},t)$ and $\Psi (\vec{r},t)$ (analogous to the stream functions in hydrodynamics) as follows:
\begin{eqnarray}\label{eq:def_decomposition_app}
&& \vec{B}_{pol}=\vec{\nabla}\times(\vec{\nabla}\times\Phi\vec{k})~,\\
&& \vec{B}_{tor}=\vec{\nabla}\times\Psi\vec{k}~,
\end{eqnarray}
where $\vec{k}$ is an arbitrary vector. This decomposition is particularly useful in situations where $\vec{k}$ is taken to be normal to one of the physical boundaries.
Therefore, for a spherical domain, and using spherical coordinates $(r,\theta,\varphi)$, a suitable choice  is $\vec{k}=\vec{r}$. In this case, $\vec{\nabla}\times\vec{r}=0$, and
we can write:
\begin{eqnarray}\label{eq:def_pot_vect_app}
&& \vec{B}_{pol}=\vec{\nabla}\times(\vec{\nabla}\Phi\times\vec{r}) = -\vec{r} ~\nabla^2 \Phi + \nabla 
\left( \frac{\partial (r\Phi)}{\partial r}\right)~,\\
&& \vec{B}_{tor}=\vec{\nabla}\Psi\times\vec{r}~.
\end{eqnarray}
Generally speaking, the radial component of the magnetic field is included in the poloidal part, while the $\theta$ and $\varphi$ components are shared between poloidal and toroidal components. In axial symmetry, $\Phi=\Phi(r,\theta)$ and $\Psi=\Psi(r,\theta)$, the expressions are further simplified: the toroidal magnetic field is directed along the azimuthal direction $\hat{\varphi}$. In this case the $\varphi -$component of the potential vector is given by $\vec{A_\varphi} = - \vec{r} \times \nabla \Phi ~,$ and the poloidal field can be directly derived from $\vec{B}_{\rm pol} = \nabla \times \vec{A_\varphi}~.$ 

Alternatively, another common notation expresses the magnetic field in terms of two other scalar functions, $P$ and $\Theta$ as:
\begin{equation}\label{eq:mf_clebsch_app}
\vec{B}=\vec{\nabla} P \times \vec{\nabla} \Theta~.
\end{equation}
In axial symmetry, and with the choice $\Theta=\varphi-\xi (r,\theta)$, the \textit{magnetic flux function} $P(r,\theta)$ is related to the $\varphi-$component of the vector potential by
\begin{equation}\label{eq:gamma_aphi_app}
P(r,\theta)= r\sin\theta ~ A_\varphi(r,\theta) ~,
\end{equation}
and the poloidal and toroidal components are 
\begin{eqnarray}\label{eq:def_poloidal_app}
&& \vec{B}_{pol}=\frac{\vec{\nabla} P(r,\theta) \times \hat{\varphi}}{r \sin\theta}~,\\
&& \vec{B}_{tor}=(\vec{\nabla} \xi)_{pol} \times (\vec{\nabla}P)_{pol} \equiv \frac{T}{r\sin\theta}\hat{\varphi}~,
\end{eqnarray}
where we have introduced the scalar stream funtion $T$ used, for instance, in \cite{akgun17} and following works (in the force-free case, $T$ is a function of $P$, see \S~\ref{sec:forcefree}). The conversion between the two formalisms in axial symmetry is shown in Table~\ref{tab:formalism}.

\begin{table}[t]
	\begin{center}
		\begin{tabular}{l r r r}
			\hline
			{\bf Formalisms}  &  \cite{akgun17} & \cite{kojima17} & \cite{geppert91} \\
			\hline
			Poloidal function 	& $P(r,\theta)$		& $G(r,\theta)$	     & $\Phi(r,\theta)$\\
			Toroidal function 	& $T(r,\theta)$           & $S(r,\theta)$    &   $\Psi(r,\theta)$\\
			Toroidal potential vector $A_\varphi$         & $P(r,\theta)/r \sin\theta$	   & $G(r,\theta)/\varpi$		    	& $-\partial_\theta\Phi$ \\
			Magnetic flux		& $2\pi P$			& $2\pi G$				& $-2\pi r\sin\theta\partial_\theta \Phi$\\
			Poloidal magnetic field $\vec{B}_{pol}$ 	& $(\vec{\nabla}P\times\hat{\varphi})/r\sin\theta$  & $(\vec{\nabla}G\times\hat{\varphi})/\varpi$	& $\vec{\nabla}\times(\vec{\nabla}\Phi\times\vec{r})$ \\
			Toroidal magnetic field $\vec{B}_{tor}$ 	& $(T/r\sin\theta)~\hat{\varphi}$    & $(S/\varpi)~\hat{\varphi}$ 	& $\vec{\nabla}\Psi\times\vec{r} $ \\
			\hline
			\label{tab:formalism}
		\end{tabular}
		\caption{Comparison between different notations in axial symmetry. \cite{pons09} used the same notation as \cite{geppert91}, and in \cite{gourgouliatos16} $\Phi$ and $\Psi$ are denominated 
		by $V_p$ and $V_t$, respectively.}		
	\end{center}
\end{table}

\section{Potential solutions with Green's method}
For potential configurations, we can express the potential magnetic field in terms of the magnetostatic potential $\chi_m$, so that
\begin{eqnarray}
&& \vec{B}=\vec{\nabla}\chi_m~, \label{eq:magnetostatic_potential}\\ 
&& \nabla^2 \chi_m=0~.\label{eq:laplace_potential}
\end{eqnarray}
The second Green's identity, applied to a volume enclosed by a surface $S$, relates the magnetostatic potential $\chi_m$ with a {\em Green's function} $G$ (see Eq.~(1.42) of \citealt{jackson91}):

\begin{equation}\label{eq:green_psi1}
2\pi \chi_m(\vec{r}) = -\int_S \frac{\partial G}{\partial n'}(\vec{r},\vec{r}') \chi_m(\vec{r}') {\rm d}S' + \int_S G(\vec{r},\vec{r}')\frac{\partial \chi_m}{\partial n'}(\vec{r}'){\rm d}S'~,
\end{equation}
where $\hat{n}'$ is the normal to the surface. Comparing with the electrostatic problem, we see that no volume integral is present, because $\dive\vec{B}\equiv \nabla^2\chi_m=0$. Note also that the factor $2\pi$ appears instead of the canonical $4\pi$, because inside the star Eq.~(\ref{eq:laplace_potential}) does not hold, thus $2\pi$ is the solid angle seen from the surface. The Green's function has to satisfy $\nabla'^2 G(\vec{r},\vec{r}')=-2\pi \delta(\vec{r}-\vec{r}')$. The functional form of $G$ is gauge dependent: given a Green's function $G$, any function $F(\vec{r},\vec{r}')$ which satisfied $\nabla'^2 F=0$ can be used to build a new Green's function $\tilde{G}=G+F$. The boundary conditions determine which gauge is more appropriate for a specific problem.

In our case the volume is the outer space, $S$ is a spherical boundary of radius $R$ (e.g., the surface of the star), and $\hat{n}'=-\hat{r}'$. We face a {\em von Neumann boundary condition problem}, because we know the form of the radial magnetic field
\begin{equation}
B_r(R,\theta) \equiv \frac{\partial\chi_m}{\partial r}(R,\theta)~.
\end{equation}
In order to reconstruct the form of
\begin{equation}
B_\theta(R,\theta)\equiv \frac{1}{R}\frac{\partial\chi_m}{\partial\theta}(R,\theta)~,
\end{equation}
we have to solve the following integral equation for $\chi_m$:

\begin{eqnarray}\label{eq:green_psi2}
2\pi \chi_m(\vec{r}) & = & R^2\left\{\int_0^\pi\int_0^{2\pi}  \frac{\partial G}{\partial r'}(\vec{r},\vec{r}') \chi_m(R,\theta')\sin\theta' {\rm d} \varphi' {\rm d} \theta' + \right.\nonumber\\
&& \left. - \int_0^\pi\int_0^{2\pi} G(\vec{r},\vec{r}') B_r(\theta')\sin\theta' {\rm d} \varphi' {\rm d} \theta'\right\}~.
\end{eqnarray}
So far, we have not specified the Green's function. In our case, the simplest Green's function is:

\begin{eqnarray}\label{eq:green0}
&& G(\vec{r},\vec{r}')=\frac{1}{|\vec{r}-\vec{r}'|}= [(r\sin\theta\cos\varphi - r'\sin\theta'\cos\varphi')^2+\nonumber\\
&& + (r\sin\theta\sin\varphi - r'\sin\theta'\sin\varphi')^2+(r\cos\theta - r'\cos\theta')^2]^{-1/2}~.
\end{eqnarray}
In axial symmetry, we can set $\varphi=0$, to obtain

\begin{equation}\label{eq:green1}
G(\vec{r},\vec{r}')=[(r\sin\theta - r'\sin\theta'\cos\varphi')^2+(r'\sin\theta'\sin\varphi')^2+(r\cos\theta - r'\cos\theta')^2]^{-1/2}~.
\end{equation}

We can evaluate $G$ and its radial derivative at $r=r'=R$
\begin{eqnarray}
&& G(R,\theta,\theta',\varphi')=\frac{1}{\sqrt{2}R}\left[1-\cos(\theta-\theta')+2\sin\theta\sin\theta'\sin^2\left(\frac{\varphi'}{2}\right)\right]^{-1/2}~, \label{green_eps0} \\
&& \derparn{r'}{G}(R,\theta,\theta',\varphi') \rightarrow -\frac{G}{2R}~. \label{dergreen_eps0}
\end{eqnarray}
Casting the two formulas above in Eq.~(\ref{eq:green_psi2}), we note that the following integral appears in the two right-hand side terms:

\begin{equation}
f(\theta,\theta')\equiv \sin\theta'\int_0^{2\pi}RG(R,\theta,\theta',\varphi'){\rm d} \varphi'~.
\end{equation}
As $G$ depends on $\varphi'$ via $\sin^2(\varphi'/2)$, we can change the integration limits to $[0,\pi/2]$, and $\varphi'\rightarrow 2\varphi'$, therefore

\begin{equation}\label{eq:green_f}
f(\theta,\theta')=\sqrt{8}\sin\theta'\int_0^{\pi/2} [1-\cos(\theta-\theta')+2\sin\theta\sin\theta'\sin^2\varphi']^{-1/2}{\rm d} \varphi'~.
\end{equation}
Casting Eq.~(\ref{eq:green_f}) in Eq.~(\ref{eq:green_psi2}), and substituting $\chi_m(\theta)=R\int_0^\theta B_\theta(R,\theta') \de\theta'$, we have

\begin{equation}\label{eq:greeneps0_final}
4\pi\int_0^\theta B_\theta(\theta') {\rm d} \theta' + \int_0^\pi  B_\theta(\theta')\left[\int_{\theta'}^{\pi}f(\theta,\theta'')\de\theta''\right]\de\theta' = -2\int_0^\pi B_r(\theta')f(\theta,\theta')\de\theta'~.
\end{equation}
In Eq.~(\ref{eq:green_f}), if $\theta=\theta'$, then $f(\theta,\theta')\rightarrow 2\int_0^{\pi/2}(\sin\varphi')^{-1}{\rm d} \varphi'$, which is not integrable because of the singularity in $\varphi'=0$ (corresponding to $\vec{r}=\vec{r}'$). However, in both terms where it appears, the function $f(\theta,\theta')$ is integrated in $\theta'$, and both terms of the equation are integrable.

For numerical purposes, we can express Eq.~(\ref{eq:greeneps0_final}) in matrix form, introducing $f_{ij}=f(\theta_i,\theta'_j)$ evaluated on two grids with vectors $\theta_i,\theta'_j$, with $m$ steps $\Delta\theta$. The coefficients of the matrix $f_{ij}$ are purely geometrical, therefore they are evaluated only once, at the beginning. The grid $\theta_i$ coincides with the locations of $B_r(R,\theta)$, while the resolution of the grid $\theta'_j$ is $M$ times the resolution of the grid $\theta_i$ ($M\gtrsim 5$) to improve the accuracy of the integral function $f_{ij}$ near the singularities $\theta_i\rightarrow\theta_j$. The resolution of the grid of $\varphi'_k$ barely affects the result, provided that it avoids the singularities $\varphi'=0,\pi/2$. We typically use $M=10$ and $n_\varphi'=1000$. The calculation of the factors $f_{ij}$ is performed just once and stored. The matrix form is:

\begin{equation}\label{eq:greeneps0_matrix}
\sum_{j=1}^m [4\pi\delta_{ij} + f_{ij}\Delta\theta] \chi_m(\theta_j)= \sum_{j=1}^m [-2f_{ij}\Delta\theta]B_r(\theta_j), \qquad i=1,m~.
\end{equation}
From this, we obtain $B_\theta$ by taking the finite difference derivative of $\chi_m(\theta)$.

\bibliographystyle{spbasic}      
\bibliography{10-references}   

\begin{thebibliography}{191}
\providecommand{\natexlab}[1]{#1}
\providecommand{\url}[1]{{#1}}
\providecommand{\urlprefix}{URL }
\expandafter\ifx\csname urlstyle\endcsname\relax
  \providecommand{\doi}[1]{DOI~\discretionary{}{}{}#1}\else
  \providecommand{\doi}{DOI~\discretionary{}{}{}\begingroup
  \urlstyle{rm}\Url}\fi
\providecommand{\eprint}[2][]{\url{#2}}

\bibitem[{{Aguilera} et~al(2008{\natexlab{a}}){Aguilera}, {Pons}, and
  {Miralles}}]{aguilera08b}
{Aguilera} DN, {Pons} JA, {Miralles} JA (2008{\natexlab{a}}) {2D Cooling of
  magnetized neutron stars}. \aap 486:255--271,
  \doi{10.1051/0004-6361:20078786}, \eprint{arXiv:0710.0854}

\bibitem[{{Aguilera} et~al(2008{\natexlab{b}}){Aguilera}, {Pons}, and
  {Miralles}}]{aguilera08a}
{Aguilera} DN, {Pons} JA, {Miralles} JA (2008{\natexlab{b}}) {The Impact of
  Magnetic Field on the Thermal Evolution of Neutron Stars}. \apjl
  673:L167--L170, \doi{10.1086/527547}, \eprint{arXiv:0712.1353}

\bibitem[{{Akg{\"u}n} et~al(2016){Akg{\"u}n}, {Miralles}, {Pons}, and
  {Cerd{\'a}-Dur{\'a}n}}]{akgun16}
{Akg{\"u}n} T, {Miralles} JA, {Pons} JA, {Cerd{\'a}-Dur{\'a}n} P (2016) {The
  force-free twisted magnetosphere of a neutron star}. \mnras 462:1894--1909,
  \doi{10.1093/mnras/stw1762}, \eprint{1605.02253}

\bibitem[{{Akg{\"u}n} et~al(2017){Akg{\"u}n}, {Cerd{\'a}-Dur{\'a}n},
  {Miralles}, and {Pons}}]{akgun17}
{Akg{\"u}n} T, {Cerd{\'a}-Dur{\'a}n} P, {Miralles} JA, {Pons} JA (2017)
  {Long-term evolution of the force-free twisted magnetosphere of a magnetar}.
  \mnras 472:3914--3923, \doi{10.1093/mnras/stx2235}, \eprint{1706.07990}

\bibitem[{{Akg{\"u}n} et~al(2018){Akg{\"u}n}, {Cerd{\'a}-Dur{\'a}n},
  {Miralles}, and {Pons}}]{akgun18b}
{Akg{\"u}n} T, {Cerd{\'a}-Dur{\'a}n} P, {Miralles} JA, {Pons} JA (2018)
  {Crust-magnetosphere coupling during magnetar evolution and implications for
  the surface temperature}. \mnras 481:5331--5338, \doi{10.1093/mnras/sty2669},
  \eprint{1807.09021}

\bibitem[{{Ant{\'o}n} et~al(2006){Ant{\'o}n}, {Zanotti}, {Miralles},
  {Mart{\'{\i}}}, {Ib{\'a}{\~n}ez}, {Font}, and {Pons}}]{anton06}
{Ant{\'o}n} L, {Zanotti} O, {Miralles} JA, {Mart{\'{\i}}} JM, {Ib{\'a}{\~n}ez}
  JM, {Font} JA, {Pons} JA (2006) {Numerical 3+1 General Relativistic
  Magnetohydrodynamics: A Local Characteristic Approach}. \apj 637:296--312,
  \doi{10.1086/498238}, \eprint{arXiv:astro-ph/0506063}

\bibitem[{{Antoniadis} et~al(2013){Antoniadis}, {Freire}, {Wex}, {Tauris},
  {Lynch}, {van Kerkwijk}, {Kramer}, {Bassa}, and {Dhillon}}]{antoniadis13}
{Antoniadis} J, {Freire} PCC, {Wex} N, {Tauris} TM, {Lynch} RS, {van Kerkwijk}
  MH, {Kramer} M, {Bassa} C, {Dhillon} VS (2013) {A Massive Pulsar in a Compact
  Relativistic Binary}. Science 340:448, \doi{10.1126/science.1233232},
  \eprint{1304.6875}

\bibitem[{{Arbona} et~al(2013){Arbona}, {Artigues}, {Bona-Casas}, {Mass{\'o}},
  {Mi{\~n}ano}, {Rigo}, {Trias}, and {Bona}}]{arbona13}
{Arbona} A, {Artigues} A, {Bona-Casas} C, {Mass{\'o}} J, {Mi{\~n}ano} B, {Rigo}
  A, {Trias} M, {Bona} C (2013) {Simflowny: A general-purpose platform for the
  management of physical models and simulation problems}. Computer Physics
  Communications 184:2321--2331, \doi{10.1016/j.cpc.2013.04.012}

\bibitem[{{Arbona} et~al(2018){Arbona}, {Minano}, {Rigo}, C., {Palenzuela},
  {Artigues}, {Bona-Casas}, and {Mass{\'o}}}]{arbona18}
{Arbona} A, {Minano} B, {Rigo} A, C B, {Palenzuela} C, {Artigues} A,
  {Bona-Casas} C, {Mass{\'o}} J (2018) {Simflowny 2: An upgraded platform for
  scientific modelling and simulation}. CoPhC 231,
  \doi{10.1016/j.cpc.2018.03.015}

\bibitem[{{Aubert} et~al(2008){Aubert}, {Aurnou}, and {Wicht}}]{aubert08}
{Aubert} J, {Aurnou} J, {Wicht} J (2008) {The magnetic structure of
  convection-driven numerical dynamos}. Geophysical Journal International
  172(3):945--956, \doi{10.1111/j.1365-246X.2007.03693.x}

\bibitem[{{Balsara}(2017)}]{balsara17}
{Balsara} DS (2017) {Higher-order accurate space-time schemes for computational
  astrophysics -- Part I: finite volume methods}. Living Rev Comput Astrophys
  3:2, \doi{10.1007/s41115-017-0002-8}, \eprint{1703.01241}

\bibitem[{{Balsara} and {Dumbser}(2015)}]{balsara15}
{Balsara} DS, {Dumbser} M (2015) {Divergence-free {MHD} on unstructured meshes
  using high order finite volume schemes based on multidimensional Riemann
  solvers}. J Comput Phys 299:687--715, \doi{10.1016/j.jcp.2015.07.012}

\bibitem[{{Barenblatt}(1952)}]{barenblatt52}
{Barenblatt} GI (1952) {On some unsteady fluid and gas motions in a porous
  medium.} Prikladnaya Matematika i Mekhanika 16

\bibitem[{{Beloborodov}(2009)}]{beloborodov09}
{Beloborodov} AM (2009) {Untwisting Magnetospheres of Neutron Stars}. \apj
  703:1044--1060, \doi{10.1088/0004-637X/703/1/1044}, \eprint{0812.4873}

\bibitem[{{Beloborodov}(2013)}]{beloborodov13}
{Beloborodov} AM (2013) {On the Mechanism of Hard X-Ray Emission from
  Magnetars}. \apj 762:13, \doi{10.1088/0004-637X/762/1/13}, \eprint{1201.0664}

\bibitem[{{Beloborodov} and {Levin}(2014)}]{beloborodov14}
{Beloborodov} AM, {Levin} Y (2014) {Thermoplastic Waves in Magnetars}. \apjl
  794:L24, \doi{10.1088/2041-8205/794/2/L24}, \eprint{1406.4850}

\bibitem[{{Belov} et~al(2017){Belov}, {Nugumanov}, and {Yakovlev}}]{Belov2017}
{Belov} PA, {Nugumanov} ER, {Yakovlev} SL (2017) {The arrowhead decomposition
  method for a block-tridiagonal system of linear equations}. In: Journal of
  Physics Conference Series, Journal of Physics Conference Series, vol 929, p
  012035, \doi{10.1088/1742-6596/929/1/012035}

\bibitem[{{Beskin} et~al(2013){Beskin}, {Istomin}, and {Philippov}}]{beskin13}
{Beskin} VS, {Istomin} YN, {Philippov} AA (2013) {Radio pulsars: the search for
  truth}. Physics Uspekhi 56:164, \doi{10.3367/UFNe.0183.201302e.0179},
  \eprint{1305.1740}

\bibitem[{{Bona} et~al(2009){Bona}, {Bona-Casas}, and {Terradas}}]{bona09}
{Bona} C, {Bona-Casas} C, {Terradas} J (2009) {Linear high-resolution schemes
  for hyperbolic conservation laws: TVB numerical evidence}. J Comput Phys
  228:2266--2281, \doi{10.1016/j.jcp.2008.12.010}, \eprint{0810.2185}

\bibitem[{{Bransgrove} et~al(2018){Bransgrove}, {Levin}, and
  {Beloborodov}}]{bransgrove18}
{Bransgrove} A, {Levin} Y, {Beloborodov} A (2018) {Magnetic field evolution of
  neutron stars - I. Basic formalism, numerical techniques and first results}.
  \mnras 473:2771--2790, \doi{10.1093/mnras/stx2508}, \eprint{1709.09167}

\bibitem[{{Burrows} and {Lattimer}(1986)}]{burrows86}
{Burrows} A, {Lattimer} JM (1986) {The birth of neutron stars}. \apj
  307:178--196, \doi{10.1086/164405}

\bibitem[{{Calabrese} et~al(2004){Calabrese}, {Lehner}, {Reula}, {Sarbach}, and
  {Tiglio}}]{Calabrese2004}
{Calabrese} G, {Lehner} L, {Reula} O, {Sarbach} O, {Tiglio} M (2004) {Summation
  by parts and dissipation for domains with excised regions}. Classical and
  Quantum Gravity 21:5735--5757, \doi{10.1088/0264-9381/21/24/004},
  \eprint{gr-qc/0308007}

\bibitem[{{Carrasco} et~al(2019){Carrasco}, {Vigan{\`o}}, {Palenzuela}, and
  {Pons}}]{carrasco19}
{Carrasco} F, {Vigan{\`o}} D, {Palenzuela} C, {Pons} JA (2019) {Triggering
  magnetar outbursts in 3D force-free simulations}. \mnras 484:L124--L129,
  \doi{10.1093/mnrasl/slz016}, \eprint{1901.08889}

\bibitem[{{Castillo} et~al(2017){Castillo}, {Reisenegger}, and
  {Valdivia}}]{castillo17}
{Castillo} F, {Reisenegger} A, {Valdivia} JA (2017) {Magnetic field evolution
  and equilibrium configurations in neutron star cores: the effect of ambipolar
  diffusion}. \mnras 471:507--522, \doi{10.1093/mnras/stx1604},
  \eprint{1705.10020}

\bibitem[{{Cerd{\'a}-Dur{\'a}n} et~al(2008){Cerd{\'a}-Dur{\'a}n}, {Font},
  {Ant{\'o}n}, and {M{\"u}ller}}]{cerdaduran08}
{Cerd{\'a}-Dur{\'a}n} P, {Font} JA, {Ant{\'o}n} L, {M{\"u}ller} E (2008) {A new
  general relativistic magnetohydrodynamics code for dynamical spacetimes}.
  \aap 492:937--953, \doi{10.1051/0004-6361:200810086}, \eprint{0804.4572}

\bibitem[{{Chamel}(2008)}]{chamel08}
{Chamel} N (2008) {Two-fluid models of superfluid neutron star cores}. \mnras
  388:737--752, \doi{10.1111/j.1365-2966.2008.13426.x}, \eprint{0805.1007}

\bibitem[{{Ciolfi} and {Rezzolla}(2013)}]{ciolfi2013}
{Ciolfi} R, {Rezzolla} L (2013) {Twisted-torus configurations with large
  toroidal magnetic fields in relativistic stars.} \mnras 435:L43--L47,
  \doi{10.1093/mnrasl/slt092}, \eprint{1306.2803}

\bibitem[{{Ciolfi} et~al(2019){Ciolfi}, {Kastaun}, {Vijay Kalinani}, and
  {Giacomazzo}}]{ciolfi2019}
{Ciolfi} R, {Kastaun} W, {Vijay Kalinani} J, {Giacomazzo} B (2019) {The first
  100 ms of a long-lived magnetized neutron star formed in a binary neutron
  star merger}. arXiv e-prints arXiv:1904.10222, \eprint{1904.10222}

\bibitem[{{Colaiuda} et~al(2008){Colaiuda}, {Ferrari}, {Gualtieri}, and
  {Pons}}]{colaiuda2008}
{Colaiuda} A, {Ferrari} V, {Gualtieri} L, {Pons} JA (2008) {Relativistic models
  of magnetars: structure and deformations}. \mnras 385:2080--2096,
  \doi{10.1111/j.1365-2966.2008.12966.x}, \eprint{0712.2162}

\bibitem[{{Colella} and {Woodward}(1984)}]{colella84}
{Colella} P, {Woodward} PR (1984) {The Piecewise Parabolic Method (PPM) for
  Gas-Dynamical Simulations}. J Comput Phys 54:174--201,
  \doi{10.1016/0021-9991(84)90143-8}

\bibitem[{{Contopoulos} et~al(1999){Contopoulos}, {Kazanas}, and
  {Fendt}}]{contopoulos99}
{Contopoulos} I, {Kazanas} D, {Fendt} C (1999) {The Axisymmetric Pulsar
  Magnetosphere}. \apj 511:351--358, \doi{10.1086/306652},
  \eprint{arXiv:astro-ph/9903049}

\bibitem[{{Coti Zelati} et~al(2018){Coti Zelati}, {Rea}, {Pons}, {Campana}, and
  {Esposito}}]{coti18}
{Coti Zelati} F, {Rea} N, {Pons} JA, {Campana} S, {Esposito} P (2018)
  {Systematic study of magnetar outbursts}. \mnras 474:961--1017,
  \doi{10.1093/mnras/stx2679}, \eprint{1710.04671}

\bibitem[{{Cromartie} et~al(2019){Cromartie}, {Fonseca}, {Ransom}, {Demorest},
  {Arzoumanian}, {Blumer}, {Brook}, {DeCesar}, {Dolch}, and
  {Ellis}}]{cromartie19}
{Cromartie} HT, {Fonseca} E, {Ransom} SM, {Demorest} PB, {Arzoumanian} Z,
  {Blumer} H, {Brook} PR, {DeCesar} ME, {Dolch} T, {Ellis} JA (2019) {A very
  massive neutron star: relativistic Shapiro delay measurements of PSR
  J0740+6620}. arXiv e-prints arXiv:1904.06759, \eprint{1904.06759}

\bibitem[{{Cumming} et~al(2004){Cumming}, {Arras}, and {Zweibel}}]{cumming04}
{Cumming} A, {Arras} P, {Zweibel} E (2004) {Magnetic Field Evolution in Neutron
  Star Crusts Due to the {Hall} Effect and Ohmic Decay}. \apj 609:999--1017,
  \doi{10.1086/421324}, \eprint{arXiv:astro-ph/0402392}

\bibitem[{{De Luca}(2017)}]{deluca17}
{De Luca} A (2017) {Central compact objects in supernova remnants}. In: Journal
  of Physics Conference Series, Journal of Physics Conference Series, vol 932,
  p 012006, \doi{10.1088/1742-6596/932/1/012006}, \eprint{1711.07210}

\bibitem[{{Dedner} et~al(2002){Dedner}, {Kemm}, {Kr{\"o}ner}, {Munz},
  {Schnitzer}, and {Wesenberg}}]{dedner02}
{Dedner} A, {Kemm} F, {Kr{\"o}ner} D, {Munz} CD, {Schnitzer} T, {Wesenberg} M
  (2002) {Hyperbolic Divergence Cleaning for the {MHD} Equations}. J Comput
  Phys 175:645--673, \doi{10.1006/jcph.2001.6961}

\bibitem[{{Demorest} et~al(2010){Demorest}, {Pennucci}, {Ransom}, {Roberts},
  and {Hessels}}]{demorest10}
{Demorest} PB, {Pennucci} T, {Ransom} SM, {Roberts} MSE, {Hessels} JWT (2010)
  {A two-solar-mass neutron star measured using Shapiro delay}. \nat
  467:1081--1083, \doi{10.1038/nature09466}, \eprint{1010.5788}

\bibitem[{{Dommes} and {Gusakov}(2017)}]{dommes}
{Dommes} VA, {Gusakov} ME (2017) {Vortex buoyancy in superfluid and
  superconducting neutron stars}. \mnras 467:L115--L119,
  \doi{10.1093/mnrasl/slx011}, \eprint{1701.06870}

\bibitem[{{Donat} and {Marquina}(1996)}]{PHM}
{Donat} R, {Marquina} A (1996) {Capturing Shock Reflections: An Improved Flux
  Formula}. J Comput Phys 125:42--58, \doi{10.1006/jcph.1996.0078}

\bibitem[{{Dormy} et~al(1998){Dormy}, {Cardin}, and {Jault}}]{dormy98}
{Dormy} E, {Cardin} P, {Jault} D (1998) {{MHD} flow in a slightly
  differentially rotating spherical shell, with conducting inner core, in a
  dipolar magnetic field}. Earth and Planetary Science Letters 160(1-2):15--30,
  \doi{10.1016/S0012-821X(98)00078-8}

\bibitem[{{Douchin} and {Haensel}(2001)}]{douchin01}
{Douchin} F, {Haensel} P (2001) {A unified equation of state of dense matter
  and neutron star structure}. \aap 380:151--167,
  \doi{10.1051/0004-6361:20011402}, \eprint{arXiv:astro-ph/0111092}

\bibitem[{{Elfritz} et~al(2016){Elfritz}, {Pons}, {Rea}, {Glampedakis}, and
  {Vigan{\`o}}}]{Elfritz2016}
{Elfritz} JG, {Pons} JA, {Rea} N, {Glampedakis} K, {Vigan{\`o}} D (2016)
  {Simulated magnetic field expulsion in neutron star cores}. \mnras
  456:4461--4474, \doi{10.1093/mnras/stv2963}, \eprint{1512.07151}

\bibitem[{{Fujisawa} and {Kisaka}(2014)}]{fuji14}
{Fujisawa} K, {Kisaka} S (2014) {Magnetic field configurations of a magnetar
  throughout its interior and exterior - core, crust and magnetosphere}. \mnras
  445:2777--2793, \doi{10.1093/mnras/stu1911}, \eprint{1409.4547}

\bibitem[{{Gabler} et~al(2014){Gabler}, {Cerd{\'a}-Dur{\'a}n}, {Stergioulas},
  {Font}, and {M{\"u}ller}}]{gabler14}
{Gabler} M, {Cerd{\'a}-Dur{\'a}n} P, {Stergioulas} N, {Font} JA, {M{\"u}ller} E
  (2014) {Modulating the magnetosphere of magnetars by internal magneto-elastic
  oscillations}. \mnras 443:1416--1424, \doi{10.1093/mnras/stu1263},
  \eprint{1407.7672}

\bibitem[{{Gavriil} et~al(2008){Gavriil}, {Gonzalez}, {Gotthelf}, {Kaspi},
  {Livingstone}, and {Woods}}]{gavriil08}
{Gavriil} FP, {Gonzalez} ME, {Gotthelf} EV, {Kaspi} VM, {Livingstone} MA,
  {Woods} PM (2008) {Magnetar-Like Emission from the Young Pulsar in {Kes 75}}.
  Science 319:1802, \doi{10.1126/science.1153465}, \eprint{0802.1704}

\bibitem[{{Geppert} and {Vigan{\`o}}(2014)}]{geppert14}
{Geppert} U, {Vigan{\`o}} D (2014) {Creation of magnetic spots at the neutron
  star surface}. \mnras 444:3198--3208, \doi{10.1093/mnras/stu1675},
  \eprint{1408.3833}

\bibitem[{{Geppert} and {Wiebicke}(1991)}]{geppert91}
{Geppert} U, {Wiebicke} HJ (1991) {Amplification of neutron star magnetic
  fields by thermoelectric effects. {I}. {G}eneral formalism}. Astron Astrophys
  Suppl Ser 87:217--228

\bibitem[{{Geppert} and {Wiebicke}(1995)}]{geppert95}
{Geppert} U, {Wiebicke} HJ (1995) {Amplification of neutron star magnetic
  fields by thermoelectric effects. {V}. {I}nduction of large-scale toroidal
  fields}. \aap 300:429

\bibitem[{{Geppert} et~al(2004){Geppert}, {K{\"u}ker}, and {Page}}]{geppert04}
{Geppert} U, {K{\"u}ker} M, {Page} D (2004) {Temperature distribution in
  magnetized neutron star crusts}. \aap 426:267--277,
  \doi{10.1051/0004-6361:20040455}, \eprint{arXiv:astro-ph/0403441}

\bibitem[{{Geppert} et~al(2006){Geppert}, {K{\"u}ker}, and {Page}}]{geppert06}
{Geppert} U, {K{\"u}ker} M, {Page} D (2006) {Temperature distribution in
  magnetized neutron star crusts. II. The effect of a strong toroidal
  component}. \aap 457:937--947, \doi{10.1051/0004-6361:20054696},
  \eprint{arXiv:astro-ph/0512530}

\bibitem[{{Giacomazzo} and {Rezzolla}(2007)}]{giacomazzo07}
{Giacomazzo} B, {Rezzolla} L (2007) {{WhiskyMHD}: a new numerical code for
  general relativistic magnetohydrodynamics}. Class Quantum Grav 24:S235--S258,
  \doi{10.1088/0264-9381/24/12/S16}, \eprint{gr-qc/0701109}

\bibitem[{{Glampedakis} et~al(2011){Glampedakis}, {Jones}, and
  {Samuelsson}}]{glampedakis11b}
{Glampedakis} K, {Jones} DI, {Samuelsson} L (2011) {Ambipolar diffusion in
  superfluid neutron stars}. \mnras 413:2021--2030,
  \doi{10.1111/j.1365-2966.2011.18278.x}, \eprint{1010.1153}

\bibitem[{{Glampedakis} et~al(2014){Glampedakis}, {Lander}, and
  {Andersson}}]{glampedakis14}
{Glampedakis} K, {Lander} SK, {Andersson} N (2014) {The inside-out view on
  neutron-star magnetospheres}. \mnras 437:2--8, \doi{10.1093/mnras/stt1814},
  \eprint{1306.6881}

\bibitem[{{Goldreich} and {Julian}(1969)}]{goldreich69}
{Goldreich} P, {Julian} WH (1969) {Pulsar Electrodynamics}. \apj 157:869,
  \doi{10.1086/150119}

\bibitem[{{Goldreich} and {Reisenegger}(1992)}]{goldreich92}
{Goldreich} P, {Reisenegger} A (1992) {Magnetic field decay in isolated neutron
  stars}. \apj 395:250--258, \doi{10.1086/171646}

\bibitem[{{Gonz{\'a}lez-Morales} et~al(2018){Gonz{\'a}lez-Morales}, {Khomenko},
  {Downes}, and {de Vicente}}]{gonzalez18}
{Gonz{\'a}lez-Morales} PA, {Khomenko} E, {Downes} TP, {de Vicente} A (2018)
  {{MHDSTS}: a new explicit numerical scheme for simulations of partially
  ionised solar plasma}. \aap 615:A67, \doi{10.1051/0004-6361/201731916},
  \eprint{1803.04891}

\bibitem[{{G{\"o}{\u{g}}{\"u}s} et~al(2016){G{\"o}{\u{g}}{\"u}s}, {Lin},
  {Kaneko}, {Kouveliotou}, {Watts}, {Chakraborty}, {Alpar}, {Huppenkothen},
  {Roberts}, and {Younes}}]{gogus16}
{G{\"o}{\u{g}}{\"u}s} E, {Lin} L, {Kaneko} Y, {Kouveliotou} C, {Watts} AL,
  {Chakraborty} M, {Alpar} MA, {Huppenkothen} D, {Roberts} OJ, {Younes} G
  (2016) {Magnetar-like X-Ray Bursts from a Rotation-powered Pulsar, PSR
  J1119-6127}. \apj 829(2):L25, \doi{10.3847/2041-8205/829/2/L25},
  \eprint{1608.07133}

\bibitem[{{Gourgouliatos} and {Cumming}(2014{\natexlab{a}})}]{gourgouliatos14b}
{Gourgouliatos} KN, {Cumming} A (2014{\natexlab{a}}) {{Hall} Attractor in
  Axially Symmetric Magnetic Fields in Neutron Star Crusts}. Phys Rev Lett
  112:171101, \doi{10.1103/PhysRevLett.112.171101}, \eprint{1311.7345}

\bibitem[{{Gourgouliatos} and {Cumming}(2014{\natexlab{b}})}]{gourgouliatos14a}
{Gourgouliatos} KN, {Cumming} A (2014{\natexlab{b}}) {{Hall} effect in neutron
  star crusts: evolution, endpoint and dependence on initial conditions}.
  \mnras 438:1618--1629, \doi{10.1093/mnras/stt2300}, \eprint{1311.7004}

\bibitem[{{Gourgouliatos} and {Cumming}(2015)}]{gourgouliatos15a}
{Gourgouliatos} KN, {Cumming} A (2015) {{Hall} drift and the braking indices of
  young pulsars}. \mnras 446:1121--1128, \doi{10.1093/mnras/stu2140},
  \eprint{1406.3640}

\bibitem[{{Gourgouliatos} and {Hollerbach}(2018)}]{gourgouliatos18}
{Gourgouliatos} KN, {Hollerbach} R (2018) {Magnetic Axis Drift and Magnetic
  Spot Formation in Neutron Stars with Toroidal Fields}. \apj 852:21,
  \doi{10.3847/1538-4357/aa9d93}, \eprint{1710.01338}

\bibitem[{{Gourgouliatos} et~al(2013){Gourgouliatos}, {Cumming}, {Reisenegger},
  {Armaza}, {Lyutikov}, and {Valdivia}}]{gourgouliatos13}
{Gourgouliatos} KN, {Cumming} A, {Reisenegger} A, {Armaza} C, {Lyutikov} M,
  {Valdivia} JA (2013) {{Hall} equilibria with toroidal and poloidal fields:
  application to neutron stars}. \mnras 434:2480--2490,
  \doi{10.1093/mnras/stt1195}, \eprint{1305.6269}

\bibitem[{{Gourgouliatos} et~al(2015){Gourgouliatos}, {Kondi{\'c}}, {Lyutikov},
  and {Hollerbach}}]{gourgouliatos15b}
{Gourgouliatos} KN, {Kondi{\'c}} T, {Lyutikov} M, {Hollerbach} R (2015)
  {Magnetar activity via the density-shear instability in {Hall}-{MHD}}. \mnras
  453:L93--L97, \doi{10.1093/mnrasl/slv106}, \eprint{1507.07454}

\bibitem[{{Gourgouliatos} et~al(2016){Gourgouliatos}, {Wood}, and
  {Hollerbach}}]{gourgouliatos16}
{Gourgouliatos} KN, {Wood} TS, {Hollerbach} R (2016) {Magnetic field evolution
  in magnetar crusts through three-dimensional simulations}. Proc Nat Acad Sci
  USA 113:3944--3949, \doi{10.1073/pnas.1522363113}, \eprint{1604.01399}

\bibitem[{{Graber} et~al(2015){Graber}, {Andersson}, {Glampedakis}, and
  {Lander}}]{Graber2015}
{Graber} V, {Andersson} N, {Glampedakis} K, {Lander} SK (2015) {Magnetic field
  evolution in superconducting neutron stars}. \mnras 453:671--681,
  \doi{10.1093/mnras/stv1648}, \eprint{1505.00124}

\bibitem[{{Guilet} et~al(2017){Guilet}, {M{\"u}ller}, {Janka}, {Rembiasz},
  {Obergaulinger}, {Cerd{\'a}-Dur{\'a}n}, and {Aloy}}]{guilet2017}
{Guilet} J, {M{\"u}ller} E, {Janka} HT, {Rembiasz} T, {Obergaulinger} M,
  {Cerd{\'a}-Dur{\'a}n} P, {Aloy} MA (2017) {How to form a millisecond
  magnetar? Magnetic field amplification in protoneutron stars}. In:
  {Marcowith} A, {Renaud} M, {Dubner} G, {Ray} A, {Bykov} A (eds) Supernova
  1987A:30 years later - Cosmic Rays and Nuclei from Supernovae and their
  Aftermaths, IAU Symposium, vol 331, pp 119--124,
  \doi{10.1017/S1743921317004732}, \eprint{1706.08733}

\bibitem[{{Gusakov} et~al(2017){Gusakov}, {Kantor}, and {Ofengeim}}]{gusakov17}
{Gusakov} ME, {Kantor} EM, {Ofengeim} DD (2017) {Evolution of the magnetic
  field in neutron stars}. \prd 96:103012, \doi{10.1103/PhysRevD.96.103012},
  \eprint{1705.00508}

\bibitem[{{Haberl}(2007)}]{haberl}
{Haberl} F (2007) {The magnificent seven: magnetic fields and surface
  temperature distributions}. \apss 308:181--190,
  \doi{10.1007/s10509-007-9342-x}, \eprint{astro-ph/0609066}

\bibitem[{{Haensel} et~al(2007){Haensel}, {Potekhin}, and
  {Yakovlev}}]{2007ASSL..326.....H}
{Haensel} P, {Potekhin} AY, {Yakovlev} DG (2007) {Neutron Stars 1: Equation of
  State and Structure}, Astrophysics and Space Science Library, vol 326.
  Springer, New York, \doi{10.1007/978-0-387-47301-7}

\bibitem[{{Heinke} and {Ho}(2010)}]{2010ApJ...719L.167H}
{Heinke} CO, {Ho} WCG (2010) {Direct Observation of the Cooling of the
  Cassiopeia A Neutron Star}. \apjl 719:L167--L171,
  \doi{10.1088/2041-8205/719/2/L167}, \eprint{1007.4719}

\bibitem[{{Helliwell}(1965)}]{helliwell65}
{Helliwell} RA (1965) {Whistlers and Related Ionospheric Phenomena}. Stanford
  University Press, Stanford, CA

\bibitem[{{Ho} et~al(2012){Ho}, {Glampedakis}, and {Andersson}}]{Ho2012}
{Ho} WCG, {Glampedakis} K, {Andersson} N (2012) {Magnetars: super(ficially) hot
  and super(fluid) cool}. \mnras 422:2632--2641,
  \doi{10.1111/j.1365-2966.2012.20826.x}, \eprint{1112.1415}

\bibitem[{{Ho} et~al(2015){Ho}, {Elshamouty}, {Heinke}, and
  {Potekhin}}]{2015PhRvC..91a5806H}
{Ho} WCG, {Elshamouty} KG, {Heinke} CO, {Potekhin} AY (2015) {Tests of the
  nuclear equation of state and superfluid and superconducting gaps using the
  Cassiopeia A neutron star}. \prc 91:015806, \doi{10.1103/PhysRevC.91.015806},
  \eprint{1412.7759}

\bibitem[{{Hollerbach}(2000)}]{Hollerbach2000}
{Hollerbach} R (2000) {A spectral solution of the magneto-convection equations
  in spherical geometry}. International Journal for Numerical Methods in Fluids
  32:773--797,
  \doi{10.1002/(SICI)1097-0363(20000415)32:7<773::AID-FLD988>3.0.CO;2-P}

\bibitem[{{Hollerbach} and {R{\"u}diger}(2002)}]{HR2002}
{Hollerbach} R, {R{\"u}diger} G (2002) {The influence of {Hall} drift on the
  magnetic fields of neutron stars}. \mnras 337:216--224,
  \doi{10.1046/j.1365-8711.2002.05905.x}, \eprint{astro-ph/0208312}

\bibitem[{{Hollerbach} and {R{\"u}diger}(2004)}]{HR2004}
{Hollerbach} R, {R{\"u}diger} G (2004) {{Hall} drift in the stratified crusts
  of neutron stars}. \mnras 347:1273--1278,
  \doi{10.1111/j.1365-2966.2004.07307.x}

\bibitem[{Hornung and Kohn(2002)}]{hornung02}
Hornung RD, Kohn SR (2002) Managing application complexity in the {SAMRAI}
  object-oriented framework. Concurrency and Computation: Practice and
  Experience 14:347--368, \doi{10.1002/cpe.652}

\bibitem[{{Horowitz} and {Kadau}(2009)}]{horowitz09}
{Horowitz} CJ, {Kadau} K (2009) {Breaking Strain of Neutron Star Crust and
  Gravitational Waves}. \prl 102:191102, \doi{10.1103/PhysRevLett.102.191102},
  \eprint{0904.1986}

\bibitem[{{Hoyos} et~al(2008){Hoyos}, {Reisenegger}, and
  {Valdivia}}]{2008A&A...487..789H}
{Hoyos} J, {Reisenegger} A, {Valdivia} JA (2008) {Magnetic field evolution in
  neutron stars: one-dimensional multi-fluid model}. \aap 487:789--803,
  \doi{10.1051/0004-6361:200809466}, \eprint{0801.4372}

\bibitem[{{Hoyos} et~al(2010){Hoyos}, {Reisenegger}, and
  {Valdivia}}]{2010MNRAS.408.1730H}
{Hoyos} JH, {Reisenegger} A, {Valdivia} JA (2010) {Asymptotic, non-linear
  solutions for ambipolar diffusion in one dimension}. \mnras 408:1730--1741,
  \doi{10.1111/j.1365-2966.2010.17237.x}, \eprint{1003.5262}

\bibitem[{{Huba}(2003)}]{huba03}
{Huba} JD (2003) {{Hall} Magnetohydrodynamics - A Tutorial}. In: {B{\"u}chner
  J, Dum C \& Scholer M} (ed) Space Plasma Simulation, Lecture Notes in
  Physics, Berlin Springer Verlag, vol 615, pp 166--192

\bibitem[{{Hurley} et~al(1999){Hurley}, {Cline}, {Mazets}, {Barthelmy},
  {Butterworth}, {Marshall}, {Palmer}, {Aptekar}, {Golenetskii}, {Il'Inskii},
  {Frederiks}, {McTiernan}, {Gold}, and {Trombka}}]{hurley99}
{Hurley} K, {Cline} T, {Mazets} E, {Barthelmy} S, {Butterworth} P, {Marshall}
  F, {Palmer} D, {Aptekar} R, {Golenetskii} S, {Il'Inskii} V, {Frederiks} D,
  {McTiernan} J, {Gold} R, {Trombka} J (1999) {A giant periodic flare from the
  soft {$\gamma$}-ray repeater SGR1900+14}. \nat 397:41--43,
  \doi{10.1038/16199}, \eprint{astro-ph/9811443}

\bibitem[{{Jackson}(1991)}]{jackson91}
{Jackson} JD (1991) {Classical Electrodynamics}. Wiley, New Jersey, USA

\bibitem[{Jiang and Shu(1996)}]{jiang96}
Jiang GS, Shu CW (1996) Efficient implementation of weighted eno schemes. J
  Comput Phys 126:202 -- 228, \doi{https://doi.org/10.1006/jcph.1996.0130}

\bibitem[{{Johnston} and {Karastergiou}(2017)}]{johnston17}
{Johnston} S, {Karastergiou} A (2017) {Pulsar braking and the
  {$P$}--{$\dot{P}$} diagram}. \mnras 467:3493--3499,
  \doi{10.1093/mnras/stx377}, \eprint{1702.03616}

\bibitem[{{Jones}(1988)}]{jones1988}
{Jones} PB (1988) {Neutron star magnetic field decay - {Hall} drift and Ohmic
  diffusion}. \mnras 233:875--885, \doi{10.1093/mnras/233.4.875}

\bibitem[{{Kaminker} et~al(2014){Kaminker}, {Kaurov}, {Potekhin}, and
  {Yakovlev}}]{2014MNRAS.442.3484K}
{Kaminker} AD, {Kaurov} AA, {Potekhin} AY, {Yakovlev} DG (2014) {Thermal
  emission of neutron stars with internal heaters}. \mnras 442:3484--3494,
  \doi{10.1093/mnras/stu1102}, \eprint{1406.0723}

\bibitem[{{Kaplan} et~al(2011){Kaplan}, {Kamble}, {van Kerkwijk}, and
  {Ho}}]{2011ApJ...736..117K}
{Kaplan} DL, {Kamble} A, {van Kerkwijk} MH, {Ho} WCG (2011) {New
  Optical/Ultraviolet Counterparts and the Spectral Energy Distributions of
  Nearby, Thermally Emitting, Isolated Neutron Stars}. \apj 736:117,
  \doi{10.1088/0004-637X/736/2/117}, \eprint{1105.4178}

\bibitem[{{Karageorgopoulos} et~al(2019){Karageorgopoulos}, {Gourgouliatos},
  and {Contopoulos}}]{karageorgopoulos19}
{Karageorgopoulos} V, {Gourgouliatos} KN, {Contopoulos} I (2019) {Current
  closure through the neutron star crust}. \mnras 487:3333--3341,
  \doi{10.1093/mnras/stz1507}, \eprint{1903.05093}

\bibitem[{{Kaspi} and {Beloborodov}(2017)}]{2017ARA&A..55..261K}
{Kaspi} VM, {Beloborodov} AM (2017) {Magnetars}. \araa 55:261--301,
  \doi{10.1146/annurev-astro-081915-023329}, \eprint{1703.00068}

\bibitem[{{Keil} and {Janka}(1995)}]{keil1995}
{Keil} W, {Janka} HT (1995) {Hadronic phase transitions at supranuclear
  densities and the delayed collapse of newly formed neutron stars.} \aap
  296:145

\bibitem[{{Kojima}(2017)}]{kojima17}
{Kojima} Y (2017) {Axisymmetric force-free magnetosphere in the exterior of a
  neutron star}. \mnras 468:2011--2016, \doi{10.1093/mnras/stx584},
  \eprint{1703.02273}

\bibitem[{{Kondi{\'c}} et~al(2011){Kondi{\'c}}, {R{\"u}diger}, and
  {Hollerbach}}]{Kondi2011}
{Kondi{\'c}} T, {R{\"u}diger} G, {Hollerbach} R (2011) {The shear-{Hall}
  instability in newborn neutron stars}. \aap 535:L2,
  \doi{10.1051/0004-6361/201116776}, \eprint{1110.3937}

\bibitem[{{Konenkov} and {Geppert}(2000)}]{konenkov2000}
{Konenkov} D, {Geppert} U (2000) {The effect of the neutron-star crust on the
  evolution of a core magnetic field}. \mnras 313:66--72,
  \doi{10.1046/j.1365-8711.2000.03188.x}, \eprint{astro-ph/9910492}

\bibitem[{{Koto}(2008)}]{IMEX}
{Koto} T (2008) {IMEX Runge-Kutta schemes for reaction-diffusion equations}.
  Journal of Computational and Applied Mathematics 215:182--195

\bibitem[{{Lander}(2016)}]{lander16}
{Lander} SK (2016) {Magnetar Field Evolution and Crustal Plasticity}. \apjl
  824:L21, \doi{10.3847/2041-8205/824/2/L21}, \eprint{1604.02972}

\bibitem[{{Lander} and {Gourgouliatos}(2019)}]{lander19}
{Lander} SK, {Gourgouliatos} KN (2019) {Magnetic-field evolution in a
  plastically-failing neutron-star crust}. \mnras \doi{10.1093/mnras/stz1042},
  \eprint{1902.02121}

\bibitem[{{Lander} et~al(2015){Lander}, {Andersson}, {Antonopoulou}, and
  {Watts}}]{lander15}
{Lander} SK, {Andersson} N, {Antonopoulou} D, {Watts} AL (2015) {Magnetically
  driven crustquakes in neutron stars}. \mnras 449:2047--2058,
  \doi{10.1093/mnras/stv432}, \eprint{1412.5852}

\bibitem[{{Lyutikov} and {Gavriil}(2006)}]{lyutikov06}
{Lyutikov} M, {Gavriil} FP (2006) {Resonant cyclotron scattering and
  Comptonization in neutron star magnetospheres}. \mnras 368:690--706,
  \doi{10.1111/j.1365-2966.2006.10140.x}, \eprint{astro-ph/0507557}

\bibitem[{{Marchant} et~al(2014){Marchant}, {Reisenegger}, {Alejandro
  Valdivia}, and {Hoyos}}]{marchant14}
{Marchant} P, {Reisenegger} A, {Alejandro Valdivia} J, {Hoyos} JH (2014)
  {Stability of {Hall} Equilibria in Neutron Star Crusts}. \apj 796:94,
  \doi{10.1088/0004-637X/796/2/94}, \eprint{1410.5833}

\bibitem[{{Margalit} and {Metzger}(2017)}]{margalit17}
{Margalit} B, {Metzger} BD (2017) {Constraining the Maximum Mass of Neutron
  Stars from Multi-messenger Observations of GW170817}. \apjl 850:L19,
  \doi{10.3847/2041-8213/aa991c}, \eprint{1710.05938}

\bibitem[{{Mart{\'\i}} and {M{\"u}ller}(2015)}]{marti2015}
{Mart{\'\i}} JM, {M{\"u}ller} E (2015) {Grid-based Methods in Relativistic
  Hydrodynamics and Magnetohydrodynamics}. Living Rev Comput Astrophys 1:3,
  \doi{10.1007/lrca-2015-3}

\bibitem[{{Mereghetti} et~al(2015){Mereghetti}, {Pons}, and
  {Melatos}}]{2015SSRv..191..315M}
{Mereghetti} S, {Pons} JA, {Melatos} A (2015) {Magnetars: Properties, Origin
  and Evolution}. \ssr 191:315--338, \doi{10.1007/s11214-015-0146-y},
  \eprint{1503.06313}

\bibitem[{{M{\"o}sta} et~al(2015){M{\"o}sta}, {Ott}, {Radice}, {Roberts},
  {Schnetter}, and {Haas}}]{mosta15}
{M{\"o}sta} P, {Ott} CD, {Radice} D, {Roberts} LF, {Schnetter} E, {Haas} R
  (2015) {A large-scale dynamo and magnetoturbulence in rapidly rotating
  core-collapse supernovae}. \nat 528:376--379, \doi{10.1038/nature15755},
  \eprint{1512.00838}

\bibitem[{{Muslimov} and {Tsygan}(1985)}]{muslimov1985}
{Muslimov} AG, {Tsygan} AI (1985) {Vortex lines in neutron star superfluids and
  decay of pulsar magnetic fields}. \apss 115:43, \doi{10.1007/BF00653825}

\bibitem[{{Nunn}(1974)}]{nunn74}
{Nunn} D (1974) {A self-consistent theory of triggered VLF emissions}. Planss
  22:349--378, \doi{10.1016/0032-0633(74)90070-1}

\bibitem[{{Obergaulinger} et~al(2015){Obergaulinger}, {Janka}, and
  {Aloy}}]{ober2015}
{Obergaulinger} M, {Janka} HT, {Aloy} MA (2015) {Magnetic Field Amplification
  in Non-Rotating Stellar Core Collapse}. In: {Pogorelov} NV, {Audit} E, {Zank}
  GP (eds) Numerical Modeling of Space Plasma Flows ASTRONUM-2014, Astronomical
  Society of the Pacific Conference Series, vol 498, p 115

\bibitem[{{Ofengeim} and {Gusakov}(2018)}]{ofengeim18}
{Ofengeim} DD, {Gusakov} ME (2018) {Fast magnetic field evolution in neutron
  stars: The key role of magnetically induced fluid motions in the core}. \prd
  98:043007, \doi{10.1103/PhysRevD.98.043007}, \eprint{1805.03956}

\bibitem[{{O'Sullivan} and {Downes}(2006)}]{osullivan06}
{O'Sullivan} S, {Downes} TP (2006) {An explicit scheme for multifluid
  magnetohydrodynamics}. \mnras 366:1329--1336,
  \doi{10.1111/j.1365-2966.2005.09898.x}, \eprint{arXiv:astro-ph/0511478}

\bibitem[{{Page}(2009)}]{2009ASSL..357..247P}
{Page} D (2009) {Neutron Star Cooling: I}. In: {Becker} W (ed) Neutron Stars
  and Pulsars, Springer, Berlin, Heidelberg, Astrophysics and Space Science
  Library, vol 357, p 247, \doi{10.1007/978-3-540-76965-1}

\bibitem[{{Page} et~al(2004){Page}, {Lattimer}, {Prakash}, and
  {Steiner}}]{2004ApJS..155..623P}
{Page} D, {Lattimer} JM, {Prakash} M, {Steiner} AW (2004) {Minimal Cooling of
  Neutron Stars: A New Paradigm}. \apjs 155:623--650, \doi{10.1086/424844},
  \eprint{astro-ph/0403657}

\bibitem[{{Page} et~al(2007){Page}, {Geppert}, and
  {K{\"u}ker}}]{2007Ap&SS.308..403P}
{Page} D, {Geppert} U, {K{\"u}ker} M (2007) {Cooling of neutron stars with
  strong toroidal magnetic fields}. \apss 308:403--412,
  \doi{10.1007/s10509-007-9316-z}, \eprint{astro-ph/0701442}

\bibitem[{{Page} et~al(2011){Page}, {Prakash}, {Lattimer}, and
  {Steiner}}]{2011PhRvL.106h1101P}
{Page} D, {Prakash} M, {Lattimer} JM, {Steiner} AW (2011) {Rapid Cooling of the
  Neutron Star in {Cassiopeia A} Triggered by Neutron Superfluidity in Dense
  Matter}. Phys Rev Lett 106:081101, \doi{10.1103/PhysRevLett.106.081101},
  \eprint{1011.6142}

\bibitem[{{Palenzuela}(2013)}]{palenzuela13}
{Palenzuela} C (2013) {Modelling magnetized neutron stars using resistive
  magnetohydrodynamics}. \mnras 431(2):1853--1865, \doi{10.1093/mnras/stt311},
  \eprint{1212.0130}

\bibitem[{{Palmer} et~al(2005){Palmer}, {Barthelmy}, {Gehrels}, {Kippen},
  {Cayton}, {Kouveliotou}, {Eichler}, {Wijers}, {Woods}, {Granot}, {Lyubarsky},
  {Ramirez-Ruiz}, {Barbier}, {Chester}, {Cummings}, {Fenimore}, {Finger},
  {Gaensler}, {Hullinger}, {Krimm}, {Markwardt}, {Nousek}, {Parsons}, {Patel},
  {Sakamoto}, {Sato}, {Suzuki}, and {Tueller}}]{palmer05}
{Palmer} DM, {Barthelmy} S, {Gehrels} N, {Kippen} RM, {Cayton} T, {Kouveliotou}
  C, {Eichler} D, {Wijers} RAMJ, {Woods} PM, {Granot} J, {Lyubarsky} YE,
  {Ramirez-Ruiz} E, {Barbier} L, {Chester} M, {Cummings} J, {Fenimore} EE,
  {Finger} MH, {Gaensler} BM, {Hullinger} D, {Krimm} H, {Markwardt} CB,
  {Nousek} JA, {Parsons} A, {Patel} S, {Sakamoto} T, {Sato} G, {Suzuki} M,
  {Tueller} J (2005) {A giant {$\gamma$}-ray flare from the magnetar SGR 1806 -
  20}. \nat 434:1107--1109, \doi{10.1038/nature03525},
  \eprint{astro-ph/0503030}

\bibitem[{{Parfrey} et~al(2013){Parfrey}, {Beloborodov}, and {Hui}}]{parfrey13}
{Parfrey} K, {Beloborodov} AM, {Hui} L (2013) {Dynamics of Strongly Twisted
  Relativistic Magnetospheres}. \apj 774:92, \doi{10.1088/0004-637X/774/2/92},
  \eprint{1306.4335}

\bibitem[{{Passamonti} et~al(2017{\natexlab{a}}){Passamonti}, {Akg{\"u}n},
  {Pons}, and {Miralles}}]{passamonti17a}
{Passamonti} A, {Akg{\"u}n} T, {Pons} JA, {Miralles} JA (2017{\natexlab{a}})
  {On the magnetic field evolution time-scale in superconducting neutron star
  cores}. \mnras 469:4979--4984, \doi{10.1093/mnras/stx1192},
  \eprint{1704.02016}

\bibitem[{{Passamonti} et~al(2017{\natexlab{b}}){Passamonti}, {Akg{\"u}n},
  {Pons}, and {Miralles}}]{passamonti17b}
{Passamonti} A, {Akg{\"u}n} T, {Pons} JA, {Miralles} JA (2017{\natexlab{b}})
  {The relevance of ambipolar diffusion for neutron star evolution}. \mnras
  465:3416--3428, \doi{10.1093/mnras/stw2936}, \eprint{1608.00001}

\bibitem[{{Pattle}(1959)}]{pattle59}
{Pattle} RE (1959) {Diffusion from an instantaneous point source with a
  concentration-dependent coefficient}. Quart J Mech Appl Math 12

\bibitem[{{P{\'e}rez-Azor{\'{\i}}n} et~al(2005){P{\'e}rez-Azor{\'{\i}}n},
  {Miralles}, and {Pons}}]{perez05}
{P{\'e}rez-Azor{\'{\i}}n} JF, {Miralles} JA, {Pons} JA (2005) {Thermal
  radiation from magnetic neutron star surfaces}. \aap 433:275--283,
  \doi{10.1051/0004-6361:20041612}, \eprint{astro-ph/0410664}

\bibitem[{{P{\'e}rez-Azor{\'{\i}}n} et~al(2006){P{\'e}rez-Azor{\'{\i}}n},
  {Miralles}, and {Pons}}]{perez06}
{P{\'e}rez-Azor{\'{\i}}n} JF, {Miralles} JA, {Pons} JA (2006) {Anisotropic
  thermal emission from magnetized neutron stars}. \aap 451:1009--1024,
  \doi{10.1051/0004-6361:20054403}, \eprint{astro-ph/0510684}

\bibitem[{{Perna} and {Pons}(2011)}]{perna11}
{Perna} R, {Pons} JA (2011) {A Unified Model of the Magnetar and Radio Pulsar
  Bursting Phenomenology}. \apjl 727:L51, \doi{10.1088/2041-8205/727/2/L51},
  \eprint{1101.1098}

\bibitem[{{Philippov} et~al(2014){Philippov}, {Tchekhovskoy}, and
  {Li}}]{philippov14}
{Philippov} A, {Tchekhovskoy} A, {Li} JG (2014) {Time evolution of pulsar
  obliquity angle from 3D simulations of magnetospheres}. \mnras
  441:1879--1887, \doi{10.1093/mnras/stu591}, \eprint{1311.1513}

\bibitem[{{Pili} et~al(2015){Pili}, {Bucciantini}, and {Del Zanna}}]{pili15}
{Pili} AG, {Bucciantini} N, {Del Zanna} L (2015) {General relativistic neutron
  stars with twisted magnetosphere}. \mnras 447:2821--2835,
  \doi{10.1093/mnras/stu2628}, \eprint{1412.4036}

\bibitem[{{Pons} and {Geppert}(2007)}]{PonsGeppert2007}
{Pons} JA, {Geppert} U (2007) {Magnetic field dissipation in neutron star
  crusts: from magnetars to isolated neutron stars}. \aap 470:303--315,
  \doi{10.1051/0004-6361:20077456}, \eprint{arXiv:astro-ph/0703267}

\bibitem[{{Pons} and {Geppert}(2010)}]{Pons2010}
{Pons} JA, {Geppert} U (2010) {Confirmation of the occurrence of the {Hall}
  instability in the non-linear regime}. \aap 513:L12,
  \doi{10.1051/0004-6361/201014197}, \eprint{1004.1054}

\bibitem[{{Pons} and {Perna}(2011)}]{pons11}
{Pons} JA, {Perna} R (2011) {Magnetars versus High Magnetic Field Pulsars: A
  Theoretical Interpretation of the Apparent Dichotomy}. \apj 741:123,
  \doi{10.1088/0004-637X/741/2/123}, \eprint{1109.5184}

\bibitem[{{Pons} et~al(1999){Pons}, {Reddy}, {Prakash}, {Lattimer}, and
  {Miralles}}]{pons99}
{Pons} JA, {Reddy} S, {Prakash} M, {Lattimer} JM, {Miralles} JA (1999)
  {Evolution of Proto-Neutron Stars}. \apj 513:780--804, \doi{10.1086/306889},
  \eprint{arXiv:astro-ph/9807040}

\bibitem[{{Pons} et~al(2009{\natexlab{a}}){Pons}, {Miralles}, and
  {Geppert}}]{Pons2009}
{Pons} JA, {Miralles} JA, {Geppert} U (2009{\natexlab{a}}) {Magneto-thermal
  evolution of neutron stars}. \aap 496:207--216,
  \doi{10.1051/0004-6361:200811229}, \eprint{0812.3018}

\bibitem[{{Pons} et~al(2009{\natexlab{b}}){Pons}, {Miralles}, and
  {Geppert}}]{pons09}
{Pons} JA, {Miralles} JA, {Geppert} U (2009{\natexlab{b}}) {Magneto-thermal
  evolution of neutron stars}. \aap 496:207--216,
  \doi{10.1051/0004-6361:200811229}, \eprint{0812.3018}

\bibitem[{{Pons} et~al(2013){Pons}, {Vigan{\`o}}, and {Rea}}]{pons13}
{Pons} JA, {Vigan{\`o}} D, {Rea} N (2013) {A highly resistive layer within the
  crust of X-ray pulsars limits their spin periods}. Nature Phys 9:431--434,
  \doi{10.1038/nphys2640}, \eprint{1304.6546}

\bibitem[{{Posselt} and {Pavlov}(2018)}]{posselt18}
{Posselt} B, {Pavlov} GG (2018) {Upper Limits on the Rapid Cooling of the
  Central Compact Object in {Cas A}}. \apj 864:135,
  \doi{10.3847/1538-4357/aad7fc}, \eprint{1808.00531}

\bibitem[{{Posselt} et~al(2007){Posselt}, {Popov}, {Haberl}, {Tr{\"u}mper},
  {Turolla}, and {Neuh{\"a}user}}]{2007Ap&SS.308..171P}
{Posselt} B, {Popov} SB, {Haberl} F, {Tr{\"u}mper} J, {Turolla} R,
  {Neuh{\"a}user} R (2007) {The {Magnificent Seven} in the dusty prairie}.
  \apss 308:171--179, \doi{10.1007/s10509-007-9344-8},
  \eprint{astro-ph/0609275}

\bibitem[{{Potekhin} and {Chabrier}(2010)}]{potekhin10}
{Potekhin} AY, {Chabrier} G (2010) {Thermodynamic Functions of Dense Plasmas:
  Analytic Approximations for Astrophysical Applications}. Contrib Plasma Phys
  50:82--87, \doi{10.1002/ctpp.201010017}, \eprint{1001.0690}

\bibitem[{{Potekhin} and {Chabrier}(2018)}]{potekhin2018}
{Potekhin} AY, {Chabrier} G (2018) {Magnetic neutron star cooling and
  microphysics}. \aap 609:A74, \doi{10.1051/0004-6361/201731866},
  \eprint{1711.07662}

\bibitem[{{Potekhin} et~al(2012){Potekhin}, {Suleimanov}, {van Adelsberg}, and
  {Werner}}]{potekhin12}
{Potekhin} AY, {Suleimanov} VF, {van Adelsberg} M, {Werner} K (2012) {Radiative
  properties of magnetic neutron stars with metallic surfaces and thin
  atmospheres}. \aap 546:A121, \doi{10.1051/0004-6361/201219747},
  \eprint{1208.6582}

\bibitem[{{Potekhin} et~al(2015{\natexlab{a}}){Potekhin}, {De Luca}, and
  {Pons}}]{potekhin_rev15b}
{Potekhin} AY, {De Luca} A, {Pons} JA (2015{\natexlab{a}}) {Neutron Stars --
  Thermal Emitters}. \ssr 191:171--206, \doi{10.1007/s11214-014-0102-2},
  \eprint{1409.7666}

\bibitem[{{Potekhin} et~al(2015{\natexlab{b}}){Potekhin}, {Pons}, and
  {Page}}]{potekhin_rev15a}
{Potekhin} AY, {Pons} JA, {Page} D (2015{\natexlab{b}}) {Neutron Stars --
  Cooling and Transport}. \ssr 191:239--291, \doi{10.1007/s11214-015-0180-9},
  \eprint{1507.06186}

\bibitem[{Press et~al(2007)Press, Teukolsky, Vetterling, and Flannery}]{NumRec}
Press WH, Teukolsky SA, Vetterling WT, Flannery BP (2007) Numerical Recipes:
  The Art of Scientific Computing, 3rd edn. Cambridge University Press

\bibitem[{{Radice} et~al(2018){Radice}, {Perego}, {Zappa}, and
  {Bernuzzi}}]{radice18}
{Radice} D, {Perego} A, {Zappa} F, {Bernuzzi} S (2018) {GW170817: Joint
  Constraint on the Neutron Star Equation of State from Multimessenger
  Observations}. \apjl 852:L29, \doi{10.3847/2041-8213/aaa402},
  \eprint{1711.03647}

\bibitem[{{R{\"a}dler} et~al(2001){R{\"a}dler}, {Fuchs}, {Geppert},
  {Rheinhardt}, and {Zannias}}]{radler2001}
{R{\"a}dler} KH, {Fuchs} H, {Geppert} U, {Rheinhardt} M, {Zannias} T (2001)
  {General-relativistic free decay of magnetic fields in a spherically
  symmetric body}. \prd 64:083008, \doi{10.1103/PhysRevD.64.083008}

\bibitem[{{Rea} and {Esposito}(2011)}]{2011ASSP...21..247R}
{Rea} N, {Esposito} P (2011) {Magnetar outbursts: an observational review}.
  Astrophysics and Space Science Proceedings 21:247,
  \doi{10.1007/978-3-642-17251-9-21}, \eprint{1101.4472}

\bibitem[{{Rea} et~al(2008){Rea}, {Zane}, {Turolla}, {Lyutikov}, and
  {G{\"o}tz}}]{rea08}
{Rea} N, {Zane} S, {Turolla} R, {Lyutikov} M, {G{\"o}tz} D (2008) {Resonant
  Cyclotron Scattering in Magnetars' Emission}. \apj 686:1245--1260,
  \doi{10.1086/591264}, \eprint{0802.1923}

\bibitem[{{Rea} et~al(2010){Rea}, {Esposito}, {Turolla}, {Israel}, {Zane},
  {Stella}, {Mereghetti}, {Tiengo}, {G{\"o}tz}, {G{\"o}{\u g}{\"u}{\c s}}, and
  {Kouveliotou}}]{2010Sci...330..944R}
{Rea} N, {Esposito} P, {Turolla} R, {Israel} GL, {Zane} S, {Stella} L,
  {Mereghetti} S, {Tiengo} A, {G{\"o}tz} D, {G{\"o}{\u g}{\"u}{\c s}} E,
  {Kouveliotou} C (2010) {A Low-Magnetic-Field Soft Gamma Repeater}. Science
  330:944, \doi{10.1126/science.1196088}, \eprint{1010.2781}

\bibitem[{{Rea} et~al(2012){Rea}, {Israel}, {Esposito}, {Pons},
  {Camero-Arranz}, {Mignani}, {Turolla}, {Zane}, {Burgay}, {Possenti},
  {Campana}, {Enoto}, {Gehrels}, {G{\"o}{\v g}{\"u}{\c s}}, {G{\"o}tz},
  {Kouveliotou}, {Makishima}, {Mereghetti}, {Oates}, {Palmer}, {Perna},
  {Stella}, and {Tiengo}}]{2012ApJ...754...27R}
{Rea} N, {Israel} GL, {Esposito} P, {Pons} JA, {Camero-Arranz} A, {Mignani} RP,
  {Turolla} R, {Zane} S, {Burgay} M, {Possenti} A, {Campana} S, {Enoto} T,
  {Gehrels} N, {G{\"o}{\v g}{\"u}{\c s}} E, {G{\"o}tz} D, {Kouveliotou} C,
  {Makishima} K, {Mereghetti} S, {Oates} SR, {Palmer} DM, {Perna} R, {Stella}
  L, {Tiengo} A (2012) {A New Low Magnetic Field Magnetar: The 2011 Outburst of
  Swift J1822.3-1606}. \apj 754:27, \doi{10.1088/0004-637X/754/1/27},
  \eprint{1203.6449}

\bibitem[{{Rea} et~al(2013){Rea}, {Israel}, {Pons}, {Turolla}, {Vigan{\`o}},
  {Zane}, {Esposito}, {Perna}, {Papitto}, {Terreran}, {Tiengo}, {Salvetti},
  {Girart}, {Palau}, {Possenti}, {Burgay}, {G{\"o}{\u g}{\"u}{\c s}},
  {Caliandro}, {Kouveliotou}, {G{\"o}tz}, {Mignani}, {Ratti}, and
  {Stella}}]{2013ApJ...770...65R}
{Rea} N, {Israel} GL, {Pons} JA, {Turolla} R, {Vigan{\`o}} D, {Zane} S,
  {Esposito} P, {Perna} R, {Papitto} A, {Terreran} G, {Tiengo} A, {Salvetti} D,
  {Girart} JM, {Palau} A, {Possenti} A, {Burgay} M, {G{\"o}{\u g}{\"u}{\c s}}
  E, {Caliandro} GA, {Kouveliotou} C, {G{\"o}tz} D, {Mignani} RP, {Ratti} E,
  {Stella} L (2013) {The Outburst Decay of the Low Magnetic Field Magnetar {SGR
  0418+5729}}. \apj 770:65, \doi{10.1088/0004-637X/770/1/65},
  \eprint{1303.5579}

\bibitem[{{Rea} et~al(2014){Rea}, {Vigan{\`o}}, {Israel}, {Pons}, and
  {Torres}}]{2014ApJ...781L..17R}
{Rea} N, {Vigan{\`o}} D, {Israel} GL, {Pons} JA, {Torres} DF (2014) {3XMM
  J185246.6+003317: Another Low Magnetic Field Magnetar}. \apjl 781:L17,
  \doi{10.1088/2041-8205/781/1/L17}, \eprint{1311.3091}

\bibitem[{{Rea} et~al(2016){Rea}, {Borghese}, {Esposito}, {Coti Zelati},
  {Bachetti}, {Israel}, and {De Luca}}]{rea16}
{Rea} N, {Borghese} A, {Esposito} P, {Coti Zelati} F, {Bachetti} M, {Israel}
  GL, {De Luca} A (2016) {Magnetar-like Activity from the Central Compact
  Object in the {SNR RCW103}}. \apj 828(1):L13,
  \doi{10.3847/2041-8205/828/1/L13}, \eprint{1607.04107}

\bibitem[{{Reisenegger} et~al(2007){Reisenegger}, {Benguria}, {Prieto},
  {Araya}, and {Lai}}]{Reise2007}
{Reisenegger} A, {Benguria} R, {Prieto} JP, {Araya} PA, {Lai} D (2007) {{Hall}
  drift of axisymmetric magnetic fields in solid neutron-star matter}. \aap
  472:233--240, \doi{10.1051/0004-6361:20077874}, \eprint{0705.1901}

\bibitem[{Richtmyer and Morton(1967)}]{bookRM67}
Richtmyer RD, Morton KW (1967) Difference methods for initial-value problems.
  Interscience Publishers

\bibitem[{{Roumeliotis} et~al(1994){Roumeliotis}, {Sturrock}, and
  {Antiochos}}]{roumeliotis94}
{Roumeliotis} G, {Sturrock} PA, {Antiochos} SK (1994) {A Numerical Study of the
  Sudden Eruption of Sheared Magnetic Fields}. \apj 423:847,
  \doi{10.1086/173862}

\bibitem[{{Ruiz} et~al(2018){Ruiz}, {Shapiro}, and {Tsokaros}}]{ruiz18}
{Ruiz} M, {Shapiro} SL, {Tsokaros} A (2018) {{GW170817}, general relativistic
  magnetohydrodynamic simulations, and the neutron star maximum mass}. \prd
  97(2):021501, \doi{10.1103/PhysRevD.97.021501}, \eprint{1711.00473}

\bibitem[{{Shalybkov} and {Urpin}(1995)}]{1995MNRAS.273..643S}
{Shalybkov} DA, {Urpin} VA (1995) {Ambipolar diffusion and anisotropy of
  resistivity in neutron star cores}. \mnras 273:643--648,
  \doi{10.1093/mnras/273.3.643}

\bibitem[{{Shternin} et~al(2011){Shternin}, {Yakovlev}, {Heinke}, {Ho}, and
  {Patnaude}}]{2011MNRAS.412L.108S}
{Shternin} PS, {Yakovlev} DG, {Heinke} CO, {Ho} WCG, {Patnaude} DJ (2011)
  {Cooling neutron star in the {Cassiopeia A} supernova remnant: evidence for
  superfluidity in the core}. \mnras 412:L108--L112,
  \doi{10.1111/j.1745-3933.2011.01015.x}, \eprint{1012.0045}

\bibitem[{Shu(1998)}]{shu98}
Shu CW (1998) Essentially non-oscillatory and weighted essentially
  non-oscillatory schemes for hyperbolic conservation laws. In: Quarteroni A
  (ed) Advanced Numerical Approximation of Nonlinear Hyperbolic Equations:
  Lectures given at the 2nd Session of the Centro Internazionale Matematico
  Estivo (C.I.M.E.) held in Cetraro, Italy, June 23--28, 1997, Springer,
  Berlin, Heidelberg, pp 325--432, \doi{10.1007/BFb0096355}

\bibitem[{{Spitkovsky}(2006)}]{spitkovsky06}
{Spitkovsky} A (2006) {Time-dependent Force-free Pulsar Magnetospheres:
  Axisymmetric and Oblique Rotators}. \apjl 648:L51--L54, \doi{10.1086/507518},
  \eprint{arXiv:astro-ph/0603147}

\bibitem[{Suresh and Huynh(1997)}]{suresh97}
Suresh A, Huynh H (1997) Accurate monotonicity-preserving schemes with
  runge--kutta time stepping. J Comput Phys 136:83 -- 99,
  \doi{https://doi.org/10.1006/jcph.1997.5745}

\bibitem[{Thomas(1949)}]{Thomas1949}
Thomas LH (1949) Elliptic problems in linear difference equations over a
  network. Watson Sci. Comput. Lab. Rept., Columbia University, New York

\bibitem[{{Thompson} and {Duncan}(1995)}]{1995MNRAS.275..255T}
{Thompson} C, {Duncan} RC (1995) {The soft gamma repeaters as very strongly
  magnetized neutron stars - I. Radiative mechanism for outbursts}. \mnras
  275:255--300

\bibitem[{{Thompson} and {Duncan}(1996)}]{1996ApJ...473..322T}
{Thompson} C, {Duncan} RC (1996) {The Soft Gamma Repeaters as Very Strongly
  Magnetized Neutron Stars. II. Quiescent Neutrino, X-Ray, and Alfven Wave
  Emission}. \apj 473:322, \doi{10.1086/178147}

\bibitem[{Toro(1997)}]{toro97}
Toro E (1997) Riemann Solvers and Numerical Methods for Fluid Dynamics: A
  Practical Introduction. Springer, Berlin, Heidelberg,
  \doi{10.1007/978-3-662-03490-3}

\bibitem[{{Toro}(2009)}]{toro09}
{Toro} EF (2009) {Riemann Solvers and Numerical Methods for Fluid Dynamics: A
  Practical Introduction}, 3rd edn. Springer, Berlin, Heidelberg,
  \doi{10.1007/b79761}

\bibitem[{{T{\'o}th}(2000)}]{toth00}
{T{\'o}th} G (2000) {The {$\nabla \cdot B=0$} Constraint in Shock-Capturing
  Magnetohydrodynamics Codes}. J Comput Phys 161:605--652,
  \doi{10.1006/jcph.2000.6519}

\bibitem[{{T{\'o}th} et~al(2008){T{\'o}th}, {Ma}, and {Gombosi}}]{Toth2008}
{T{\'o}th} G, {Ma} Y, {Gombosi} TI (2008) {{Hall} magnetohydrodynamics on
  block-adaptive grids}. \jcp 227:6967--6984, \doi{10.1016/j.jcp.2008.04.010}

\bibitem[{{Tsuruta}(1964)}]{tsuruta64}
{Tsuruta} S (1964) {Neutron star models}. PhD thesis, Columbia University

\bibitem[{{Tsuruta}(2009)}]{2009ASSL..357..289T}
{Tsuruta} S (2009) {Neutron Star Cooling: II}. In: {Becker} W (ed) Neutron
  Stars and Pulsars, Springer, Berlin, Heidelberg, Astrophysics and Space
  Science Library, vol 357, p 289, \doi{10.1007/978-3-540-76965-1}

\bibitem[{{Turolla} et~al(2004){Turolla}, {Zane}, and {Drake}}]{turolla04}
{Turolla} R, {Zane} S, {Drake} JJ (2004) {Bare Quark Stars or Naked Neutron
  Stars? The Case of RX J1856.5-3754}. \apj 603:265--282, \doi{10.1086/379113},
  \eprint{astro-ph/0308326}

\bibitem[{{Turolla} et~al(2015){Turolla}, {Zane}, and
  {Watts}}]{2015RPPh...78k6901T}
{Turolla} R, {Zane} S, {Watts} AL (2015) {Magnetars: the physics behind
  observations. A review}. Rep Progr Phys 78:116901,
  \doi{10.1088/0034-4885/78/11/116901}, \eprint{1507.02924}

\bibitem[{{Urpin} and {Yakovlev}(1980)}]{1980SvA....24..425U}
{Urpin} VA, {Yakovlev} DG (1980) {Thermogalvanomagnetic Effects in White Dwarfs
  and Neutron Stars}. \sovast 24:425

\bibitem[{{Vainshtein} et~al(2000){Vainshtein}, {Chitre}, and
  {Olinto}}]{Vai2000}
{Vainshtein} SI, {Chitre} SM, {Olinto} AV (2000) {Rapid dissipation of magnetic
  fields due to the {Hall} current}. \pre 61:4422--4430,
  \doi{10.1103/PhysRevE.61.4422}, \eprint{arXiv:astro-ph/9911386}

\bibitem[{{van Adelsberg} et~al(2005){van Adelsberg}, {Lai}, {Potekhin}, and
  {Arras}}]{vanadelsberg05}
{van Adelsberg} M, {Lai} D, {Potekhin} AY, {Arras} P (2005) {Radiation from
  Condensed Surface of Magnetic Neutron Stars}. \apj 628:902--913,
  \doi{10.1086/430871}, \eprint{astro-ph/0406001}

\bibitem[{{van Haarlem}(2013)}]{lofar13}
{van Haarlem} MPea (2013) {{LOFAR}: The {LOw-Frequency ARray}}. \aap 556:A2,
  \doi{10.1051/0004-6361/201220873}, \eprint{1305.3550}

\bibitem[{{van Leer}(1977)}]{vanleer77}
{van Leer} B (1977) {Towards the Ultimate Conservative Difference Scheme. {IV}.
  {A} New Approach to Numerical Convection}. J Comput Phys 23:276,
  \doi{10.1016/0021-9991(77)90095-X}

\bibitem[{{van Riper}(1991)}]{vanriper91}
{van Riper} KA (1991) {Neutron star thermal evolution}. \apjs 75:449--462,
  \doi{10.1086/191538}

\bibitem[{{Vigan{\`o}} et~al(2011){Vigan{\`o}}, {Pons}, and
  {Miralles}}]{vigano11}
{Vigan{\`o}} D, {Pons} JA, {Miralles} JA (2011) {Force-free twisted
  magnetospheres of neutron stars}. \aap 533:A125,
  \doi{10.1051/0004-6361/201117105}, \eprint{1106.5934}

\bibitem[{{Vigan{\`o}} et~al(2012){Vigan{\`o}}, {Pons}, and
  {Miralles}}]{vigano12a}
{Vigan{\`o}} D, {Pons} JA, {Miralles} JA (2012) {A new code for the
  {Hall}-driven magnetic evolution of neutron stars}. CoPhC 183:2042--2053,
  \doi{10.1016/j.cpc.2012.04.029}, \eprint{arXiv:astro-ph/1204.4707}

\bibitem[{{Vigan{\`o}} et~al(2013){Vigan{\`o}}, {Rea}, {Pons}, {Perna},
  {Aguilera}, and {Miralles}}]{vigano13}
{Vigan{\`o}} D, {Rea} N, {Pons} JA, {Perna} R, {Aguilera} DN, {Miralles} JA
  (2013) {Unifying the observational diversity of isolated neutron stars via
  magneto-thermal evolution models}. \mnras 434:123--141,
  \doi{10.1093/mnras/stt1008}, \eprint{1306.2156}

\bibitem[{{Vigan{\`o}} et~al(2015){Vigan{\`o}}, {Torres}, and
  {Mart{\'{\i}}n}}]{vigano15}
{Vigan{\`o}} D, {Torres} DF, {Mart{\'{\i}}n} J (2015) {A systematic
  synchro-curvature modelling of pulsar {$\gamma$}-ray spectra unveils hidden
  trends}. \mnras 453:2599--2621, \doi{10.1093/mnras/stv1582},
  \eprint{1507.04021}

\bibitem[{{Vigan{\`o}} et~al(2019){Vigan{\`o}}, {Mart{\'\i}nez-G{\'o}mez},
  {Pons}, {Palenzuela}, {Carrasco}, {Mi{\~n}ano}, {Arbona}, {Bona}, and
  {Mass{\'o}}}]{vigano19}
{Vigan{\`o}} D, {Mart{\'\i}nez-G{\'o}mez} D, {Pons} JA, {Palenzuela} C,
  {Carrasco} F, {Mi{\~n}ano} B, {Arbona} A, {Bona} C, {Mass{\'o}} J (2019) {A
  Simflowny-based high-performance 3D code for the generalized induction
  equation}. Computer Physics Communications 237:168--183,
  \doi{10.1016/j.cpc.2018.11.022}, \eprint{1811.08198}

\bibitem[{{Wiebicke} and {Geppert}(1991)}]{wiebicke91}
{Wiebicke} HJ, {Geppert} U (1991) {Amplification of neutron star magnetic
  fields by thermoelectric effects. II - Linear approximation}. \aap
  245:331--340

\bibitem[{{Wiebicke} and {Geppert}(1992)}]{wiebicke92}
{Wiebicke} HJ, {Geppert} U (1992) {Amplification of neutron star magnetic
  fields by thermoelectric effects. III - Growth limits in nonlinear
  calculations}. \aap 262:125--130

\bibitem[{{Wiebicke} and {Geppert}(1995)}]{wiebicke95}
{Wiebicke} HJ, {Geppert} U (1995) {Amplification of neutron star magnetic
  fields by thermoelectric effects. IV. Averaged small-scale modes and
  selection rules for large-scale modes.} \aap 294:303--312

\bibitem[{{Wiebicke} and {Geppert}(1996)}]{wiebicke96}
{Wiebicke} HJ, {Geppert} U (1996) {Amplification of neutron star magnetic
  fields by thermoelectric effects. {VI}. {A}nalytical approach}. \aap
  309:203--212

\bibitem[{{Wijngaarden} et~al(2019){Wijngaarden}, {Ho}, {Chang}, {Heinke},
  {Page}, {Beznogov}, and {Patnaude}}]{wijngaarden19}
{Wijngaarden} MJP, {Ho} WCG, {Chang} P, {Heinke} CO, {Page} D, {Beznogov} M,
  {Patnaude} DJ (2019) {Diffusive nuclear burning in cooling simulations and
  application to new temperature data of the Cassiopeia A neutron star}. \mnras
  484:974--988, \doi{10.1093/mnras/stz042}, \eprint{1901.01012}

\bibitem[{{Wood} and {Hollerbach}(2015)}]{wood15}
{Wood} TS, {Hollerbach} R (2015) {Three Dimensional Simulation of the Magnetic
  Stress in a Neutron Star Crust}. Phys Rev Lett 114(19):191101,
  \doi{10.1103/PhysRevLett.114.191101}, \eprint{1501.05149}

\bibitem[{{Yakovlev} and {Pethick}(2004)}]{2004ARA&A..42..169Y}
{Yakovlev} DG, {Pethick} CJ (2004) {Neutron Star Cooling}. \araa 42:169--210,
  \doi{10.1146/annurev.astro.42.053102.134013}, \eprint{astro-ph/0402143}

\bibitem[{{Yakovlev} and {Shalybkov}(1990)}]{1990SvAL...16...86Y}
{Yakovlev} DG, {Shalybkov} DA (1990) {Electrical Conductivity and Resistivity
  in Magnetized Cores of Neutron Stars}. Soviet Astronomy Letters 16:86

\bibitem[{{Yakovlev} et~al(2008){Yakovlev}, {Gnedin}, {Kaminker}, and
  {Potekhin}}]{2008AIPC..983..379Y}
{Yakovlev} DG, {Gnedin} OY, {Kaminker} AD, {Potekhin} AY (2008) {Theory of
  cooling neutron stars versus observations}. In: {Bassa} C, {Wang} Z,
  {Cumming} A, {Kaspi} VM (eds) 40 Years of Pulsars: Millisecond Pulsars,
  Magnetars and More, American Institute of Physics, AIP Conference Series, vol
  983, pp 379--387, \doi{10.1063/1.2900259}, \eprint{0710.2047}

\bibitem[{Yamaleev and Carpenter(2009)}]{yamaleev09}
Yamaleev NK, Carpenter MH (2009) Third-order energy stable {WENO} scheme. J
  Comput Phys 228:3025--3047, \doi{https://doi.org/10.1016/j.jcp.2009.01.011}

\bibitem[{{Yang} et~al(1986){Yang}, {Sturrock}, and {Antiochos}}]{yang86}
{Yang} WH, {Sturrock} PA, {Antiochos} SK (1986) {Force-free Magnetic Fields:
  The Magneto-frictional Method}. \apj 309:383, \doi{10.1086/164610}

\bibitem[{{Zhang} and {Cheng}(1997)}]{zhang97}
{Zhang} L, {Cheng} KS (1997) {High-Energy Radiation from Rapidly Spinning
  Pulsars with Thick Outer Gaps}. \apj 487:370, \doi{10.1086/304589}

\end{thebibliography}


\end{document}